\documentclass[preprint,12pt,sort&compress]{elsarticle}
\usepackage{amsmath}
\usepackage{mathtools}
\usepackage{amssymb}
\usepackage[mathscr]{eucal}
\usepackage{amscd}
\usepackage{color}
\usepackage{verbatim}
\usepackage{sistyle}
\usepackage{subfigure}
\usepackage{makecell}
\usepackage[usenames,dvipsnames]{xcolor}
\usepackage{tikz-cd}
\usepackage{algorithm}
\usepackage{xfrac}
\usepackage[permil]{overpic}

\DeclareMathOperator{\SO3}{{{\rm SO(3)}}} 
\DeclareMathOperator{\SE3}{{{\rm SE(3)}}} 
\DeclareMathOperator{\so3}{{{\rm so(3)}}}

\newcommand{\f}[1]{{\boldsymbol{#1}}}
\newcommand{\baf}[1]{{\bar{\boldsymbol{#1}}}}

\newcommand{\sub}{\subset}
\newcommand{\bEq}{\begin{equation}}
\newcommand{\eEq}{\end{equation}}
\newcommand{\beq}{\begin{equation*}}
\newcommand{\eeq}{\end{equation*}}

\newcommand{\car}{\times}
\newcommand{\mto}{\mapsto}

\newcommand{\byd}{\,{\raisebox{.092ex}{\rm :}{\rm =}}\,}

\newcommand{\fR}[1]{{\mathbf{#1}}}
\newcommand{\hfR}[1]{\hat{\mathbf{#1}}}

\newcommand{\sepr}[1]{\,\,\,\,\textnormal{{#1}}\,\,\,\,}

\newcommand{\lf}{\left}
\newcommand{\rg}{\right}

\newcommand{\sS}[1]{{\scriptscriptstyle {#1}}}
\newcommand{\Tra}{^{\mathsf{\sS\!T}}}
\newcommand{\Rn}{\text{I\!R}}
\newcommand{\tif}[1]{{\widetilde{\boldsymbol{#1}}}}

\newcommand{\bAl}{\begin{align}}

\newcommand{\ome}{\omega}
\newcommand{\Gam}{\varGamma}
\newcommand{\B}[1]{{\mathbb{#1}}}
\newcommand{\veps}{\varepsilon}
\newcommand{\gam}{\gamma}
\newcommand{\del}{\delta}

\newcommand{\vTht}{\varTheta}

\newcommand{\fr}[2]{\frac{#1}{#2}\,}

\newcommand{\wti}[1]{{\widetilde{#1}}}
\newcommand{\htif}[1]{\hat{\widetilde{\boldsymbol{#1}}}} 
\newcommand{\hf}[1]{\hat{\boldsymbol{#1}}} 
\newcommand{\hti}[1]{\hat{\widetilde{#1}}}
\newcommand{\haf}[1]{{\hat{\boldsymbol{#1}}}}
\newcommand{\ha}[1]{{\hat{#1}}}
\newcommand{\chf}[1]{{\check{\f{#1}}}}
\newcommand{\alp}{\alpha}
\newcommand{\bet}{\beta}
\newcommand{\bdg}{\beq\begin{diagram}}
\newcommand{\edg}{\end{diagram}\eeq}

\newcommand{\CNinf}{\B C_{N\infty}}
\newcommand{\CMinf}{\B C_{M\infty}}
\newcommand{\CNalp}{\B C^v_{N\alp}}
\newcommand{\CMalp}{\B C^v_{M\alp}}
\newcommand{\CNz}{\B C_{N0}}
\newcommand{\CMz}{\B C_{M0}}
\newcommand{\CNb}{\bar{\B C}_{N}}
\newcommand{\CMb}{\bar{\B C}_{M}}
\newcommand{\betgp}{\f\bet^{n}_{\Gam\alp} }

\newcommand{\betkp}{\f\bet^{n}_{K \alp } }

\newcommand{\betgps}{\f\bet^{n}_{\Gam\alp,s} }
\newcommand{\betkps}{\f\bet^{n}_{K \alp ,s} }

\renewcommand{\r}[1]{{\color{black}{#1}}} 
 
\usepackage{amssymb}
\usepackage{amsthm}

\newproof{prf}{Proof}[section]
\newproof{rmk}{Remark}[section]

\usepackage{lineno}
\usepackage[top=1in,bottom=1in,left=1in,right=1in]{geometry}

\journal{Computer Methods in Applied Mechanics and Engineering}

\begin{document}

\begin{frontmatter}

\title{An isogemetric analysis formulation for the dynamics of geometrically exact viscoelastic beams and beam systems with arbitrarily curved initial geometry}

\author[fi]{Giulio Ferri}
\author[fi]{Enzo Marino\corref{cor1}}
\ead{enzo.marino@unifi.it}
\cortext[cor1]{Corresponding author}

\address[fi]{Department of Civil and Environmental Engineering, University of Florence\\ Via di S. Marta 3, 50139 Firenze, Italy}

\begin{abstract}
We present a novel formulation for the dynamics of geometrically exact Timoshenko beams and beam structures made of viscoelastic material featuring complex, arbitrarily curved initial geometries. 
An $\SO3$-consistent and second-order accurate time integration scheme for accelerations, velocities and rate-dependent viscoelastic strain measures is adopted. 
To achieve high efficiency and geometrical flexibility, the spatial discretization is carried out with the isogemetric collocation (IGA-C) method, which permits bypassing elements integration keeping all the advantages of the isogeometric analysis (IGA) in terms of high-order space accuracy and geometry representation. Moreover, a primal formulation guarantees the minimal kinematic unknowns. 
The generalized Maxwell model is deployed directly to the one-dimensional beam strain and stress measures. This allows to express the internal variables in terms of the same kinematic unknowns, as for the case of linear elastic rate-independent materials \r{bypassing the complexities introduced by the viscoelastic material.} As a result, existing SO(3)-consistent linearizations of the governing equations in the strong form (and associated updating formulas) can straightforwardly be used. 
Through a series of numerical tests, the attributes and potentialities of the proposed formulation are demonstrated. In particular, we show the capability to accurately simulate beams and beam systems featuring complex initial geometry and topology, opening interesting perspectives in the inverse design of programmable mechanical meta-materials and objects.
\end{abstract}


\end{frontmatter}


\section{Introduction}
Accurately predicting the dynamics of one-dimensional solids made of viscoelastic materials featuring complex, arbitrarily curved geometries is a rather difficult task that is made even more challenging when finite rotations and displacements are involved in the deformation process. 
Addressing a nonlinear dynamics problem such as this, where geometrical and constitutive complexities combine, can have a significant impact on several engineering sectors, and in particular on the rapidly emerging field of mechanical meta-materials \cite{Bertoldi_etal2017}. Dynamic simulations of meta-materials are still often based on simplified models preventing a faithful representation of their rich nonlinear response. Common approaches, for example, employ discrete models that typically comprise stiffer elements connected by flexible hinges~\cite{Xue_etal2023,Deng_etal2021JAP,Deng_etal2021JMPS}. Meta-materials are made of periodic (or non-) assembly of unit cells that need to be carefully designed as they determine the bulk mechanical properties of the meta-material. The possibility to control the geometry of these pivotal elements, using arbitrarily curved elements tuned by a minimal number of parameters, is fundamental for the design of mechanical meta-materials, especially for inverse-based design approaches~\cite{Zheng_etal2023}. More evolved computational tools for the nonlinear dynamics of meta-materials can also contribute to a better understanding and design of some of their specific capabilities, such as waves control~\cite{Karathanasopoulos_etal2017} and energy dissipation~\cite{Zhu_etal2021,Portela_etal2021}. 

The study of geometrically exact beams, namely one-dimensional elements that can undergo spatial finite displacements and rotations without any kinematic approximation, was initiated by J.~C.~Simo in~\cite{Simo1985}. The first quasi-static finite element (FE) formulation was proposed in~\cite{SimoVu-Quoc1986}, then the finite-rotation dynamic problem was addressed in \cite{Simo&Vu-Quoc1988,Cardona1988}. 
After these seminal works, different key aspects of the dynamics of geometrically exact beams have been investigated in a number of valuable contributions~\cite{IbraMikdad1998,Jelenic1998,Jelenic1999,Makinen2001,Romero&Armero2002,Makinen2007,Makinen2008,PimentaCampelloWriggers2008,Lang_etal2011,Bruls_etal2012,Zupan_etal2012,Zupan_etal2013,Sonneville_etal2014,Thanh-Nam_etal2014,Almonacid2015,Weeger_etal2017,Zupan&Zupan2018,Marino2019a,Marino2019b,Chen_etal2022}. In particular, \cite{IbraMikdad1998,Makinen2001,Makinen2007,Makinen2008,Marino2019a,Marino2019b} focused on establishing proper $\SO3$-consistent explicit and implicit Newmark time integration schemes, including new space discretization techniques; \cite{Jelenic1998,Bruls_etal2012} proposed versions of $\alp$-methods for finite rotations; objectivity aspects were addressed in \cite{Jelenic1999,Romero&Armero2002,Makinen2008}; conserving algorithms were developed in \cite{Romero&Armero2002,Leyendecker_etal2006,PimentaCampelloWriggers2008,Almonacid2015,Zupan&Zupan2018}; quaternion-based Runge Kutta and Newmark methods were presented in \cite{Zupan_etal2012} and \cite{Zupan_etal2013}, respectively, whereas quaternion-based Euler and Crank–Nicolson schemes in \cite{Weeger_etal2017}. Special Euclidean Lie group, $\SE3$, integrators were used in \cite{Sonneville_etal2014,Chen_etal2022}; corotational approaches in \cite{Galvanetto1996,Thanh-Nam_etal2014}.
Applications of geometrically exact beam formulations to cable structures were made in~\cite{Boyer2011,Arena_2016}. Methods for inverse dynamics problems involving geometrically exact strings and beams have been more recently proposed in \cite{Stroehle&Betsch2022} and \cite{Schubert_etal2023}, respectively. 
Analytical and standard FE approaches to nonlinear free vibrations of Timoshenko beams have also been proposed in \cite{Firouzi_etal2024}. 
 
All the above works, however, are restricted to linear elastic materials. Only in one case damping is considered \cite{Weeger_etal2017}. 

In the framework of geometrically nonlinear beam dynamics, viscoelastic materials are modeled in~\cite{Lang_etal2011,Lang&Arnold2012,Linn_etal2013} within a finite difference scheme and a quaternion-based parametrization of finite rotations. In~\cite{Giusteri_etal2021}, $\SE3$ is used for the beam kinematics with a finite difference scheme. FE formulations with applications to multi-body dynamics are proposed in \cite{Zhang_etal2009,Mohamed&Shabana_2011,Bauchau&Nemani2021}. In~\cite{Audoly_etal2013,Lestringant_etal2020}, a discrete geometrically exact formulation encompassing constitutive models for incompressible viscous fluids and viscoelastic one-dimensional solids is proposed and used for simulating active and shape-morphing structures. Linear viscoelastic materials in a co-rotational setting are modeled in \cite{Glaesener_etal2021}. 
A third-order nonlinear model for a viscoelastic self-healing composite beam is used in \cite{Amabili_etal2022}. 

It is noted that none of the above cited papers specifically address the dynamic problem of inelastic geometrically exact beams with arbitrarily curved initial geometries. 

In the present paper, we propose a novel approach to the dynamics of shear-deformable geometrically exact viscoelastic beams, featuring arbitrarily curved initial geometries \cite{Marino_etal2020,Ignesti_etal2023}. We extend the quasi-static model proposed in \cite{Ferri_etal2023} to dynamics by proposing an $\SO3$-consistent, high-order space and second-order time accurate formulation. 
\r{We demonstrate that the complexities introduced by the rate-dependent material can be effectively neutralized by using the trapezoidal rule for the discretized viscous strains~\cite{Marino2019b}}. The new formulation we propose is designed to achieve very high efficiency due to the combination of some key ingredients. First, the spatial discretization is performed using the isogeometric collocation (IGA-C) method. IGA-C \cite{Auricchio2010,Auricchio2012,Fahrendorf_etal2022} keeps the attributes of classical isogeometric analysis \cite{Hughes2005a,Cottrell2009} in terms of capability to reconstruct ``exactly'' complex geometries. Moreover, being based on the discretization of the strong form of the governing equations, the method completely bypasses the cost for elements integration. It requires only one evaluation point per degree of freedom, regardless of the approximation degree \cite{Schillinger2013}. IGA-C proved excellent performances for a wide range of problems \cite{Schillinger2013,Gomez2014,DeLorenzis2015,Kruse2015,Gomez&DeLorenzis2016,Auricchio2013,Kiendl2015a,KiendlAuricchioReali2017,Reali2015,Kiendl2015, KiendlMarinoDeLorenzis2017,Maurin_elal2018,Maurin_etal2018b,Evans_etal2018,Fahrendorf_etal2020,Marino_etal2020,Torre_etal2023}, including the geometrically exact beam problems \cite{Marino2016a,Weeger2017,Marino2017b,Marino2019a,Marino2019b}. Exsisting elasto-visco-plastic \cite{Weeger_etal2022} and viscoelastic~\cite{Ferri_etal2023} IGA-C beam formulations are restricted to the quasi-static problem. 

In addition to the IGA-C attributes, we demonstrate that the same desirable features holding for the static case are here preserved. Namely, 
i) It is possible to set a primal formulation, where incremental displacements and rotations are the only needed unknowns---neither a mixed formulation as in \cite{Weeger_etal2022} nor other additional unknowns are required due to internal variables; 
ii) Finite rotations are updated using the incremental rotation vector (together with the exponential map for $\SO3$), that means minimal rotation unknowns (i.e., the three components of the incremental rotation vector only); iii) The evolution law can be integrated using the same $\SO3$-consistent (implicit) method used for linear elastic materials~\cite{Simo&Vu-Quoc1988,Marino2019b}, with almost no conceptual differences in the construction of the tangent matrix. A tangible advantage of this feature is that the same $\SO3$-consistent linearization available in for rate-independent material~\cite{Marino2019b} can straightforwardly be used.

The outline of the paper is as follows. In Section~\ref{sec:bal_eqs}, the governing equations for the dynamic problem of geometrically exact Timoshenko beams are recalled together with the rate-dependent material model. Section~\ref{sec:time_stepping_algo} focuses on the time discretization, whereas the $\SO3$-consistent linearization and space discretization of the governing equations are discussed in Section~\ref{sec:time_space_disc}. Numerical applications involving a series of increasingly challenging problems to prove the expected capabilities of the formulation are presented and discussed in Section~\ref{sec:num_app}. Finally, a summary and the main conclusions of the work are drawn in Section~\ref{sec:conclusions}.

\section{Governing equations\label{sec:bal_eqs}}
In this section, we first briefly recall the governing equations for geometrically exact beams and then formulate the linear viscoelastic problem in a displacement-based material form. 

\subsection{Strong form of the balance equations}
Consider a spatial curve $s\mto \f c(s) \in \Rn^3$, where $s\in [0, L]\sub\Rn$ is the arc length parameterization, with $L$ denoting the length of the beam in the material setting. From now on, we assume that the curve $\f c$ represents the axis (centroid line) of a beam whose balance equations are given as follows \cite{Simo1985} 
\bAl
 \f n,_{s} + \baf n &= \mu \f a 	\,,\label{eq:ge_f}\\
 \f m,_{s} + \f c,_{s} \car \f n + \baf m &= \f j \f \alp + \tif \ome \f j \f\ome \,. \label{eq:ge_m}
\end{align}

In the above equations, 
$\f n$ and $\f m$ are the internal forces and moments in the spatial form, respectively;
$\baf n$ and $\baf m$ are the distributed external forces and moments per unit length;
$\mu$ is the mass per unit length of the beam;
$\f j$ is the spatial inertia tensor;
$\tif\ome = \dot{\fR R} \fR R\Tra$ is the spatial skew-symmetric angular velocity tensor and $\f \ome = \rm axial(\tif \ome)$ its axial vector, where $\fR R\in \SO3$ is the rotation operator mapping the (rigid) rotation of the beam cross section at $s$ from the material to the current configuration. 
$\f a$ and $\f \alp$ are the spatial linear and angular accelerations of the beam cross section at $s$.
All quantities involved in the governing equations depend on space $s\in(0,L)$ and time $t\in(0,T]$, where $T$ is the length of the time domain. 

To manage any arbitrarily curved initial geometries, the robust technique proposed in \cite{Ignesti_etal2023}, based on the (local) Bishop frame \cite{Bishop1975}, is here adopted. 

For a correct deployment of the constitutive equations and time stepping algorithm \cite{Simo&Vu-Quoc1988}, it is convenient to pullback the governing equations to the material setting as follows 
\bAl
\tif K \f N +\f N,_s+\fR R\Tra \baf n & = \mu \fR R\Tra \f a \,, \label{mat_f} \\ 
\tif K \f M +\f M,_s+ \lf( \fR R\Tra \f{c},_{s}\rg) \car \, \f N + \fR R\Tra \baf m &= \f J \f A + \tif W \f J \f W\,, \label{mat_m}
\end{align}
where $\f N = \fR R\Tra \f n$ and $\f M = \fR R\Tra \f m$ are the internal forces and couples per unit length in the material form, respectively; $\f J$ is the material (time-independent) inertia tensor;
$\tif W = \fR R\Tra\dot{\fR R} $ is the material skew-symmetric angular velocity tensor and $\f W = \rm axial(\tif W) $ its axial vector;
$\f A$ is the material angular acceleration vector. $\tif K$ the material beam curvature, a skew-symmetric tensor defined as $\tif K \byd \fR R\Tra \fR R,_{s}\in\so3$\footnote{With the symbol $\sim$ we mark elements of $\so3$, that is the set of $3 \times 3$ skew-symmetric matrices. In this context, they are used to represent curvature matrices and infinitesimal incremental rotations. Furthermore, we recall that for any skew-symmetric matrix $\tif a\in\so3$, $\f a = \textnormal{axial}(\tif a)$ indicates the axial vector of $\tif a$ such that $\tif a \f h= \f a \car \f h$, for any $\f h\in \Rn^3$, where $\car$ is the cross product.

Moreover, note that all the quantities involved in the problem, like $\f c$, $\fR R$, $\f K$, etc., depend both on space and time, whereas initial quantities, like $\f{c}_0$, $\fR{R}_0$, $\tif K_0$, depend only on space.}.

Boundary (i.e., $s = \{0,L\}$) and initial (i.e., $t = 0$) conditions are given as follows
\bAl
\f \eta & =\baf \eta_c \sepr{or} \f N = \fR R\Tra \baf n_c\,, \label{eq:bcseta}\\
\f \vTht & =\baf \vTht_c \sepr{or} \f M = \fR R\Tra \baf m_c\,, \label{eq:bcsTht}\\
\f v &=\f v_0\,,\label{eq:icv}\\
\f W &=\f W_0\,,\label{eq:icW}
\end{align}
where $\baf n_c$ and $\baf m_c$ are the external concentrated forces and moments applied to any of the beam ends in the current configuration;
$\baf \eta_c$ and $\baf \vTht_c$ are the prescribed displacement (spatial) and rotation (material) vectors at any of the beam ends.
$\f v$ is the spatial velocity vector of the cross section centroid;
$\f v_0$ and $\f W_0$ are the initial linear and angular velocity vectors of the cross section centroid.

\subsection{A brief review of the generalized Maxwell model for one-dimensional problems}
The viscoelastic material behaviour is modeled through the generalized Maxwell model \cite{Christensen2013} directly deployed on the one-dimensional strain measures, \r{$\f \Gam_N=\fR{R}\Tra\f{c},_s- \fR{R}_0\Tra\f{c}_0,_s$} and \r{$\f K_M=\f K - \f K_0$}, as done for the quasi-static case (see \cite{Ferri_etal2023}). \r{$\fR{R}_0$ and $\f{c}_0$ define the beam initial configuration with curvature $\tif K_0 = \fR R_0\Tra \fR R_{0,s}$.}

Assuming \textit{m} spring-dashpot elements in the rheological model, the material form of the total internal forces and couples can be written as
\bEq\label{stress_viscoel}
\f N = \f N_{\infty} +\sum_{\alp=1}^{m}\f N_{\alp}\sepr{and} \f M = \f M_{\infty} +\sum_{\alp=1}^{m}\f M_{\alp}\,,
\eEq
where $\f N_{\infty}$ and $\f M_{\infty}$ are the long-term elastic internal forces and couples, whereas $\f N_{\alp}$ and $\f M_{\alp}$ are the viscous contributions related to the $\alp$th Maxwell element. The evolution laws for the $\alp$th internal variables are 
\bEq\label{evol_f_m}
\dot{\f\Gam}_{N\alp}=\fr {1} {\tau_\alp}(\f\Gam_N-\f\Gam_{N\alp})
\sepr{and}
\dot{\f K}_{M\alp}=\fr {1} {\tau_\alp}(\f K_M-\f {K}_{M\alp})\,,
\eEq
where we have introduced the relaxation times $\tau_\alp$ with $\alp = 1,\ldots,m$ and have assumed, without loss of generality, that they are the same for both strain measures. 

Finally, following the standard approach in linear viscoelasticity, the total material internal forces and couples become 
\bAl
\f N &= \CNinf\f\Gam_N +\sum_{\alp=1}^{m}\CNalp(\f\Gam_N-\f\Gam_{N\alp})\,,\label{N_constitutive}\\
\f M &= \CMinf\f{K}_M +\sum_{\alp=1}^{m}\CMalp(\f{K}_M-\f{K}_{M\alp})\,,\label{M_constitutive}
\end{align}
with 
\beq\label{Cinf}
\CNinf=\mathrm{diag}(E_{\infty}A,G_{\infty}A_2,G_{\infty}A_3) \sepr{and} \CMinf=\mathrm{diag}(G_{\infty}J_t,E_{\infty}J_2,E_{\infty}J_3)\,,
\eeq
\beq\label{CNalp}
\CNalp=\B H^v_{N\alp}/\tau_\alp = \mathrm{diag}(E_{\alp}A,G_{\alp}A_2,G_{\alp}A_3)
\sepr{and}
\CMalp= \B H^v_{M\alp}/\tau_\alp =\mathrm{diag}(G_{\alp}J_t,E_{\alp}J_2,E_{\alp}J_3)\,,
\eeq
where $E_{\infty}$ and $G_{\infty} = E_{\infty}/2(1+\nu)$ are the long-term Young's and shear moduli, 
whereas $E_{\alp}$ and $G_{\alp} = E_{\alp}/2(1+\nu)$ are the Young's and shear moduli associated with the $\alp$th Maxwell element. 
We assume a constant Poisson's ratio, $\nu$, for the material.
For additional details on the constitutive formulation we refer to \r{Appendix A} and to reference~\cite{Ferri_etal2023}.

\subsection{Displacement-based strong form of the governing equations}
The governing equations expressed in terms of kinematic quantities only can now be recast by substituting Eqs.~\eqref{N_constitutive} and \eqref{M_constitutive} into \eqref{mat_f} and \eqref{mat_m} as follows 
\begin{gather}
\tif K \CNz\f\Gam_N - \tif K\sum_{\alp=1}^{m} \CNalp \f\Gam_{N\alp} + \CNz\f\Gam_{N,s} - \sum_{\alp=1}^{m}\CNalp\f\Gam_{N\alp,s}+\fR R\Tra \baf n = \mu \fR R\Tra \f a \,,\label{eq:mat_f_strain}\\
\tif K \CMz\f{K}_M-\tif K\sum_{\alp=1}^{m}\CMalp\f{K}_{M\alp}+\CMz\f K_{M,s} - \sum_{\alp=1}^{m}\CMalp\f K_{M\alp,s} + \fR R\Tra\f{c},_{s}\car{\CNz\f\Gam_N}+ \nonumber \\
 - \fR R\Tra\f{c},_{s}\car{\sum_{\alp=1}^{m}\CNalp\f\Gam_{N\alp}}+\fR R\Tra \baf m = \f J \f A + \tif W \f J \f W\,.\label{eq:mat_m_strain}
\end{gather}

The Neumann boundary conditions become
\bEq\label{eq:mat_bcsf_strain} 
\CNz\f\Gam_N-\sum_{\alp=1}^{m}\CNalp\f\Gam_{N\alp}=\fR R\Tra \baf n_c\,,
\eEq
\bEq\label{eq:mat_bcsmm_strain} 
\CMz\f{K}_M-\sum_{\alp=1}^{m}\CMalp\f{K}_{M\alp}=\fR R\Tra \baf m_c\,,
\eEq
where $\CNz = \CNinf + \sum_{\alp=1}^m \CNalp$ and $\CMz = \CMinf + \sum_{\alp=1}^m \CMalp$ are the instantaneous elasticity tensors.

\section{Time discretization of the governing equations}\label{sec:time_stepping_algo}
In this section we start with a review of the trapezoidal rule for time integration on $\SO3$ and then proceed with the time discretization of the governing equations. 

\subsection{Trapezoidal time integration scheme}
Starting from the Newmark time integration scheme for $\SO3$ proposed in \cite{Simo&Vu-Quoc1988}, here, by setting $\bet = 1/4$ and $\gam = 1/2$, we adopt the $\SO3$-consistent trapezoidal form that is written as follows 
\bAl
\fR R^{n+1} &= \fR R^n \exp(\tif \vTht^{n})\,,		\label{eq:NmkR}	\\ 
\f\vTht^{n} & = h \f W^n + \fr{h^2}{4}(\f A^n + \f A^{n+1})\,, 		\label{eq:NmkTht} \\
\f W^{n+1} & = \f W^n + \fr{h}{2}(\f A^n + \f A^{n+1})\,. \label{eq:NmkW}	 
\end{align}

It is noted that the above integration scheme has already exhibited excellent performances for linear elastic geometrically exact beams in an IGA-C context \cite{Marino2019b}. \r{Here we show how it can naturally be extend to evolution laws preserving the geometric consistency.} 
In Eqs.~\eqref{eq:NmkR}--\eqref{eq:NmkW}, the superscript $n = 0,1,\ldots$ is used to denote any temporal discrete quantity at time $t^n = nh$, where $h$ is the time step size. Note that for the rotation update we adopt the right multiplication, namely we use the material incremental rotation vector $\f\vTht^n = \textnormal{axial}(\tif \vTht^n)$. Following the same approach of \cite{Marino2019b}, we express angular acceleration and velocity at time $t^{n+1}$ in terms of quantities at $t^n$. By exploiting Eqs.~\eqref{eq:NmkTht} and \eqref{eq:NmkW}, we have
\bAl
\f A^{n+1} & = \fr{4}{h^2}\f\vTht^n - \f A^n_\ast\,,	\label{eq:NmkAn+1} \\
\f W^{n+1} & = \fr{2}{h}\f\vTht^n -\f W^n_\ast\,, \label{eq:NmkWn+1}
\end{align}
where we have set
\bAl 
\f A^n_\ast &= \fr{4}{h}\f W^n + \f A^n\,,\\
\f W^n_\ast &= \f W^n\,.
\end{align}

For the sake of completeness, we report also the standard trapezoidal rule for the translational motion of the beam cross section centroid 
\bAl
\f c^{n+1} &= \f c^n + \f \eta^n\,, 	\label{eq:Nmkc}	\\ 
\f\eta^n & = h \f v^n +\fr{h^2}{4}(\f a^n + \f a^{n+1})\,, 		\label{eq:Nmketa} \\
\f v^{n+1} & = \f v^n + \fr{h}{2}(\f a^n + \f a^{n+1})\,. \label{eq:Nmkv}	 
\end{align}

As done for the rotational quantities, using Eqs.~\eqref{eq:Nmketa} and \eqref{eq:Nmkv}, acceleration and velocity at time $t^{n+1}$ can be expressed as follows 
\bAl
\f a^{n+1} & = \fr{4}{h^2}\f\eta^n - \f a^n_\ast\,, \label{eq:Nmkan+1} \\
\f v^{n+1} & = \fr{2}{h}\f\eta^n - \f v^n_\ast \,, \label{eq:Nmkvn+1}
\end{align}
where we have set
\bAl 
\f a^n_\ast &= \fr{4}{h}\f v^n + \f a^n\,,\\
\f v^n_\ast &= \f v^n\,.
\end{align}

Time discretization of the internal forces is performed as in~\cite{Ferri_etal2023} and here we briefly summarize the procedure. By exploiting the above trapezoidal rule \r{(see more details in Appendix A)}, the time discretized $\alp$th strain measures write as follows
\bEq\label{g_trapz}
\f\Gam^{n+1}_{N\alp}=\fr {h} {(2\tau_\alp+h)}\f\Gam^{n+1}_N+\betgp\,,
\eEq
\bEq\label{k_trapz}
\f {K}^{n+1}_{N\alp}=\fr {dt} {(2\tau_\alp+h)}\f{K}^{n+1}_M+\betkp\,,
\eEq
where $\betgp$ and $\betkp$ are unknowns-free terms since they refer to the previous time step $t^n$~\cite{Ferri_etal2023}. For the sake of completeness we report the definition of these terms
\bEq\label{betg}
\betgp=\fr {h} {(2\tau_\alp+h)}\f\Gam^{n}_N+\fr{2\tau_\alp-h} {(2\tau_\alp+h)}\f\Gam^{n}_{N\alp}\,,
\eEq
\bEq\label{betk}
\betkp=\fr {h} {(2\tau_\alp+h)}\f{K}^{n}_M+\fr{2\tau_\alp-h} {(2\tau_\alp+h)}\f{K}^{n}_{M\alp}\,.
\eEq



Substituting Eqs.~\eqref{g_trapz} and~\eqref{k_trapz} into Eqs.~\eqref{eq:mat_f_strain} and \eqref{eq:mat_m_strain} evaluated at time $t^{n+1}$, the nonlinear time discretized governing equations become 
\begin{gather} 
\tif K^{n+1}\CNb\f\Gam^{n+1}_N - \tif K^{n+1}\sum_{\alp=1}^{m}\CNalp\betgp +\CNb\f\Gam^{n+1}_{N,s}+\nonumber\\
-\sum_{\alp=1}^{m}\CNalp\betgps+\fR {R\Tra}^{n+1} \baf n^{n+1} = \mu \fR {R\Tra}^{n+1} \f a^{n+1} \,,\label{eq:mat_f_def}
\end{gather}

\begin{gather} 
\tif K^{n+1}\CMb\f{K}^{n+1}_M+\CMb\f{K}^{n+1}_{M,s}+ 
-\tif K^{n+1}\sum_{\alp=1}^{m}\CMalp\betkp+\nonumber\\-\sum_{\alp=1}^{m}\CMalp\betkps+\fR R\Tra{^{n+1}}\f c,_{s}^{n+1}\car{\CNb\f\Gam^{n+1}_N}+ \nonumber\\
-\fR R\Tra{^{n+1}}\f c,_{s}^{n+1}\car{\sum_{\alp=1}^{m}\CNalp\betgp}+\fR R\Tra{^{n+1}} \baf m^{n+1} = \f J \f A^{n+1} + \tif W^{n+1} \f J \f W^{n+1}\,.\label{eq:mat_m_def}
\end{gather}

The same can be done for the Neumann boundary conditions 
\bEq\label{eq:mat_bcsf_def} 
 \CNb\f\Gam^{n+1}_N=\sum_{\alp=1}^{m}\CNalp\betgp+\fR {R\Tra}^{n+1} \baf n^{n+1}_c\,,
\eEq
\bEq\label{eq:mat_bcsmm_def} 
\CMb\f{K}^{n+1}_M=\sum_{\alp=1}^{m}\CMalp\betkp+\fR R\Tra{^n} \baf m^{n+1}_c\,,
\eEq
where we have set $\CNb=[\CNz-\sum_{\alp=1}^{m}\CNalp\fr {h} {(2\tau_\alp+h)}]$ and $\CMb=[\CMz-\sum_{\alp=1}^{m}\CMalp\fr {h} {(2\tau_\alp+dt)}]$~~\cite{Ferri_etal2023}.

\section{Consistent linearization of the governing equations\label{sec:time_space_disc}}
In this section, we present the consistent linearization of the governing equations and the space discretization using NURBS basis functions. 

\subsection{$\SO3$-consistent linearization of the governing equations}
A very appealing feature of the present formulation is that the same $\SO3$-consistent linearization rules derived for static \cite{Ferri_etal2023} or dynamic \cite{Marino2019b} formulations involving rate-independent materials can be directly used. This advantage originates from the fact that the terms in Eqs.~\eqref{betg} and \eqref{betk} do not affect the tangent operator. Since here we use the same lineariazation rules derived in~\cite{Marino2019b}, in the following we restrict the discussion only to the fundamental geometric facts on which the consistent linearization is based on. For more details, we refer to \cite{Marino2019b}. 

The kinematic of a shear-deformable beam is fully determined by the pairs $(t,s) \mto\lf(\f c(t,s), \fR R(t,s)\rg)$ belonging to the configuration manifold, a subset of $\Rn^3\car\SO3$. 
The tangent space at $(\f c,\fR R)$ is given by $T_{\f c}\Rn^3 \car T_{\fR R}\SO3$. 
Tangent vectors belonging to $T_{\f c}\Rn^3 = \Rn^3$ are denoted by $\del\f\eta$ and represent the standard incremental displacements at $\f c$, whereas $T_{\fR R}\SO3 = \{\fR R\del\tif\vTht\,\,|\,\,\del\tif\vTht \in \so3, \fR R \in \SO3\}$, where $\del\tif\vTht$ is the skew-symmetric matrix associated with the incremental rotation vector $\del\f\vTht$ at $\fR R$ \cite{Simo&Vu-Quoc1988,Makinen2007,Makinen2008}. 
The vectors $\del\f\eta$ and $\del\f\vTht$ represent the primal kinematic unknowns of our problem. 

With these fundamental geometric notions at hand, the linearization formulas for all the terms of the governing equations can be derived. \r{In Appendix A we report the linearizations of the viscous strains, all the others can be found in~\cite{Marino2019b}}. By using these formulas, the linearized version of Eqs.~\eqref{eq:mat_f_def}--\eqref{eq:mat_bcsmm_def} is obtained as follows 
\begin{gather}
[\CNb\wti{(\hat{\fR R}{\Tra{^{n+1}}}\hat{\f{c}}^{n+1}_{,s})}-\wti{(\CNb\hat{\f\Gam}^{n+1}_N)}+\sum_{\alp=1}^{m}\wti{(\CNalp\betgp)}]\del\f\vTht,_s^{n+1}+\nonumber\\
+[\ha{\tif K}^{n+1}\CNb\wti{(\hat{\fR R}{\Tra{^{n+1}}}\hat{\f{c}}^{n+1}_{,s})}-\wti{(\CNb\hat{\f\Gam}^{n+1}_N)}\ha{\tif K}^{n+1}+\ha{\tif K}^{n+1}\sum_{\alp=1}^{m}\wti{(\CNalp\betgp)}+\nonumber\\-\wti{(\ha{\tif K}^{n+1}\sum_{\alp=1}^{m}{\CNalp\betgp})}+\CNb\wti{(\hat{\fR R}{\Tra{^{n+1}}}\hat{\f{c}}^{n+1}_{,ss})}-\CNb\wti{(\tif K^{n+1}\hat{\fR R}{\Tra{^{n+1}}}\hat{\f{c}}^{n+1}_{,s})}+\nonumber\\+\wti{(\hat{\fR R}{\Tra{^{n+1}}}{\baf n}^{n+1})}-\mu\wti{(\hat{\fR R}{\Tra{^{n+1}}}\hat{\f{a}}^{n+1})}]\del\f\vTht^{n+1}
+\CNb\hat{\fR R}{\Tra{^{n+1}}}\del\f\eta_{,ss}^{n+1}+\nonumber\\
+[\tif K^{n+1}\CNb\hat{\fR R}\Tra{^{n+1}}-\CNb\tif K^{n+1}\hat{\fR R}{\Tra{^{n+1}}}]\del\boldsymbol\eta_{,s}^{n+1}-\fr{4}{h^2}\mu\hat{\fR R}\Tra{^{n+1}}\del\boldsymbol\eta^{n+1}+\hfR F^{n+1} = \f 0\,,\label{traslin}
\end{gather}

\begin{gather}
\CMb\del\f\vTht,_{ss}^{n+1}+[\CMb\ha{\tif K}^{n+1}+\ha{\tif K}^{n+1}\CMb-\wti{(\CMb\hat{\f{K}}^{n+1}_M)}+\sum_{\alp=1}^{m}\wti{(\CMalp\betkp)}]\del\f\vTht,_s^{n+1}+\nonumber\\
+\{\ha{\tif K}^{n+1}\CMb\ha{\tif K}^{n+1}-\wti{(\CMb\hat{\f{K}}^{n+1}_M)}\ha{\tif K}^{n+1}+\ha{\tif K}^{n+1}\sum_{\alp=1}^{m}\wti{(\CMalp\betkp)}-\wti{(\ha{\tif K}^{n+1}\sum_{\alp=1}^{m}{\CMalp\betkp})}+\nonumber\\
+\CMb\ha{\tif K_{,s}^{n+1}}+[\wti{(\hat{\fR R}\Tra{^{n+1}}\hat{\f{c}}^{n+1}_{,s})}\CNb-\wti{(\CNb\hat{\f\Gam}^{n+1}_N)}+\sum_{\alp=1}^{m}\wti{(\CNalp\betgp)}]\wti{(\hat{\fR R}\Tra{^{n+1}}\hat{\f{c}}^{n+1}_{,s})}+\nonumber\\
+\wti{(\hat{\fR R}\Tra{^{n+1}}\baf m^{n+1})}-[\fr{4}{h^2}\f J+\fr{2}{h}(\ha{\tif W}^{n+1}\f J-\wti{(\f J\ha{\tif W}^{n+1})})]\fR T^{-1}(\hf\vTht^n)\}\del\f\vTht^{n+1}+\nonumber\\
+[\wti{(\hfR R\Tra{^{n+1}}\haf c,_s^{n+1}})\CNb-\wti{(\CNb\hat{\f\Gam}^{n+1}_N)}+\sum_{\alp=1}^{m}\wti{(\CNalp\betgp)}]\hat{\fR R}\Tra{^{n+1}}\del\boldsymbol\eta^{n+1}_{,s}+\hfR V^{n+1} =\f 0\,,\label{rotlin}
\end{gather}
where we have defined
\begin{gather} 
\hfR F^{n+1} =\htif K^{n+1}\CNb\hat{\f\Gam}^{n+1}_N-\ha{\tif K}^{n+1}\sum_{\alp=1}^{m}\CNalp\betgp+\CNb\hat{\f\Gam}^{n+1}_{N,s}+\nonumber\\-\sum_{\alp=1}^{m}\CNalp\betgps+\hat{\fR R}\Tra{^{n+1}} \baf n^{n+1}-\mu\hat{\fR R}{\Tra{^{n+1}}\hat{\f{a}}^{n+1}}\,, \label{F_trasl} \\
\hfR V^{n+1} = \ha{\tif K}^{n+1}\CMb\hat{\f{K}}^{n+1}_M - \ha{\tif K}^{n+1}\sum_{\alp=1}^{m}\CMalp\betkp+\CMb\hat{\f{K}}^{n+1}_{M,s}-\sum_{\alp=1}^{m}\CMalp\betkps+\nonumber\\
+\hat{\fR R}\Tra{^{n+1}}\haf c,_s^{n+1} \car{[\CNb\hat{\f\Gam}^{n+1}_N-\sum_{\alp=1}^{m}\CNalp\betgp]}+\hat{\fR R}\Tra{^{n+1}} \baf m^{n+1}-\f J\hat{\f{A}}^{n+1}- \hat{\tif {W}}^{n+1}\f J \hat{\f {W}}^{n+1}\,,
\end{gather}

Similarly, for the boundary conditions we have 
\bEq 
 [\CNb\wti{(\hat{\fR R}\Tra{^{n+1}}\hat{\f{c}}^{n+1},_s)}-\wti{(\hat{\fR R}\Tra{^{n+1}}{\baf n}^{n+1}_c)}]\del\f\vTht^{n+1}+\CNb\hat{\fR R}\Tra{^{n+1}}\del\f\eta^{n+1}_{,s}=\sum_{\alp=1}^{m}\CNalp\betgp-\CNb\hat{\f\Gam}^{n+1}_N+\hat{\fR R}\Tra{^{n+1}} \baf n^{n+1}_c\,,\label{bcsf_lin}
\eEq
and 
\bEq
\CMb\del\f\vTht,_s^{n+1}+[\CMb\ha{\tif K^{n+1}}-\wti{(\hat{\fR R}\Tra{^{n+1}}{\baf m}^{n+1}_c)}]\del\f\vTht^{n+1}=\sum_{\alp=1}^{m}\CMalp\betkp-\CMb\hat{\f{K}}^{n+1}_M+\hat{\fR R}\Tra{^{n+1}} \baf m^{n+1}_c\,. \label{bcsmm_lin} 
\eEq

In the above linearized equations, with the symbol $\hat{(\cdot)}$ we denote any quantity evaluated at the time $t^{n+1}$ around which the linearization takes place. 

\begin{figure}
\centering
\includegraphics[width=0.6\textwidth]{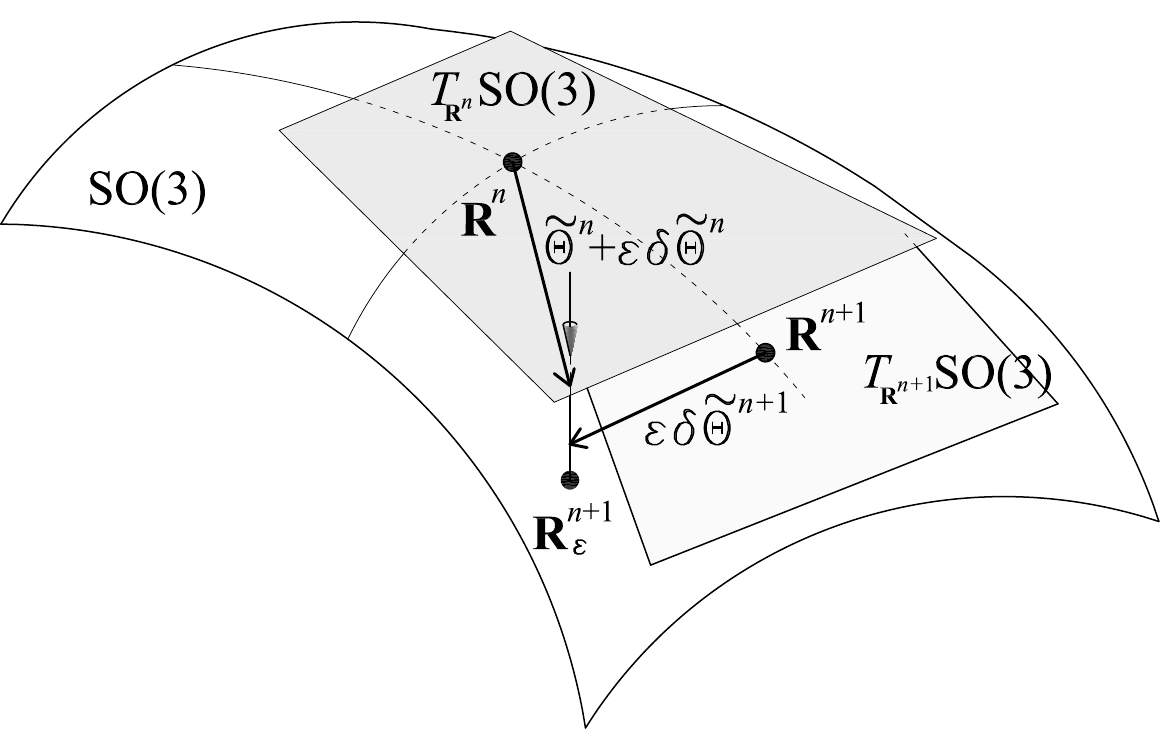}
\caption{Relation between incremental rotations belonging to different tangent spaces to $\SO3$.\label{fig:Tmap}}
\end{figure}

Importantly, we note that in performing the linearization of the angular acceleration, the rotation increment at $t^n$ appears. But since our rotational unknowns are needed at $t^{n+1}$ (to be consistent with the internal force linearized terms), in Eqs.~\eqref{traslin} and \eqref{rotlin} 
we need to make use of an operator, $\fR T^{-1}$, to map the incremental rotation $\del\f\vTht^{n}$ at $\fR R^n$ into the corresponding incremental rotation $\del\f\vTht^{n+1}$ at $\fR R^{n+1}$, namely $\del\f\vTht^n = \fR T^{-1}(\f \vTht^n) \del\f\vTht^{n+1}$.
A pictorial representation of the meaning of $\fR T^{-1}$ is given in Figure~\ref{fig:Tmap}. This specific geometric aspect is discussed in~\cite{Cardona1988,Simo&Vu-Quoc1988,Simo&Wong1991,Marino2019b} to which we refer for additional details.

\subsection{Space discretization}
The linearized governing equations, written in terms of the unknown fields $\del\f\vTht^{n+1}$ and $\del\f\eta^{n+1}$ at time $t^{n+1}$, are spatially discretized by using NURBS basis functions $R_{j,p}$ with $j = 1,\ldots,\rm n$ of degree $p$. In the isoparametric formulation, we use the same basis functions to represent the beam centroid curve. Thus, we have
\bAl
\del\f\vTht^{n+1} (u) &= \sum_{j= 1}^{\rm n} R_{j,p}(u) \del \chf\vTht^{n+1}_j\sepr{with} u\in [0,\,1]\,,\label{eq:Thtu}\\
\del\f\eta^{n+1} (u) &= \sum_{j= 1}^{\rm n} R_{j,p}(u)\del \chf\eta^{n+1}_j\sepr{with} u\in [0,\,1]\,,\label{eq:etau}\\
\f c^{n+1} (u) & = \sum_{j= 1}^{\rm n} R_{j,p}(u) \chf p^{n+1}_j\sepr{with} u\in [0,\,1]\,,\label{eq:cu}
\end{align}
where $\del \chf\vTht^{n+1}_j$ and $\del\chf\eta^{n+1}_j$ are the primal ($2\car3\car \rm n$) unknowns, namely the $j$th incremental control rotation and translation, respectively; $\chf p^{n+1}_j$ is the $j$th control point defining the beam centroid curve. 
Eqs. \eqref{eq:Thtu} and \eqref{eq:etau} are substituted into \eqref{traslin} and \eqref{rotlin} and in the boundary conditions \eqref{bcsf_lin} and \eqref{bcsmm_lin}, where the differentiations must be properly done considering the Jacobian relating the parametric $u\in[0,1]$ and arc length $s\in[0,L]$ coordinate systems.
A square system is finally built by collocating the field equations at the internal $\rm n-2$ collocation points and the Dirichlet or Neumann boundary conditions at the boundaries $u = 0$ and $u = 1$. Standard Greville abscissae \cite{Auricchio2010} are chosen as collocation points. 
At a given Newton-Raphson iteration, once solved the linear system for the primal unknowns, we follow the same updating procedure discussed in~\cite{Marino2019b} and make use Eqs.~\eqref{g_trapz}--\eqref {betk} to update the viscous terms which are only evaluated at the collocation points.

\section{Numerical applications\label{sec:num_app}}
In this section, some numerical applications are presented and discussed in order to demonstrate the capabilities of the proposed method. 
The complexity of the tests is increased gradually in order to verify all the attributes of the formulation, in particular the high-order accuracy in space, the capacity to model very complex geometries both in plane and three-dimensional motions, the potentialities towards modeling and design of mechanical meta-materials featuring complex-shaped cells. 
Odd degrees are not considered since, as it is well known, in IGA-C they do not normally improve the rates compared to the smaller even degrees.
\r{Shear-locking is prevented thanks to smoothness of the basis functions and to the slenderness ratios of the tested beams~\cite{Marino2017b}.
In some cases, we compare our results with Abaqus, using B31 elements (3D Timoshenko beam with linear shape functions) and an implicit time stepping algorithm. Given the very different technologies behind the two formulations (IGA-C and FEM), showing both results serves only the purpose of verifying the correctness of our model since no reference results are available.}

\subsection{Swinging flexible pendulum\label{sec:Pendulum}}
\begin{figure}[h]
\centering
\begin{overpic}[width=0.8\textwidth]{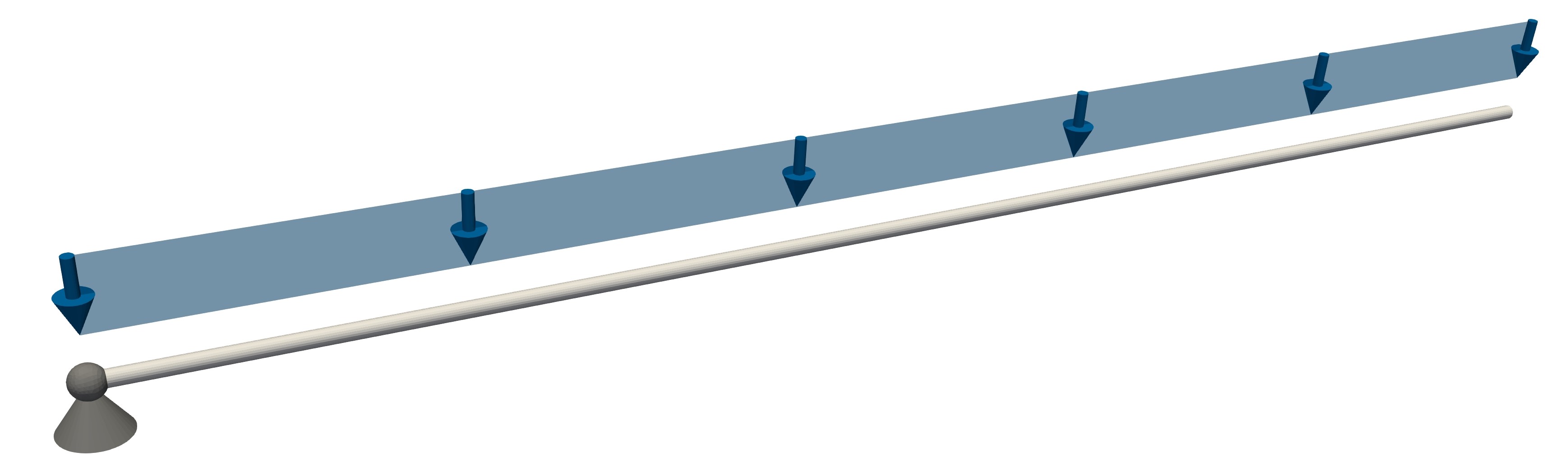}
\put(1000,250){$q_3$}
\end{overpic}
\caption{Swinging flexible pendulum subjected to a distributed vertical load.\label{fig:swing_pendulum}}
\end{figure}

The first numerical application is a swinging flexible pendulum \cite{Lang_etal2011,Marino2019a,Marino2019b} with length of $\SI{1}{m}$, lying in the $(x_2,x_3)$ plane. The beam, featuring a circular cross section with diameter $d=\SI{0.01}{m}$, is hinged at one end and subjected to a distributed load $q_3$ (self-weight) directed along $x_3$ (see Figure~\ref{fig:swing_pendulum}). 
The material density is $\rho=\SI{1100}{kg/m^3}$ leading to $q_3=\SI{-0.8475}{N/m}$. 
The instantaneous Young's modulus is $E_0=\SI{5e6}{N/m^2}$, whereas the Poisson's ratio is $\nu=0.5$. A single Maxwell element is adopted with $E_1=0.9E_0=\SI{4.5e6}{N/m^2}$, $G_1=E_1/2(1+\nu)$, and $\tau_1=\SI{0.1}{s}$. 
The duration of the simulation is $T=\SI{2}{s}$ and the time step is $h=\SI{5e-3}{s}$. 

\begin{figure}
\centering
\begin{overpic}[width=0.8\textwidth]{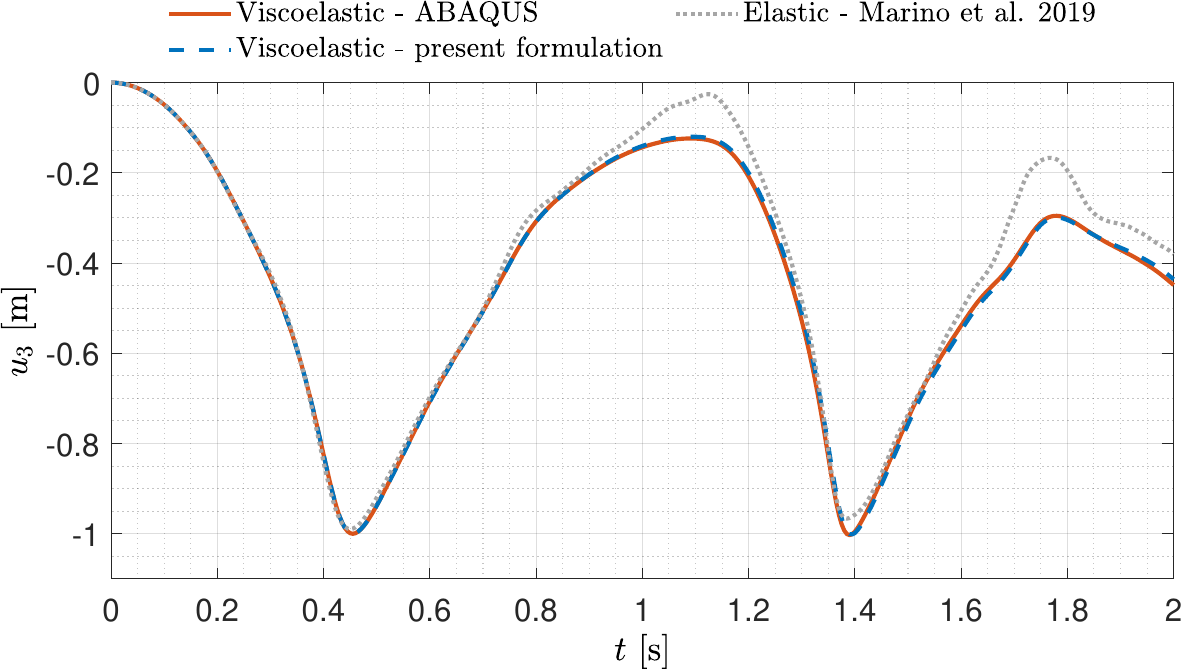}
\end{overpic}
\caption{Swinging flexible pendulum: comparison of the tip vertical displacements, $u_3$, obtained with the present formulation, with Abaqus and with the linear elastic formulation ($E=E_0=\SI{5e6}{N/m^2}$, $\nu=0.5$) in in \cite{Marino2019b}.\label{fig:swing_pendulum_u3}}
\end{figure}

Results in terms of tip displacement time history $u_3$ are shown in Figure~\ref{fig:swing_pendulum_u3}. An excellent agreement is observed with Abaqus.

\begin{figure}
\centering
\begin{overpic}[width=1\textwidth]{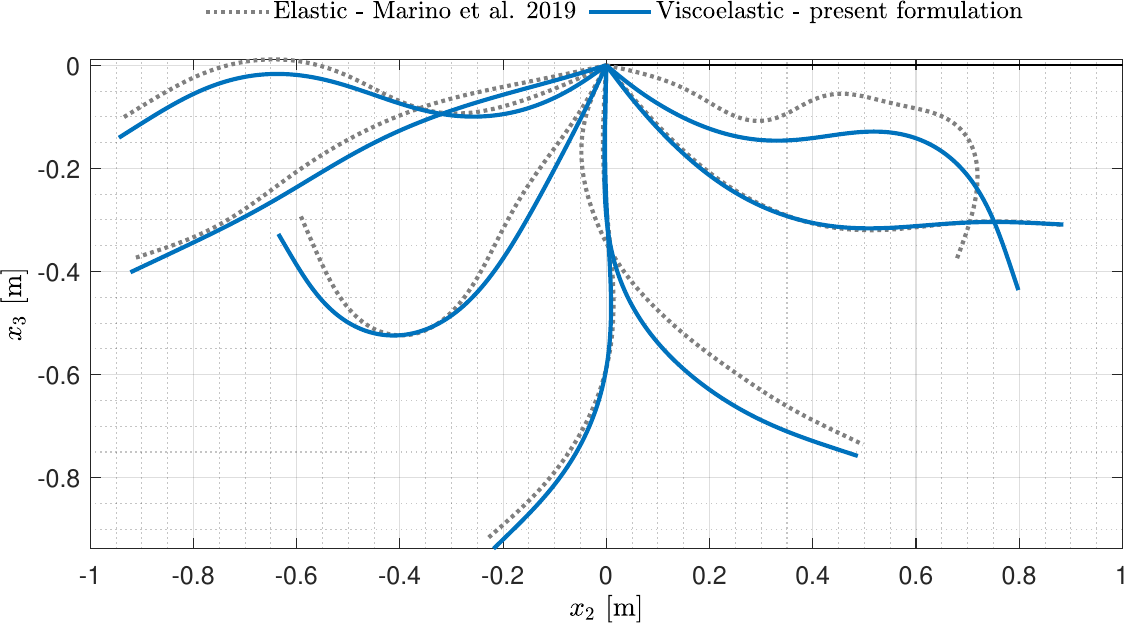}
\put(700,320){$t=\SI{0.25}{s}$}
\put(340,90){$t=\SI{0.5}{s}$}
\put(110,280){$t=\SI{0.75}{s}$}
\put(100,400){$t=\SI{1}{s}$}
\put(310,320){$t=\SI{1.25}{s}$}
\put(780,150){$t=\SI{1.5}{s}$}
\put(870,270){$t=\SI{2}{s}$}
\end{overpic}
\caption{Swinging flexible pendulum: comparison of the deformed shape with respect to the linear elastic formulation as in \cite{Marino2019b}.\label{fig:swing_pendulum_snapshots}}
\end{figure}
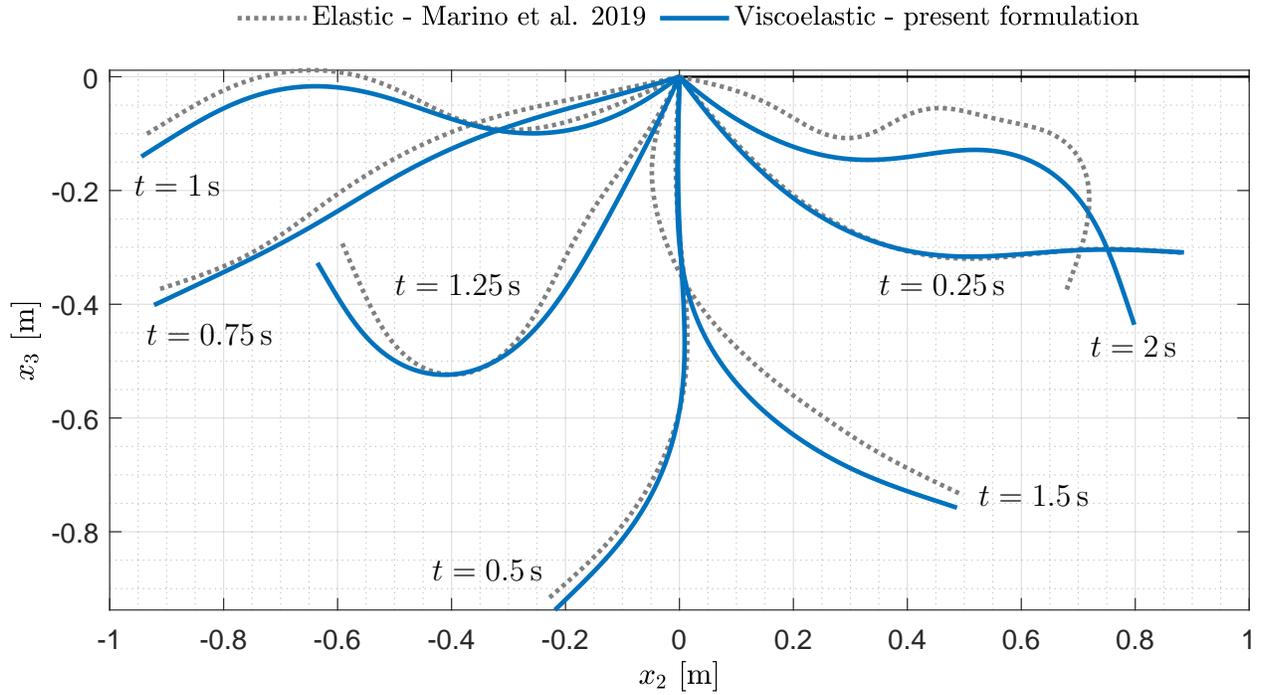

Some snapshots of the deformed beam are shown in Figure~\ref{fig:swing_pendulum_snapshots}, where blue solid line refers to the present formulation, whereas grey dotted line is associated with a linear elastic material as in \cite{Marino2019b} with $E=E_0=\SI{5e6}{N/m^2}$, $\nu=0.5$. The viscous effects are clearly visible, the deformed viscous configurations are significantly smoother compared to the linear elastic ones since oscillations are particularly damped out. 

\begin{figure}
\centering
\begin{overpic}[width=0.6\textwidth]{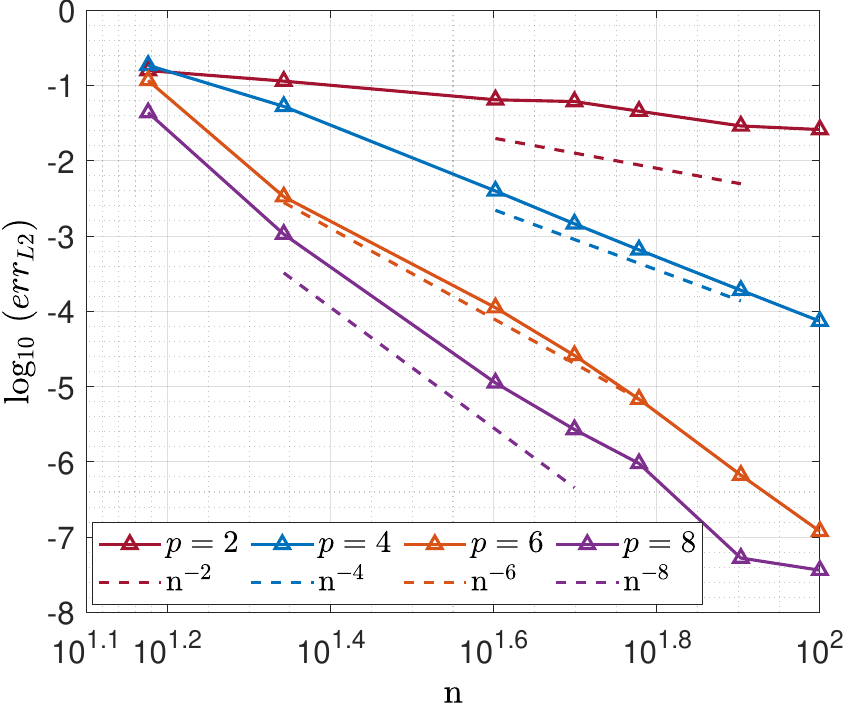}
\end{overpic}
\caption{Swinging flexible pendulum: spatial convergence plots for $p=2,\ldots, 8$ vs. the number of collocation points $\rm n$\label{fig:pndulum_conv_rates}.}
\end{figure}

We also investigate the high-order accuracy in space of the proposed formulation. We study the convergence of the $L_2$-norm of the error at $t=\SI{0.75}{s}$ for different degrees of the B-Splines, $p$, and the number of collocation points, $\rm n$. 
The error, computed over a grid of points, is given by $err_{L_2}=||\f{u}^r-\f{u}^h||/||\f{u}^r||$, where $\f{u}^h$ is the approximated displacement, while $\f{u}^r$ is the reference one, obtained with $p=8$ and $\rm n=200$. 
Results are presented in Figure~\ref{fig:pndulum_conv_rates}.

Overall, very good convergence rates are achieved. It is remarked that, due to the temporal error and the tolerance in the Newton-Raphson algorithm, it is not possible to achieve an error smaller that $1\times{10^{-8}}$.

\subsection{Cantilever beam under impulsive load\label{sec:cantilever straight}}
\begin{figure}[h]
\centering
\begin{overpic}[width=0.8\textwidth]{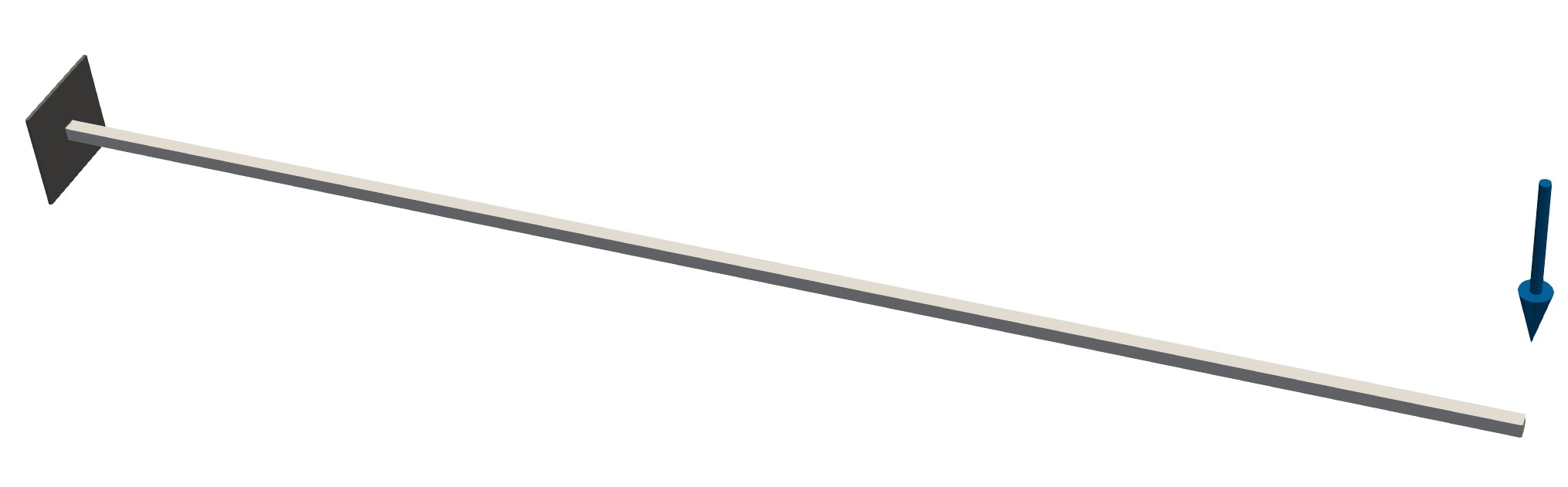}
\put(925,200){$F_3$}
\end{overpic}
\caption{Cantilever beam under impulsive load: geometry and applied force.\label{fig:cantilever_beam}}
\end{figure}

The second numerical example is a straight cantilever beam (see Figure~\ref{fig:cantilever_beam}) similar to the one studied in \cite{Gravouil&Combescure2001,Marino2019b} which deforms in the plane $(x_2,x_3)$ under the effect of a constant tip force $F_3=\SI{-100}{N}$ directed along $x_3$ applied impulsively. The beam, of length $\SI{1}{m}$ and with a square cross-section of side $\SI{0.01}{m}$, is initially aligned with the $x_2$-axis. The challenge of this test is associated with its very fast nonlinear dynamics. 

The instantaneous elastic properties are assumed to be the same as in \cite{Marino2019b}, namely the Young's modulus is $E_0=\SI{210E9}{N/m^2}$, the Poisson's ratio is $\nu=0.2$, and the material density is $\rho=\SI{7800}{kg/m^3}$. We furthermore assume that the viscoelastic behaviour is captured by one Maxwell element ($\alp = 1$) with Young's modulus $E_1=0.9E_0=\SI{189e9}{N/m^2}$ (leading to $E_\infty=\SI{210e8}{N/m^2}$) and relaxation time $\tau_1=\SI{0.1}{s}$. The total simulation time is $T=\SI{0.5}{s}$, with a time step $h=\SI{5e-4}{s}$. 

\begin{figure}
\centering
\begin{overpic}[width=0.8\textwidth]{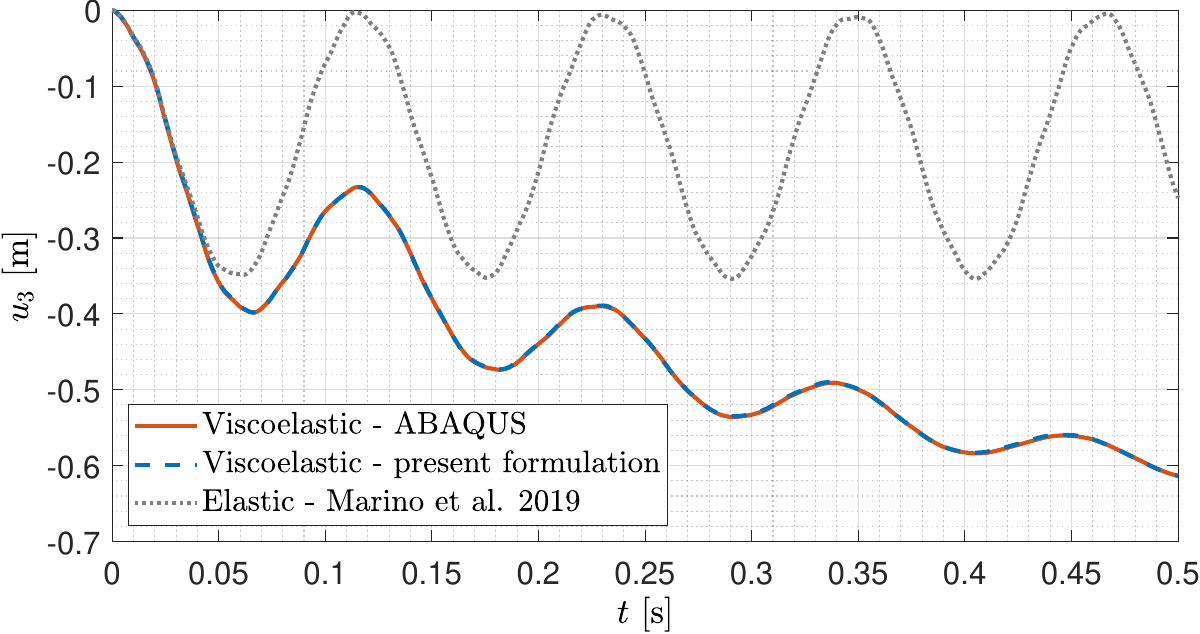}
\end{overpic}
\caption{Cantilever beam: comparison of the tip displacement $u_3$ obtained with the present formulation and Abaqus. Grey dotted line refers to the linear elastic material as in Marino at al. 2019~\cite{Marino2019b}.\label{fig:cantilever_beam_u3}}
\end{figure}

Figure~\ref{fig:cantilever_beam_u3} shows the time history of the tip displacement $u_3$ in the $x_3$ direction. An excellent agreement is observed comparing with Abaqus. To better capture the viscoelastic effects, we report in the figure also the linear elastic response (grey dotted line) \cite{Marino2019b}.

\subsection{Spivak beam\label{sec:Spivak}}
This numerical test, the ``Spivak beam'', is meant to verify the capability of the proposed model to correctly simulate structures with complex initial geometry, characterized by points with vanishing curvature. It is a rather critical aspect since formulations based on the more common Frenet-Serret local frame would fail in this case, as shown in \cite{Ignesti_etal2023}. 

\begin{figure}[h]
\centering
\begin{overpic}[width=0.7\textwidth]{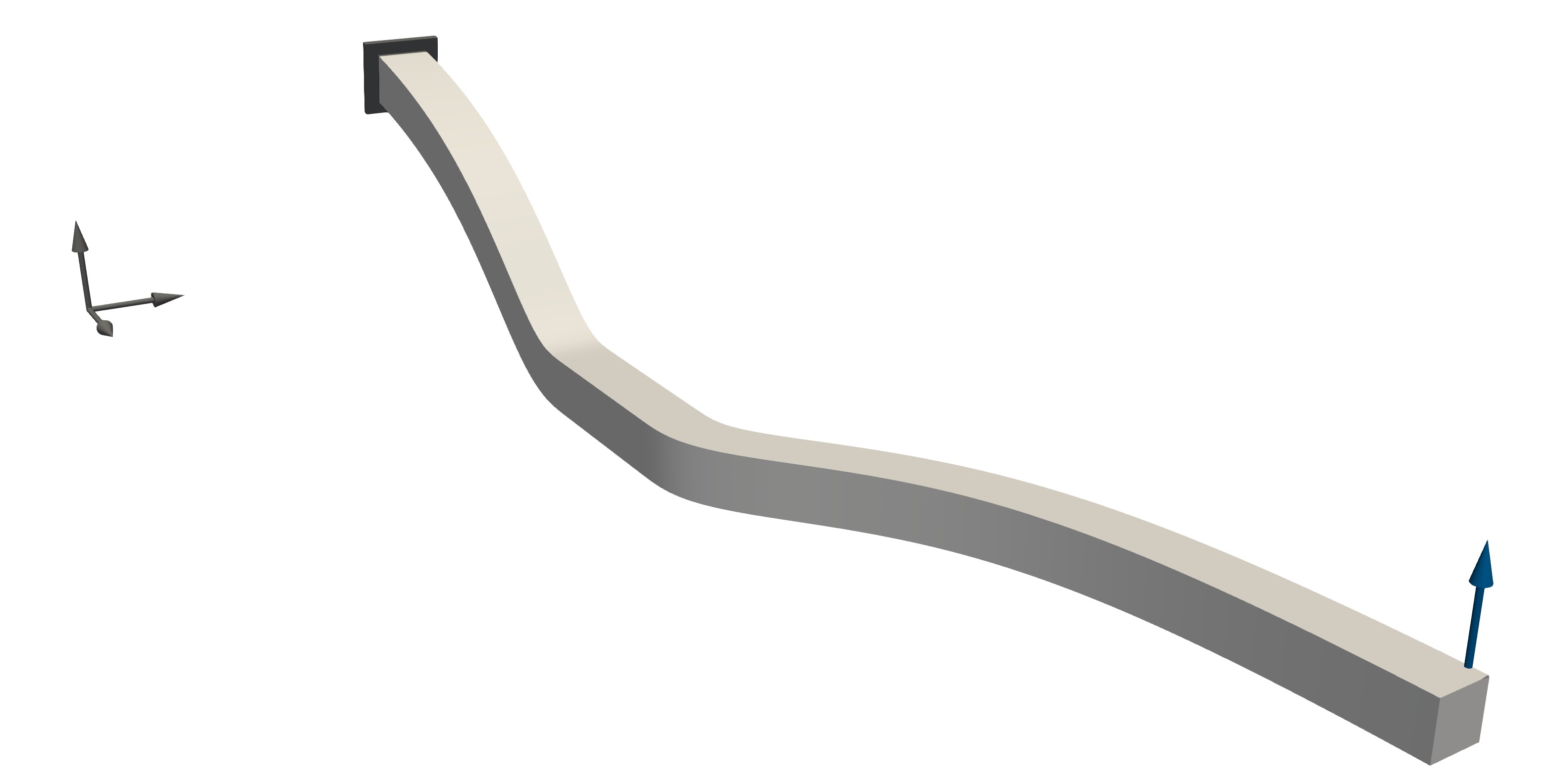}
\put(50,265){$x_1$}
\put(120,330){$x_2$}
\put(10,370){$x_3$}
\put(890,140){$F_3$}
\end{overpic}
\caption{The Spivak beam clamped at one end and subjected to a tip force along $x_3$ at the other end.\label{fig:Spivak_beam}}
\end{figure}
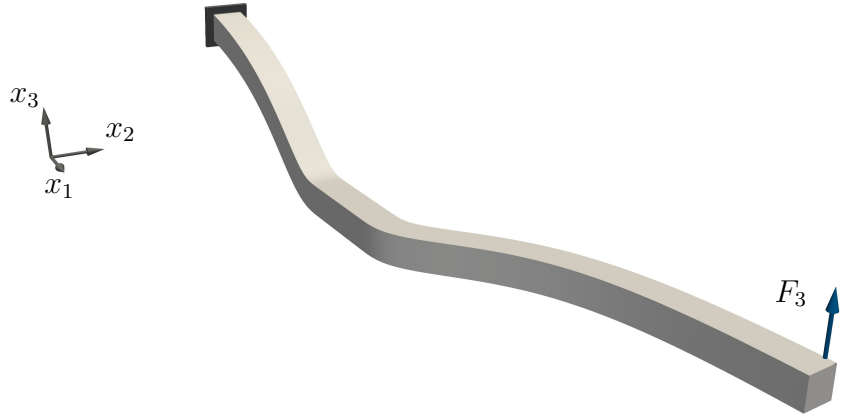

The centerline of the beam is represented by the following piecewise-defined (Spivak) curve \cite{Ignesti_etal2023} (see Figure~\ref{fig:Spivak_beam})

\bEq
\left\{
\begin{alignedat}{3}
&\f {c}(s)=[s,\, 0,\, e^{-1/{s^2}}]\Tra \,, 	&\quad 	& s\in [-2,0)\,,\\
&\f {c}(s)=[0,\, 0,\, 0]\Tra\,,			&\quad	& s=0\,,\\
 &\f {c}(s)=[s,\, e^{-1/{s^2}},\, 0]\Tra\,,	&\quad 	& s\in (0,3].
\end{alignedat}
\right.
\eEq

\begin{figure}
\centering
\subfigure[Load time history.\label{fig:spivak_load_u_a}]{\begin{overpic}[clip,width=0.8\textwidth]{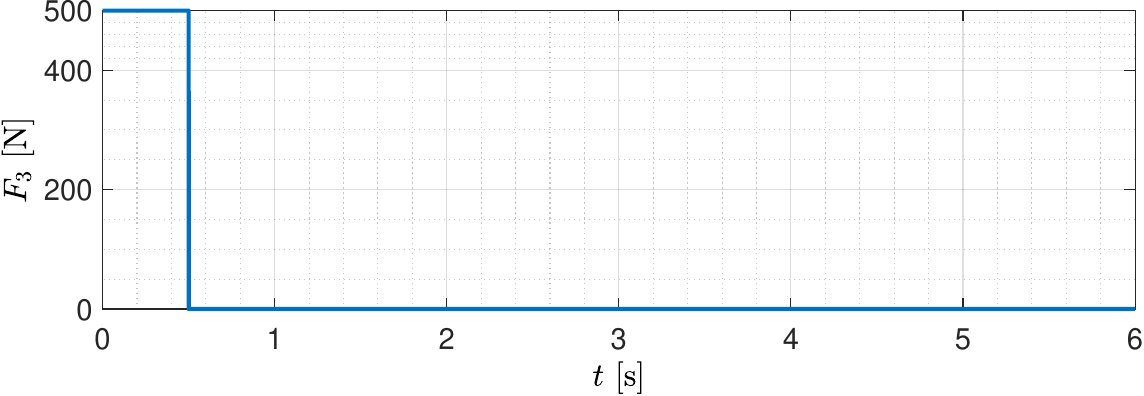}
\end{overpic}}
\subfigure[Tip displacement along the $x_1$-axis.\label{fig:spivak_load_u_b}]{\begin{overpic}[clip,width=0.8\textwidth]{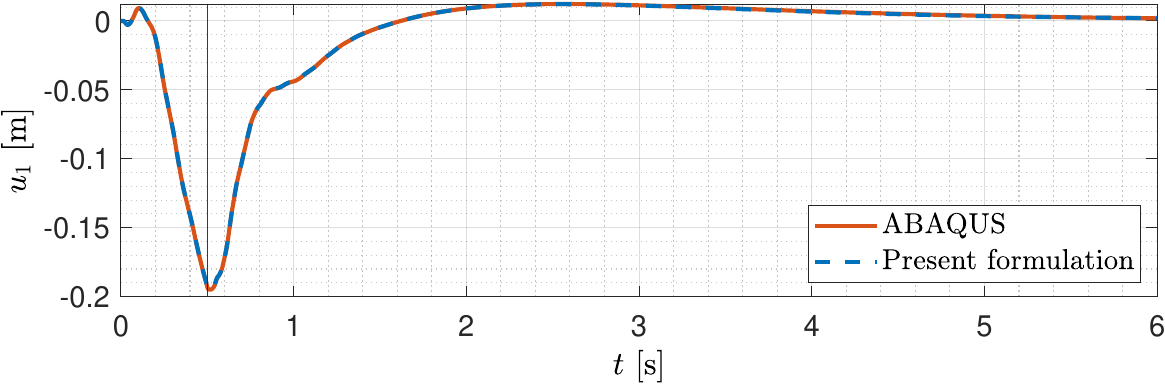}
\end{overpic}}
\subfigure[Tip displacement along the $x_2$-axis.\label{fig:spivak_load_u_c}]{\begin{overpic}[width=0.8\textwidth]{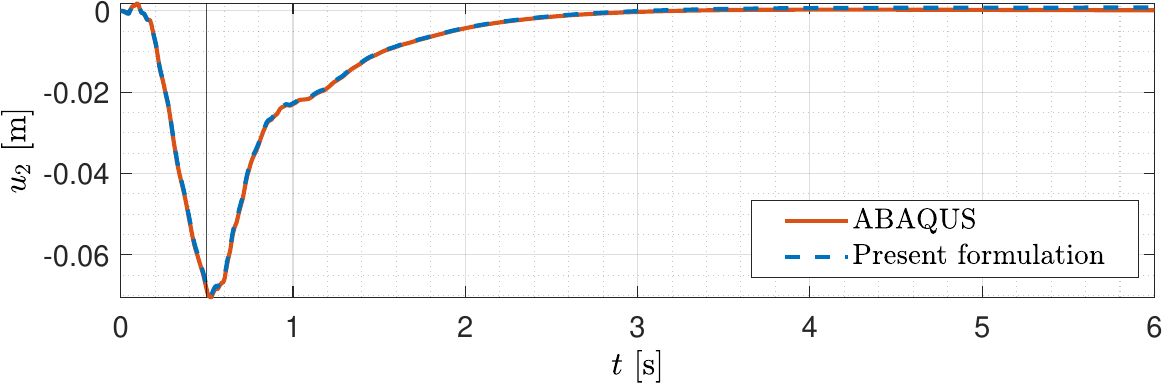}
\end{overpic}}
\subfigure[Tip displacement along the $x_3$-axis.\label{fig:spivak_load_u_d}]{\begin{overpic}[width=0.8\textwidth]{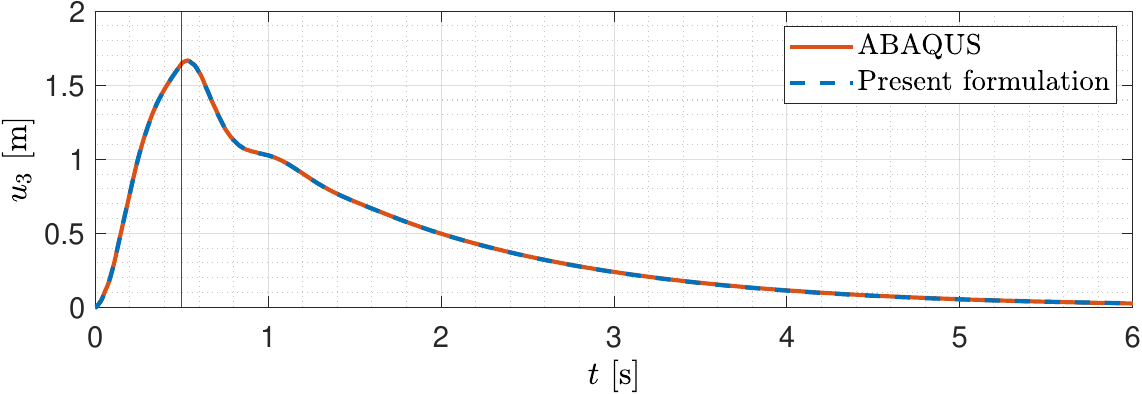}
\end{overpic}}
\caption{Spivak beam: time histories of the tip displacement components.\label{fig:spivak_load_u}}
\end{figure}

\r{The beam has a total length of $\SI{5.50}{m}$, with a square cross section of side \SI{0.1}{m}. The rheological model comprises one Maxwell element ($\alp = 1$) with $E_1 = \SI{7.439e9}{N/m^2}$ and $\tau_1 =\SI{0.1}{s}$; whereas the long term modulus is $E_\infty =\SI{5.614E8}{N/m^2}$. Density and Poisson's ration are $\rho=\SI{700}{kg/m^3}$ and $\nu=0$, respectively}. A concentrated force directed along $x_3$, $F_3=\SI{500}{N}$, is instantaneously applied at the tip of the beam, and it is kept constant for $\SI{0.5}{s}$; after that it is removed. The simulation time is $T = \SI{6}{s}$ with a time step $h = \SI{1E-3}{s}$, see Figure~\ref{fig:spivak_load_u_a}. 

The three components of the displacement over time are shown in Figures~\ref{fig:spivak_load_u_b}-~\ref{fig:spivak_load_u_d}. Again, an excellent agreement with Abaqus is observed for the all the displacement components. Inertia effects are observed since the maximum displacement occurs for $t>\SI{0.5}{s}$ (see black vertical lines in Figures 10b-10d), when the beam has already been unloaded. \r{Viscoelastic effects are clearly noticed as, after removing the force at $\SI{.5}{s}$, the free vibrations are completely damped out.}

\subsection{Spiral spring\label{sec:Spiral}}
\begin{figure}
\centering
\subfigure[3D view of the spiral beam.\label{fig:Spiral_beam_a}]
{
\begin{overpic}[width=0.5\textwidth]{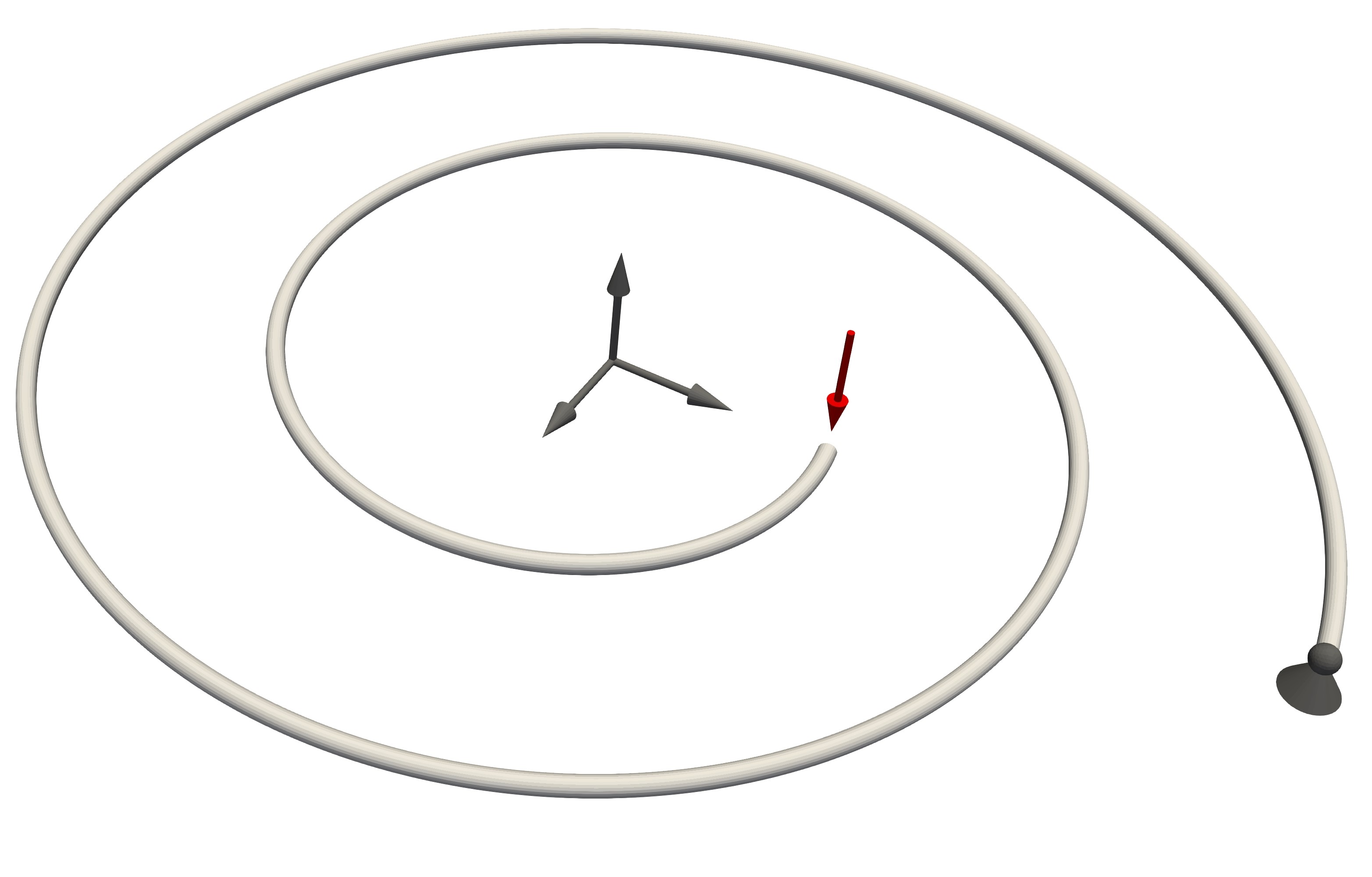}
\put(350,340){$x_1$}\put(520,370){$x_2$}\put(400,460){$x_3$}\put(630,370){$F_3$}
\end{overpic}
}
\subfigure[Plane view of the spiral beam (units in $\SI{}{cm}$).\label{fig:Spiral_beam_b}]
{
\begin{overpic}[width=0.4\textwidth]{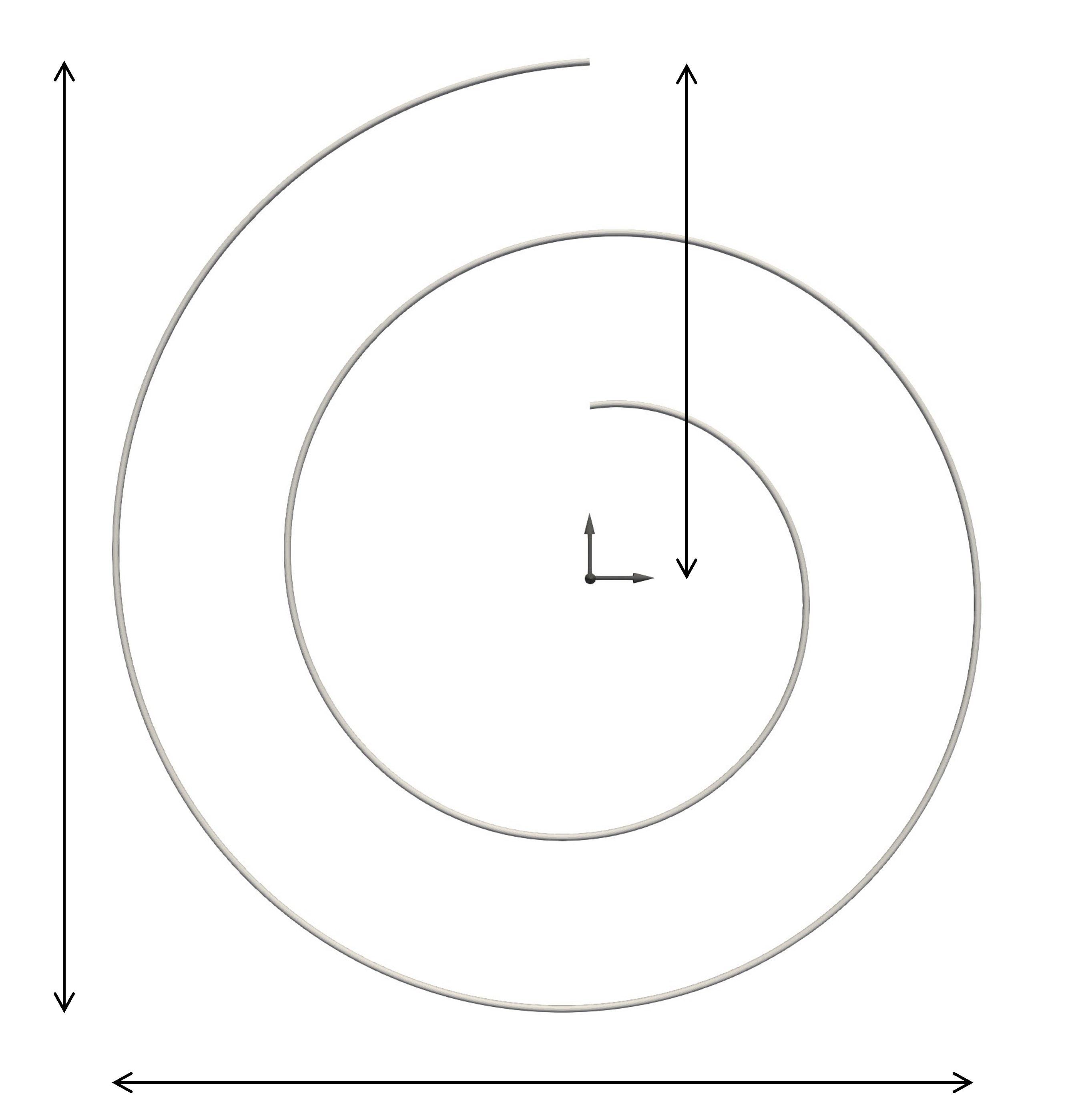}
\put(10,500){\colorbox{white}{\makebox(40,50){\textcolor{black}{34.6}}}}\put(450,-10){\colorbox{white}{\makebox(100,50){\textcolor{black}{31.4}}}}\put(580,660){\colorbox{white}{\makebox(50,50){\textcolor{black}{18.9}}}}
\end{overpic}
}
\subfigure[$F_3$ load time histiry.\label{fig:Spiral_beam_c}]
{
\begin{overpic}[width=0.9\textwidth]{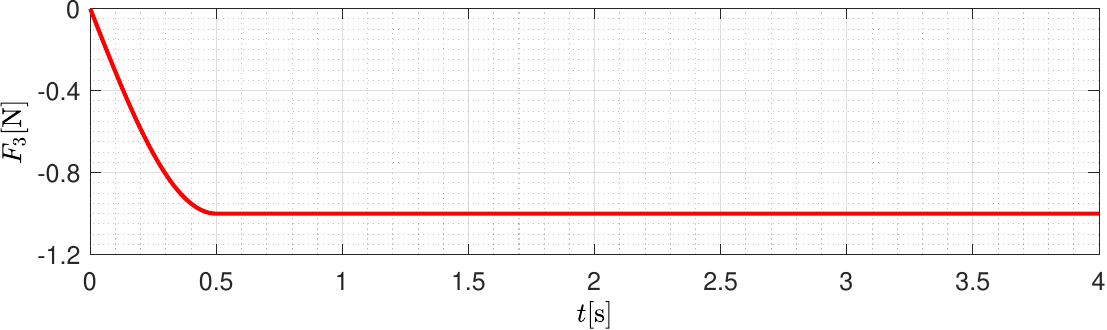}
\end{overpic}
}
\caption{Circular spiral beam:geometry and load.\label{fig:Spiral_beam}}
\end{figure}
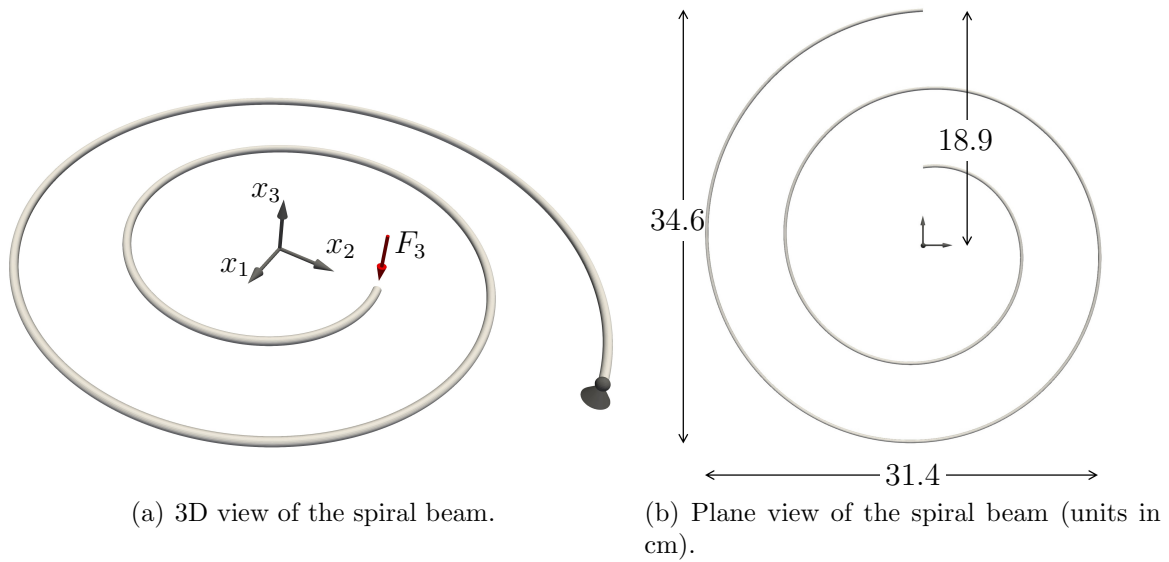
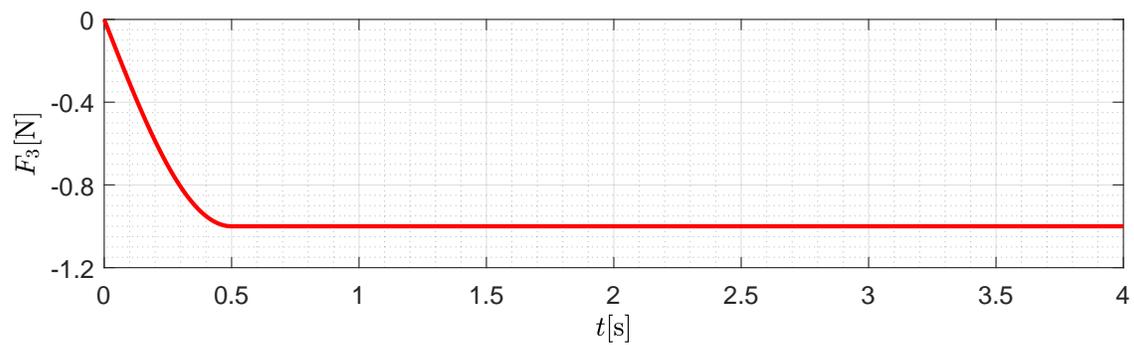

We present here the case of a spiral beam whose initial geometry is shown in Figures~\ref{fig:Spiral_beam_a} and~\ref{fig:Spiral_beam_b}. The beam is hinged at one end and subjected to its self-weight plus a concentrated force in the negative $x_3$ direction. The intensity of $F_3$ varies with time as shown in Figure~\ref{fig:Spiral_beam_c}. Firstly, it is applied following a sinusoidal ramp, $F_3(t)=-\sin{(\pi t)}$ for $t\in \SI{[0, 0.5]}{s}$, reaching the maximum absolute value of $\SI{-1}{N}$ at $t=\SI{0.5}{s}$. Then, it is kept constant for the remaining simulation time. 

The analytical expression of the beam axis in its initial configuration is given by
\bEq
\f {c}(s)=[s\sin{(s)},\, s\cos{(s)},\, 0]\Tra\sepr{with} s\in [2\pi,\,6\pi]\,.\label{eq:spiral}\\
\eEq
The length of the beam is \SI{158.46}{cm}.
The beam features a circular cross section of radius $r=\SI{0.25}{cm}$ and represents a Polylactic Acid (PLA) filament suitable for 3D printing, whose properties are taken from \cite{Wan_etal2022}. 
8 Maxwell elements are employed and the corresponding shear moduli $G_\alp$ are computed adopting a Poisson's ratio $\nu=\SI{0.4}{}$. For the sake of completeness, the Young's moduli and the corresponding relaxation times are reported in Table~\ref{tab:PLA}. The material density is $\rho=\SI{1250}{kg/m^3}$. 

\begin{table}
\centering
	\begin{tabular}{ccccc}
$\alp$&$E_\alp$ [$\SI{}{{N}/{m^2}}$] 	& $\tau_\alp$ [$\SI{}{s}$] \\
	\hline
$\infty$&\SI{2.80e5}{} &- \\
$1$&\SI{1.577e8}{} &\SI{0.02}{}\\
$2$&\SI{3.610e7}{} &\SI{0.18}{}\\
$3$&\SI{4.095e8}{} &\SI{17}{}\\
$4$&\SI{7.580e8}{} &\SI{117}{}\\
$5$&\SI{1.200e6}{} &\SI{1000}{}\\
$6$&\SI{5.800e6}{} &\SI{1600}{}\\
$7$&\SI{5.500e5}{} &\SI{1e4}{}\\
$8$&\SI{1.600e5}{} &\SI{1e5}{}\\
 \hline
	\end{tabular}

		\caption{Mechanical properties of PLA modelled with 8 Maxwell elements \cite{Wan_etal2022}.}\label{tab:PLA}
\end{table}

Due to the combination of challenging geometry and loading condition, the beam is discretized with basis functions of degree $p=6$ and $\rm n = 250$ collocation points. 
The total simulation time is $T=\SI{4}{s}$ with a time step size $h=\SI{0.005}{s}$. 
To properly capture the viscoelastic effects on the dynamic response of the beam, we carried out also a simulation considering a linear elastic material characterized by $E=E_0=E_{\infty}+\sum_{\alp=1}^8{E_{\alp}}$ and $\nu=0.4$. 

\begin{figure}
\centering
\subfigure[Free end displacement along $x_1$-axis\label{fig:Spiral_beam_u_1}]{\begin{overpic}[width=0.8\textwidth]{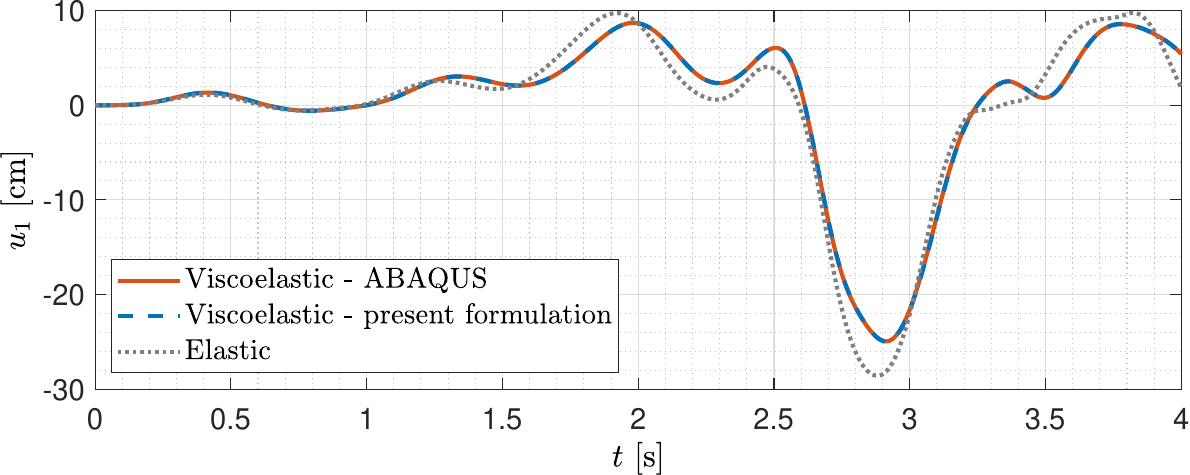}\end{overpic}}
\subfigure[Free end displacement along $x_2$-axis.\label{fig:Spiral_beam_u_2}]{\begin{overpic}[width=0.8\textwidth]{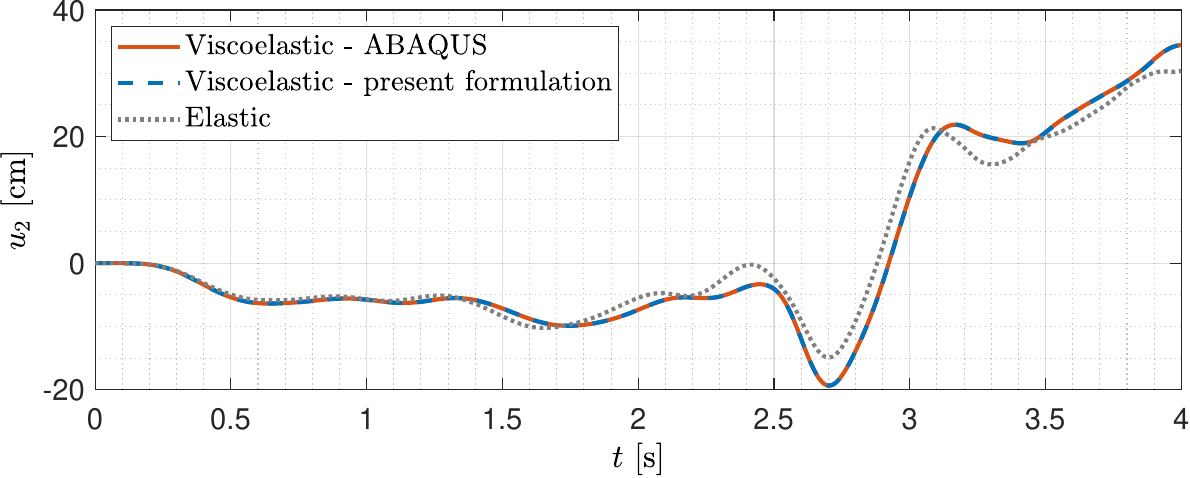}\end{overpic}}
\subfigure[Free end displacement along $x_3$-axis.\label{fig:Spiral_beam_u_3}]{\begin{overpic}[width=0.8\textwidth]{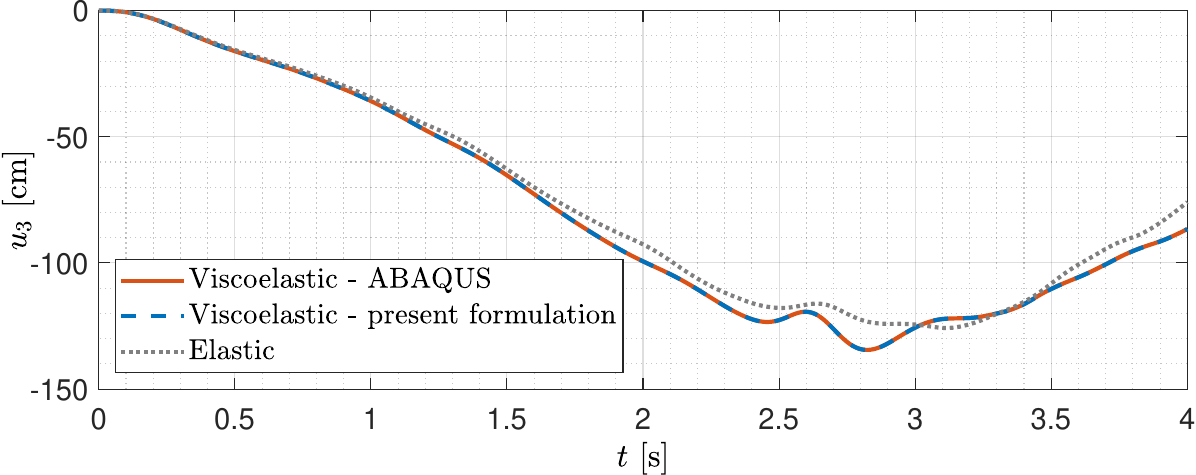}\end{overpic}}
\caption{\r{Spiral beam: free end displacements time histories.}\label{fig:Spiral_beam_u}}
\end{figure}

The three components of the free end displacement are reported in Figure~\ref{fig:Spiral_beam_u}. \r{Orange solid lines refers to the Abaqus solution, blue dashed lines refer to the viscoelastic case,} whereas grey dotted lines to the linear elastic one. \r{Again, an excellent agreement with Abaqus is observed for all the displacement components}. 

\r{Concerning the elastic and viscoelastic responses}, the differences between the two materials tend to progressively increase from $t=\SI{0.5}{s}$ on. Large peaks/crests differences are observed, see for example at $t=\SI{1.9}{s}$, $t=\SI{2.3}{s}$ and $t=\SI{2.8}{s}$ for $u_1$ (Figure~\ref{fig:Spiral_beam_u_1}). 
Furthermore, after the bounce occurring around $\SI{2.5}{s}$ the deformation becomes extremely complex and the viscous effects become significant for all the three components.

\begin{figure}
\vspace{-1.4cm}
\centering
\subfigure[Front ($x_1$-$x_3$) view of the deformed configurations for $t=\SI{0.5}{s}$ (left), $t=\SI{1.9}{s}$ (center-left), $t=\SI{2.8}{s}$ (center-right) and $t=\SI{4}{s}$ (right).\label{fig:Spiral_beam_snapshots_a}]
{
\begin{overpic}
[clip,width=.85\textwidth]{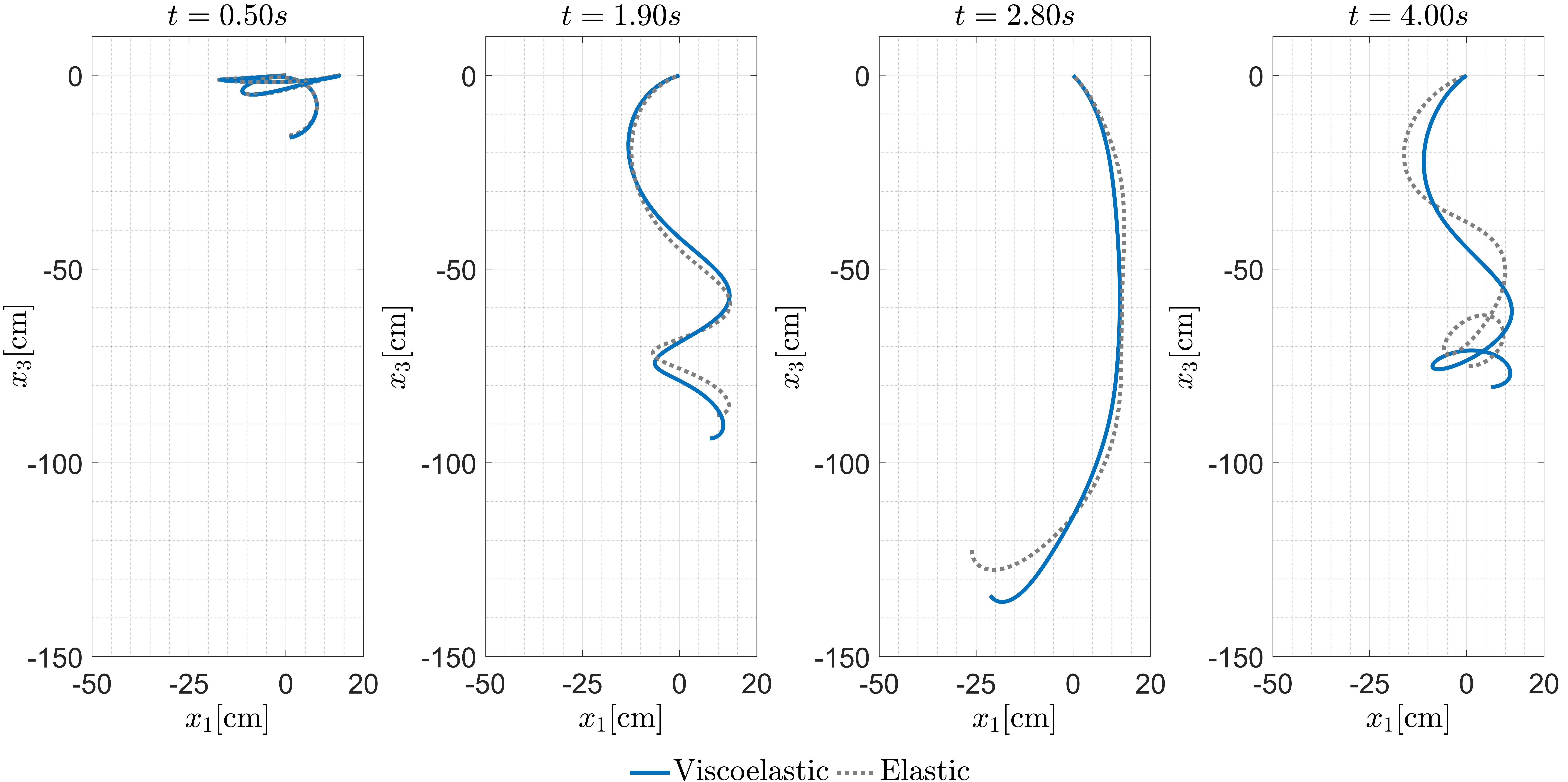}
\end{overpic}
}
\subfigure[Lateral ($x_2$-$x_3$) view of the deformed configurations for $t=\SI{0.5}{s}$ (right), $t=\SI{1.9}{s}$ (center-right), $t=\SI{2.8}{s}$ (center-left) and $t=\SI{4}{s}$ (left).\label{fig:Spiral_beam_snapshots_b}]
{
\begin{overpic}
[clip,width=.85\textwidth]{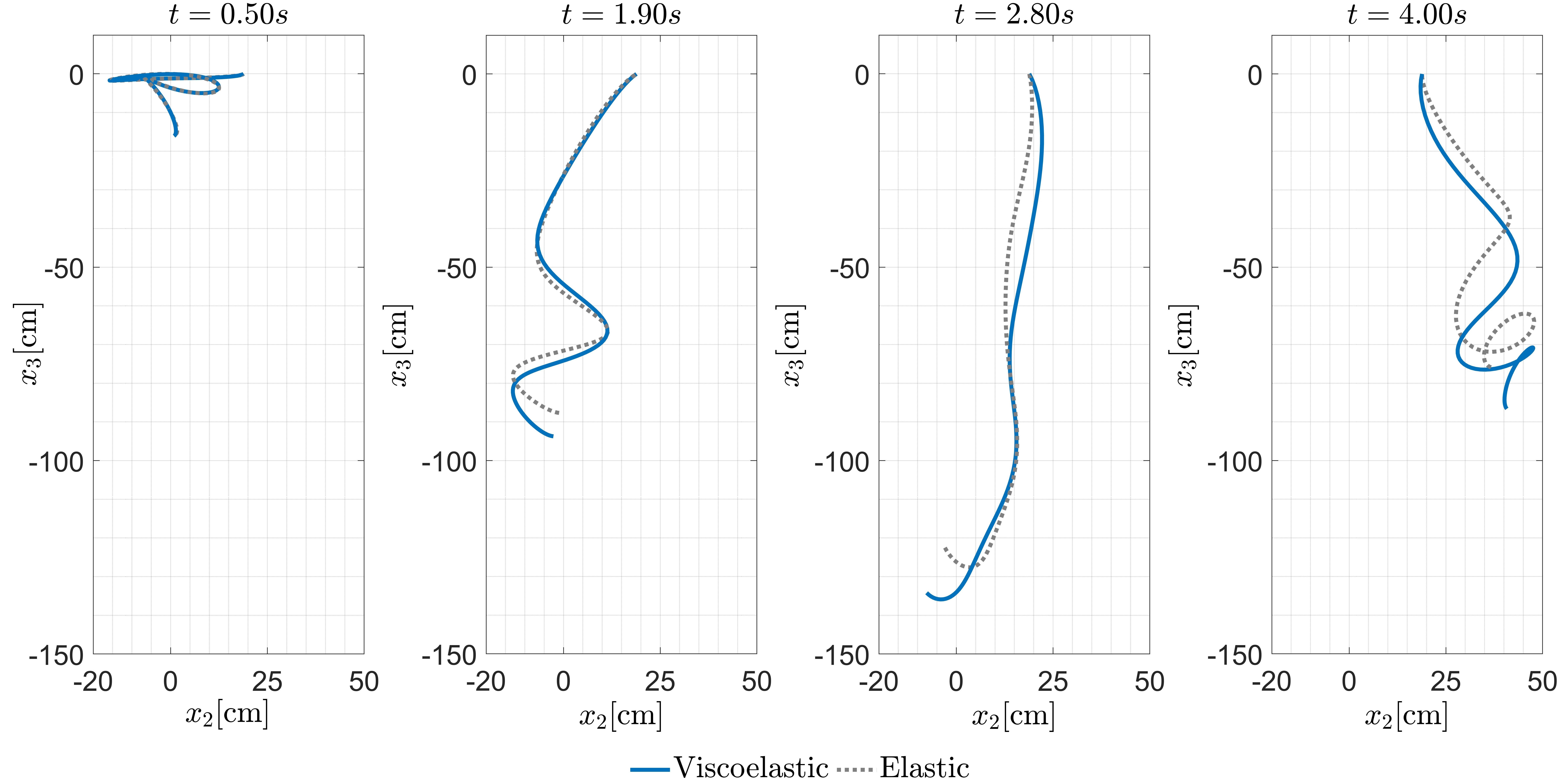}
\end{overpic}
}
\subfigure[Planar ($x_1$-$x_2$) view of the deformed configurations for $t=\SI{0.5}{s}$ (right), $t=\SI{1.9}{s}$ (center-right), $t=\SI{2.8}{s}$ (center-left) and $t=\SI{4}{s}$ (left).\label{fig:Spiral_beam_snapshots_c}]
{
\begin{overpic}[width=.85\textwidth]{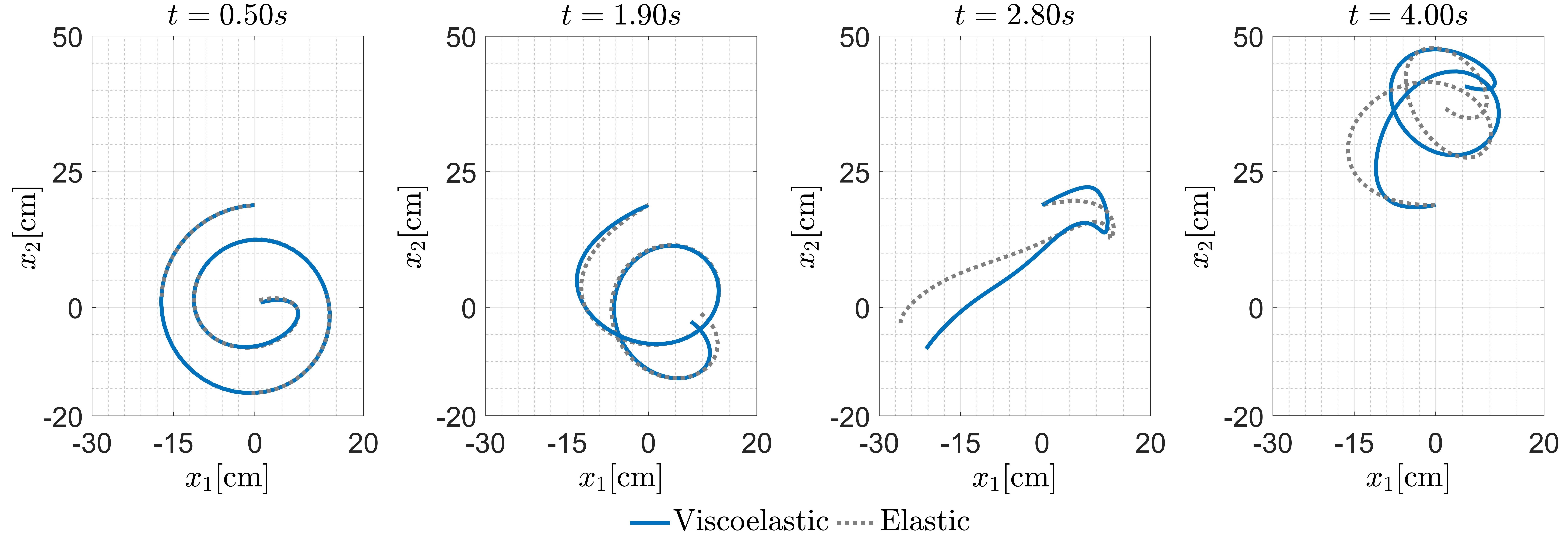}
\end{overpic}
}
\vspace{-.25cm}
\caption{Plane views of the spiral displacements.\label{fig:Spiral_beam_snapshots}}
\end{figure}

\begin{figure}
\centering
\subfigure[Deformed shape, $t=\SI{0.5}{s}$.\label{fig:15a}]{\begin{overpic}[clip,width=0.4\textwidth]{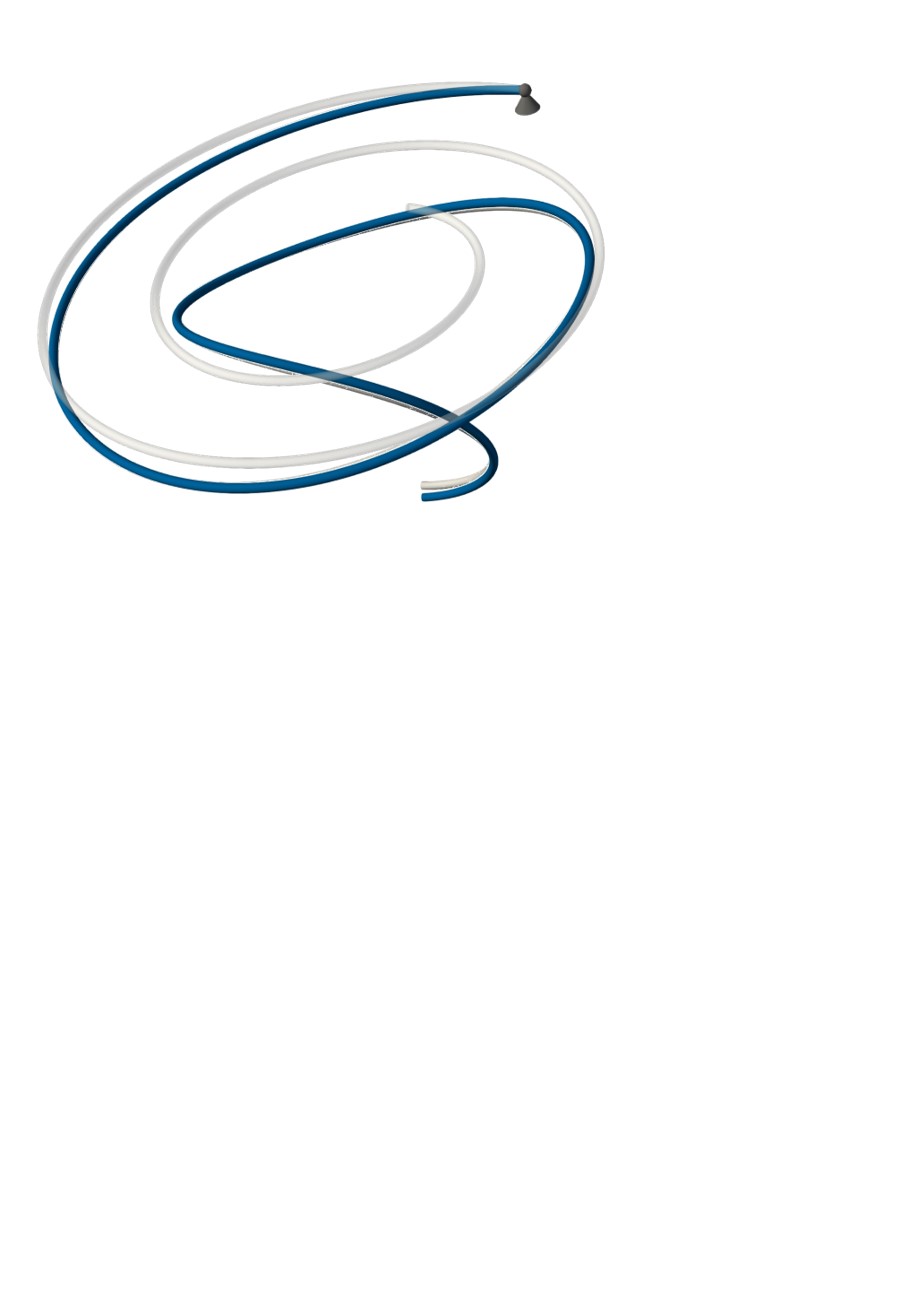}
\end{overpic}}\hspace{0.5cm}
\subfigure[Deformed shape, $t=\SI{1.9}{s}$.\label{fig:15b}]{\begin{overpic}[clip,width=0.4\textwidth]{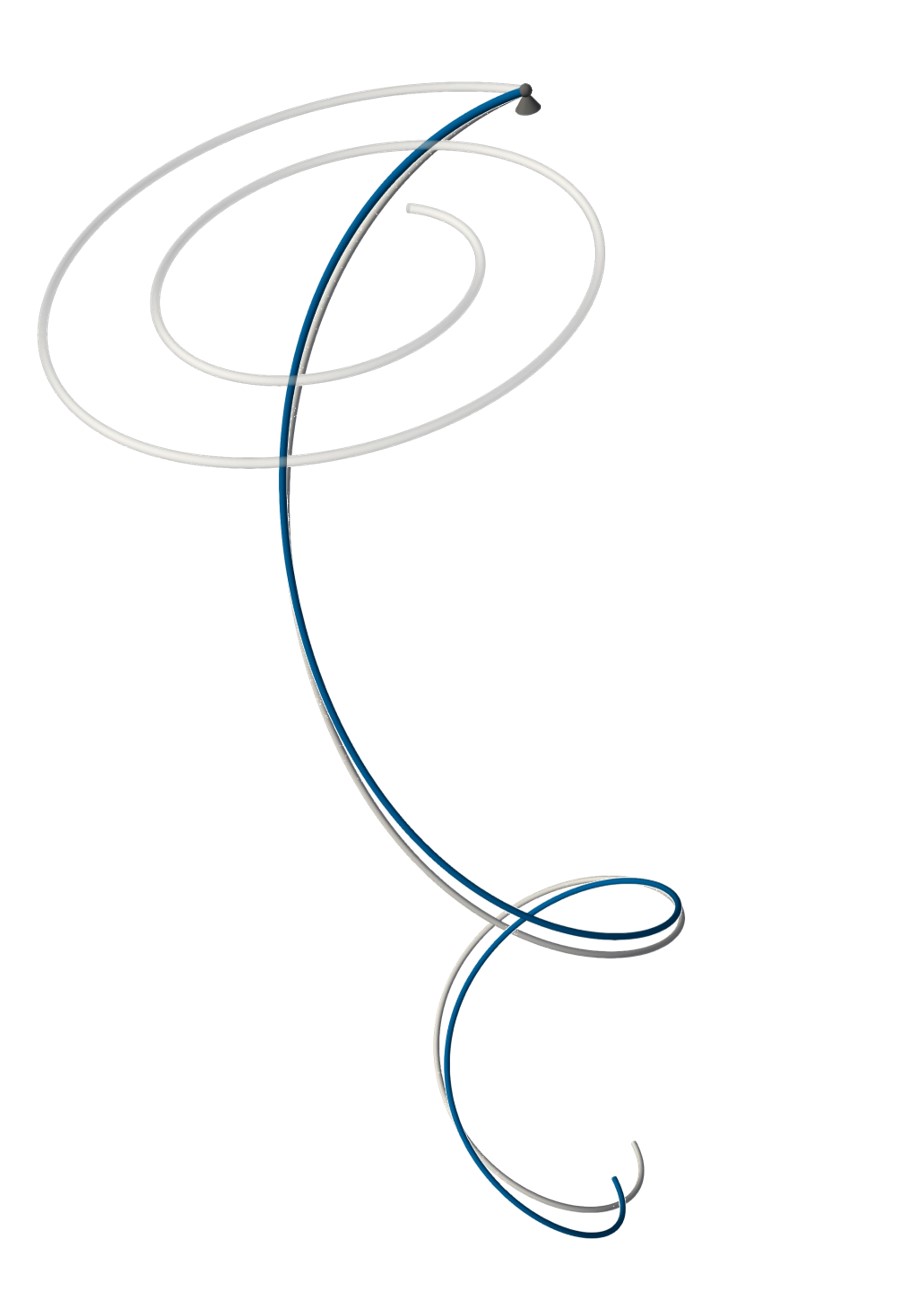}
\end{overpic}}
\subfigure[Deformed shape, $t=\SI{2.8}{s}$.\label{fig:15c}]{\begin{overpic}[clip,width=0.4\textwidth]{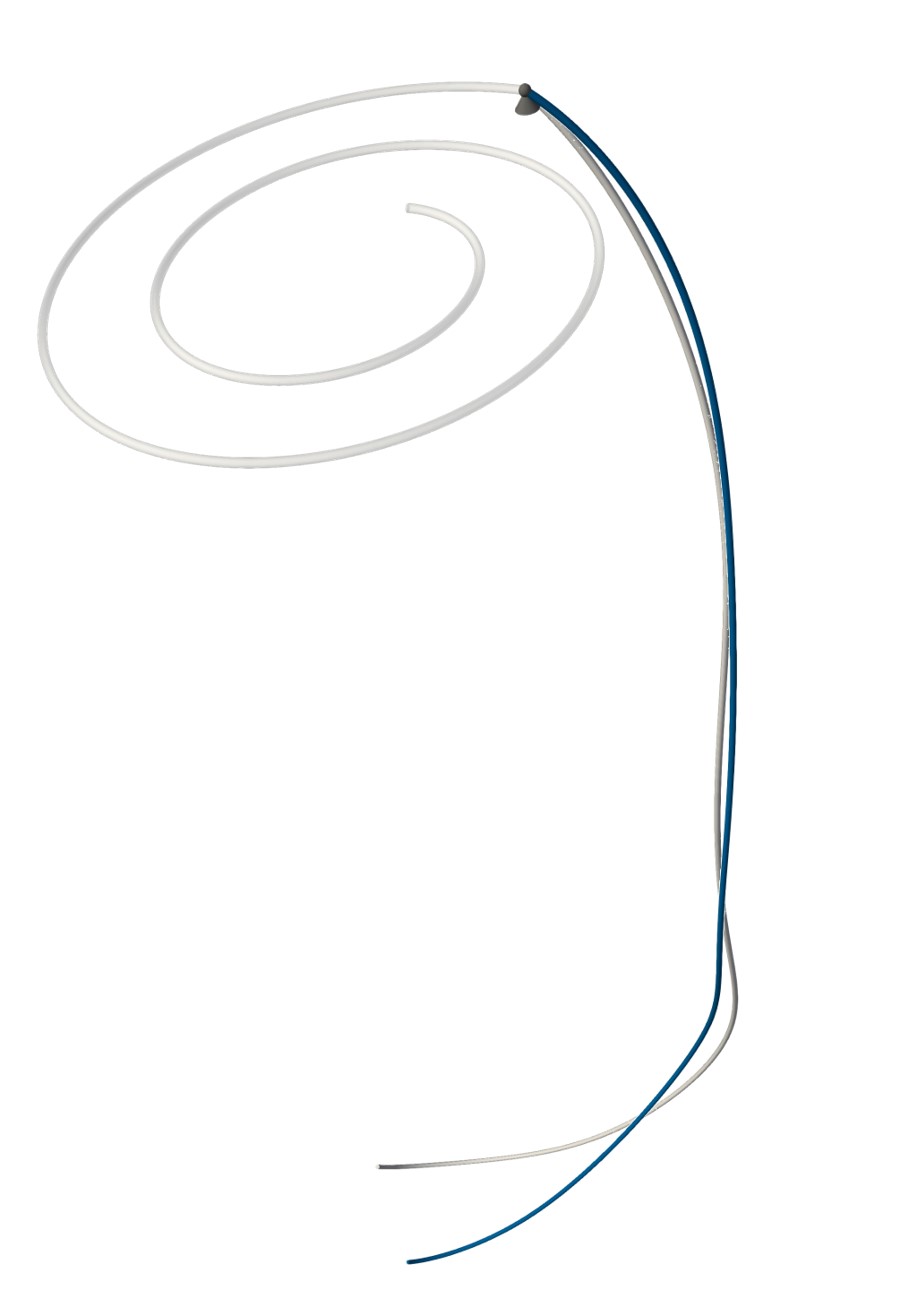}
\end{overpic}}\hspace{0.5cm}
\subfigure[Deformed shape, $t=\SI{4.0}{s}$.\label{fig:15d}]{\begin{overpic}[clip,width=0.4\textwidth]{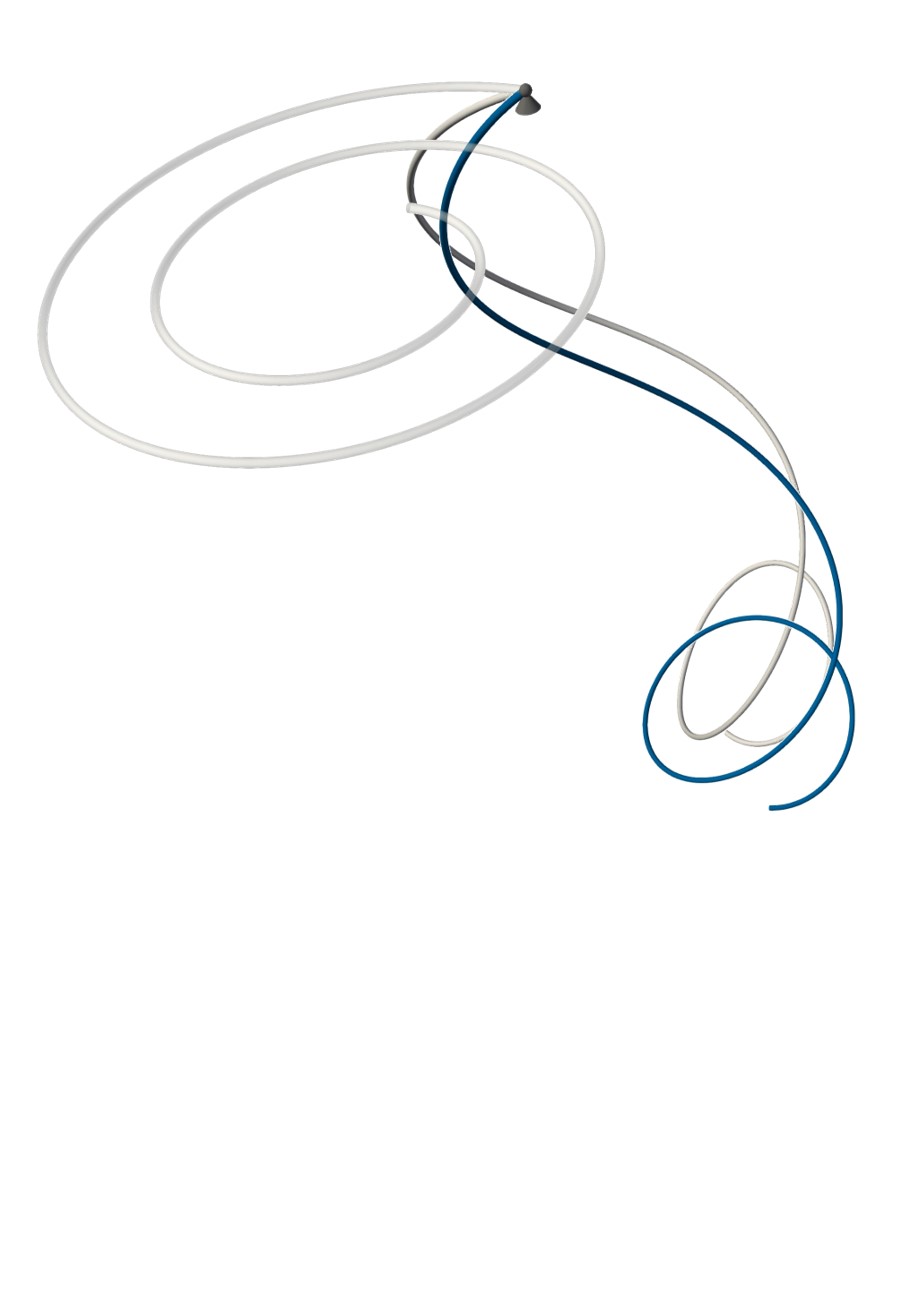}
\end{overpic}}
\caption{Spiral spring: Three-dimensional views of the deformed configurations modelled with elastic (solid grey) and viscoelastic (solid blue) materials.\label{fig:15}}
\end{figure}

To help capture the complexity of the deformation process, in Figure~\ref{fig:Spiral_beam_snapshots} we report some views of four significant snapshots. Front view of the deformed configurations are reported in Figure~\ref{fig:Spiral_beam_snapshots_a} for $t=\SI{0.5}{s}$, $t=\SI{1.9}{s}$, $t=\SI{2.8}{s}$ and $t=\SI{4}{s}$, while lateral and planar views are shown in Figure~\ref{fig:Spiral_beam_snapshots_b} and~\ref{fig:Spiral_beam_snapshots_c}. The maximum elongation of the spiral occurs at $t=\SI{2.8}{s}$. After that, the recoil internal forces cause the typical bouncing of the beam upwards. As expected, such an effect is more pronounced for the elastic case (grey dotted lines) compared to the viscoelastic one (blue solid lines). Extruded three-dimensional views of the deformed configurations are reported in Figure~\ref{fig:15}. 

\subsection{Curved lattice}\label{sec:lattice}
\begin{figure}
\centering
\subfigure[Initial geometry (units are in $\SI{}{mm}$).\label{fig:Curved_net_geom_a}]
{
\begin{overpic}
[clip,width=.45\textwidth]{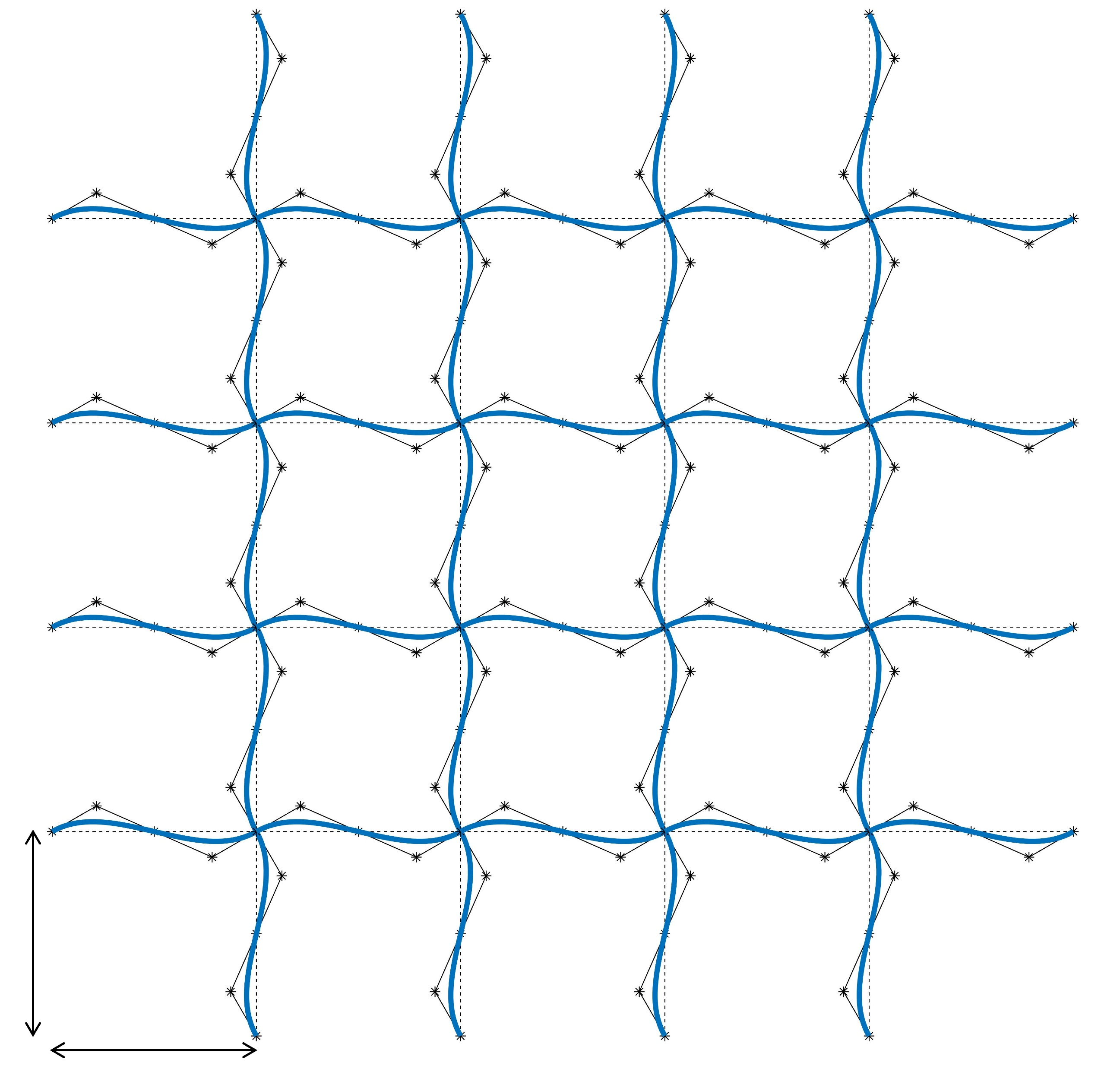}
\put(-5,90){\colorbox{white}{\makebox(40,50){\textcolor{black}{12}}}}\put(100,-5){\colorbox{white}{\makebox(40,50){\textcolor{black}{12}}}}
\end{overpic}
}\hspace{0.5cm}
\subfigure[Construction of the curved lattice by rotation of the beams control points.\label{fig:Curved_net_geom_b}]
{
\begin{overpic}
[clip,width=0.4
\textwidth]{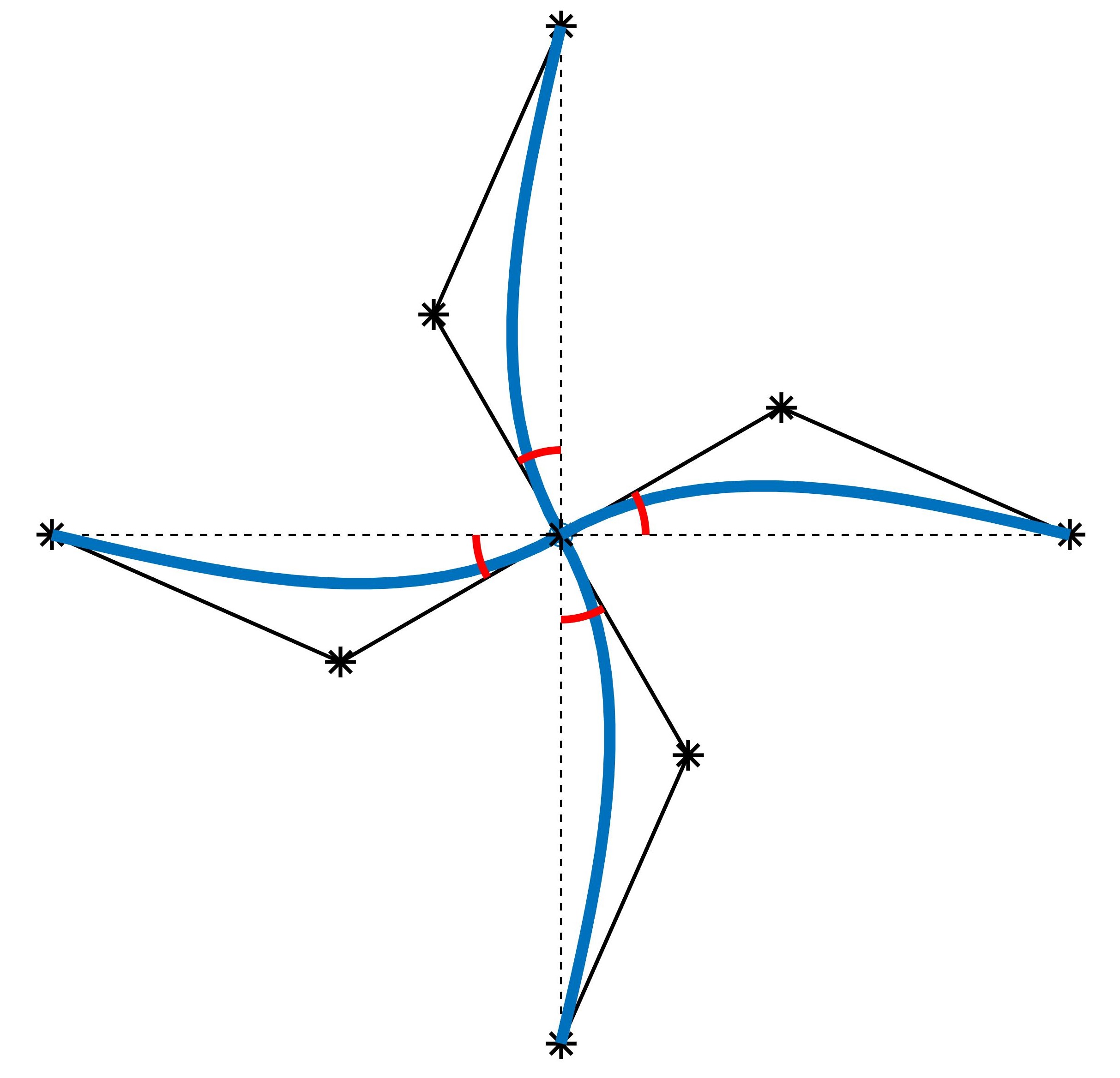}
\put(425,380){{\textcolor{red}{\small{$\psi$}}}}
\put(800,410){I}\put(510,800){II}\put(390,100){III}\put(100,510){IV}
\end{overpic}
}
\caption{Plane lattice with parametric curved nodes.\label{fig:Curved_net_geom}}
\end{figure}
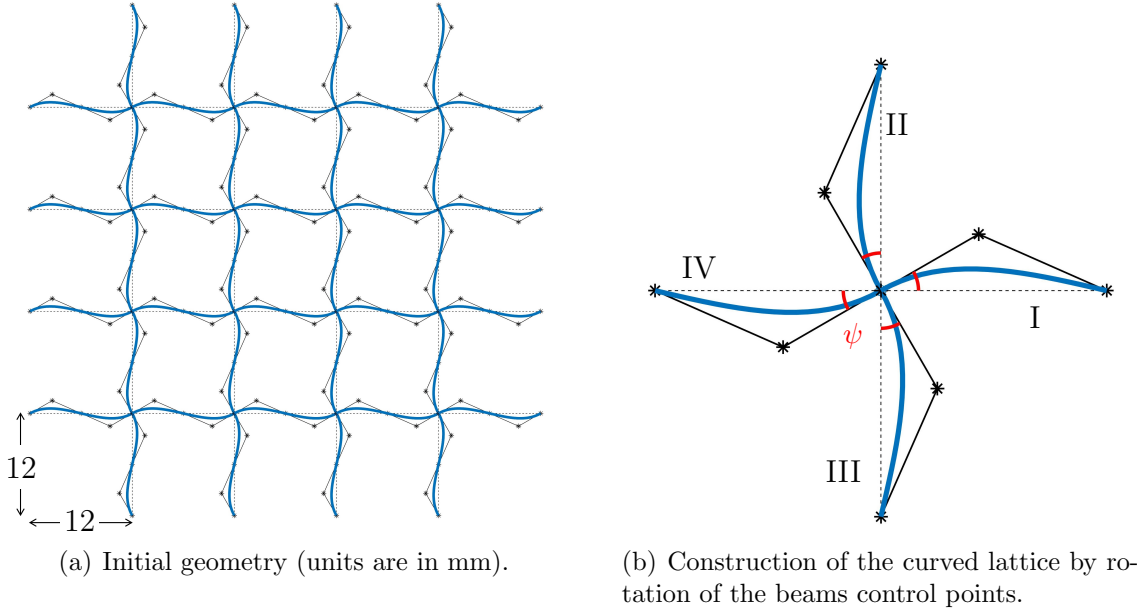

To show the potentialities of the proposed model to simulate and design mechanical meta-materials, we present a curved multi-patch net structure. The system consists of a two-dimensional lattice initially laying in the plane $(x_1,x_2)$ whose unit cell has a tunable curved geometry (see Figure~\ref{fig:Curved_net_geom_a}). Namely, the lattice initial shape is controlled by the rigid rotation of the crosses at the grid vertexes by an angle $\psi$ (see Figure~\ref{fig:Curved_net_geom_b}). 
Leveraging the IGA potentialities, this is straightforwardly done by collectively rotating the four control points surrounding the grid vertexes of the same quantity. 
The net is hinged at the external nodes. The same material adopted for the spiral beam is here employed (see Table~\ref{tab:PLA}). 
\r{A distributed load, $q_3(t)=-0.05(1-\textnormal{sin}(4\pi t+\pi/2))$, orthogonal to the plane of the initial net is applied to the four beams composing the central cell of the lattice for $t\in \SI{[0, 2]}{s}$, and then removed for $t>\SI{2}{s}$ (see the load time history in Figure~\ref{fig:Net_load})}. Such a load pattern is meant to resemble, to some extent, two impacts of a mass that is approximately 1000 times heavier than the structure. 

\begin{figure}
\centering
\includegraphics[width=1\textwidth]{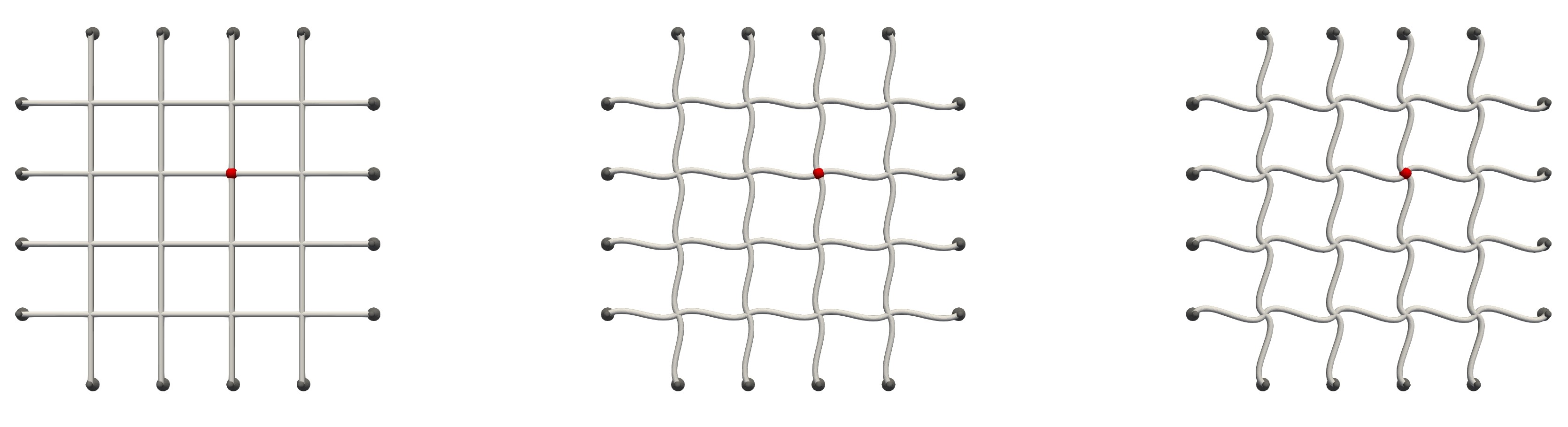}
\caption{Curved lattices analysed: $\psi=0$ (left), $\psi=\pi/6$ (centre) and $\psi=\pi/3$ (right).\label{fig:Net_configs}}
\end{figure}

\begin{figure}
\centering
\includegraphics[width=0.9\textwidth]{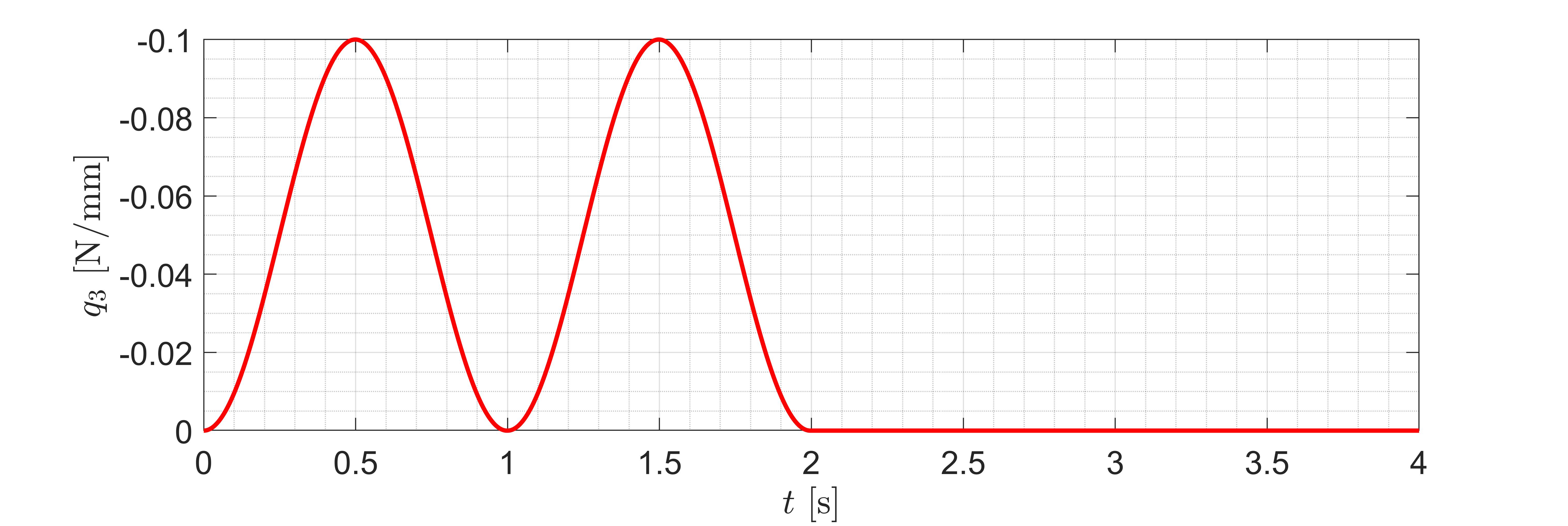}
\caption{Curved lattice: distributed load time history applied on the four beams of the central cell.\label{fig:Net_load}}
\end{figure}

\begin{figure}
\centering
\subfigure[$\psi=0$.\label{fig:curved_lattice_u_a}]
{\includegraphics[width=0.9\textwidth]{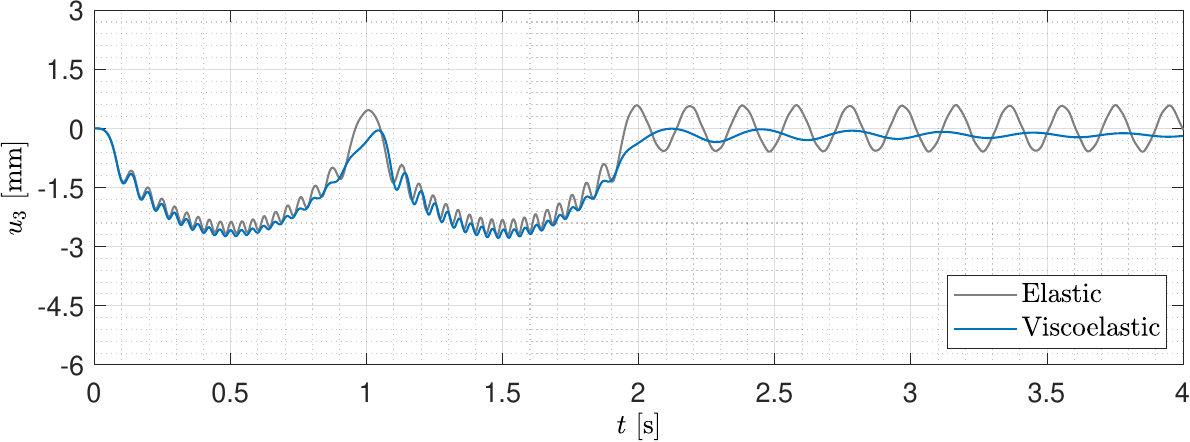}}
\subfigure[$\psi=\pi/6$.\label{fig:curved_lattice_u_b}]
{\includegraphics[width=0.9\textwidth]{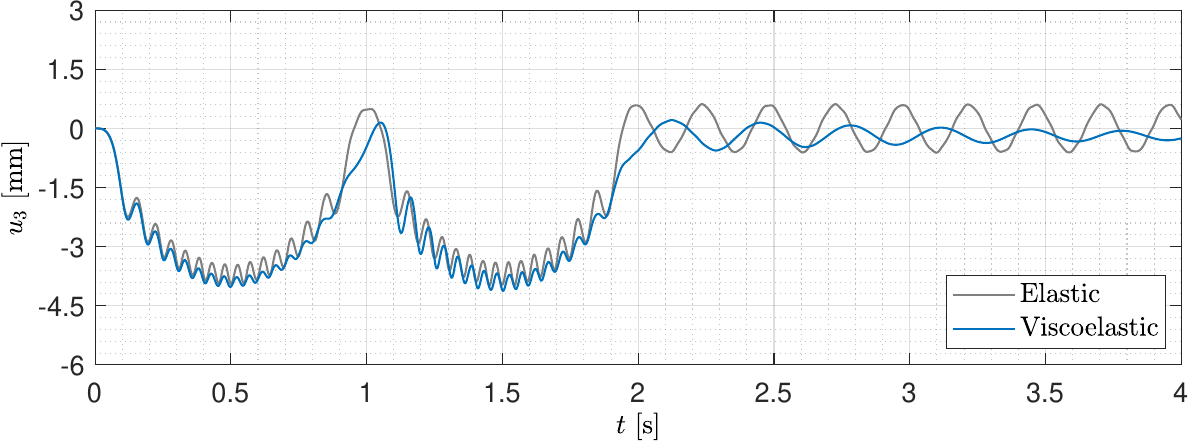}}
\subfigure[$\psi=\pi/3$.\label{fig:curved_lattice_u_c}]
{\includegraphics[width=0.9\textwidth]{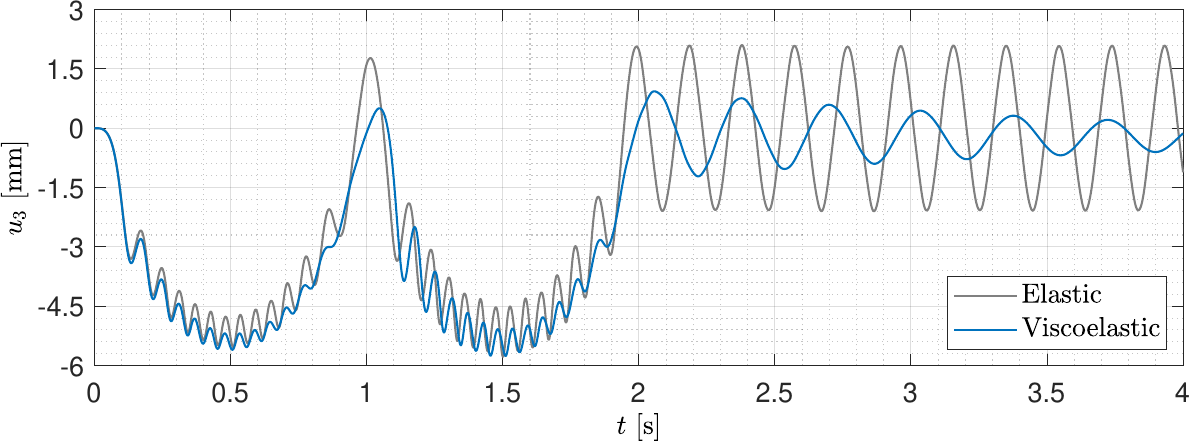}}
\caption{Comparison of the deflection of the curved lattice made of elastic (solid grey) and viscoelastic (solid blue) materials.\label{fig:curved_lattice_u}}
\end{figure}

\begin{figure}
\centering
\subfigure[$\psi=0$.\label{fig:curved_lattice_3D_a}]
{\includegraphics[width=0.9\textwidth]{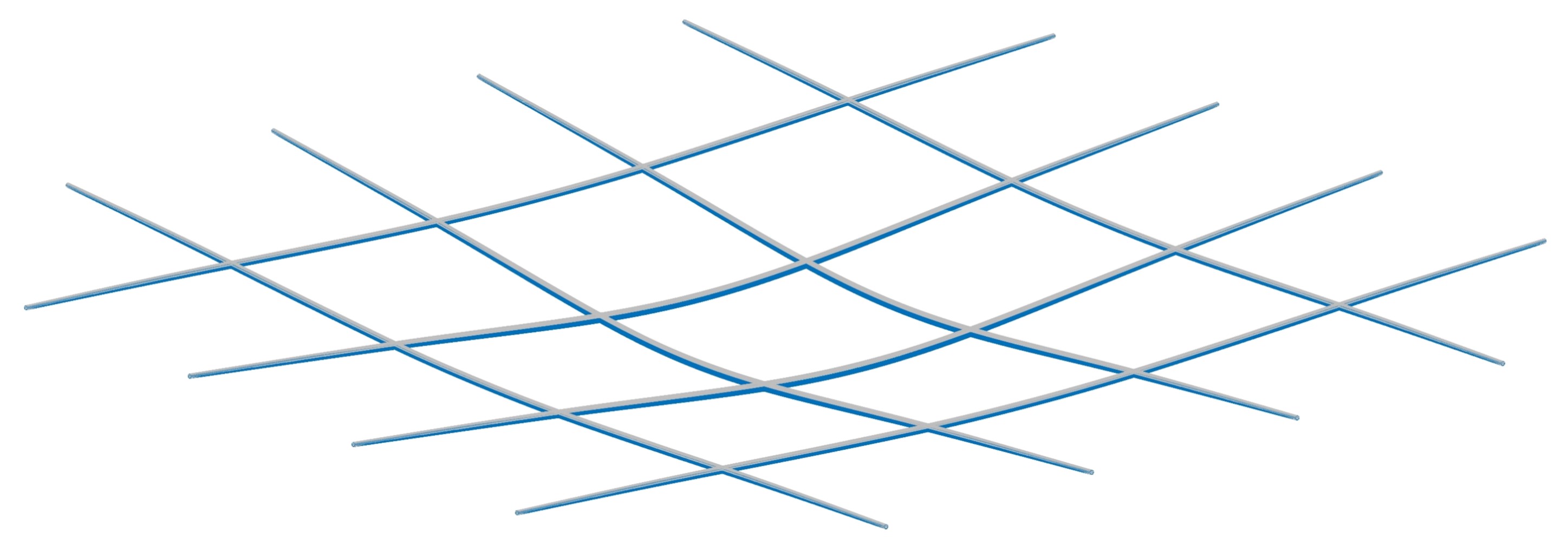}}
\subfigure[$\psi=\pi/6$.\label{fig:curved_lattice_3D_b}]
{\includegraphics[width=0.9\textwidth]{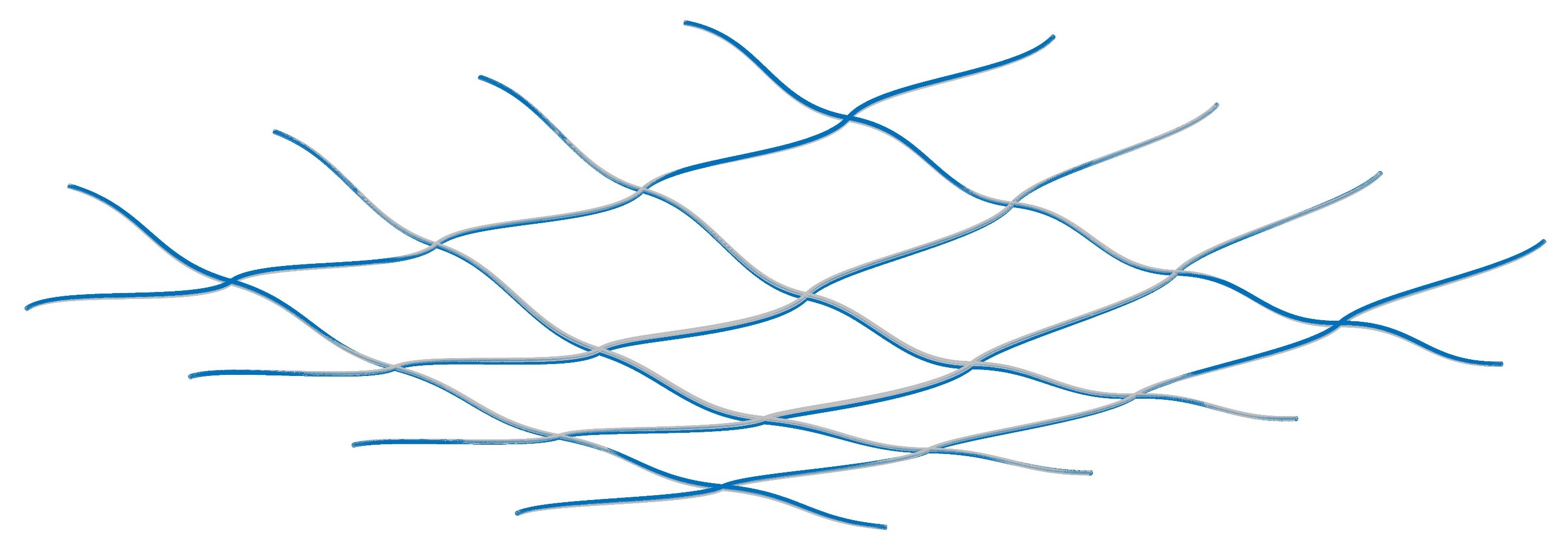}}
\subfigure[$\psi=\pi/3$.\label{fig:curved_lattice_3D_c}]
{\includegraphics[width=0.9\textwidth]{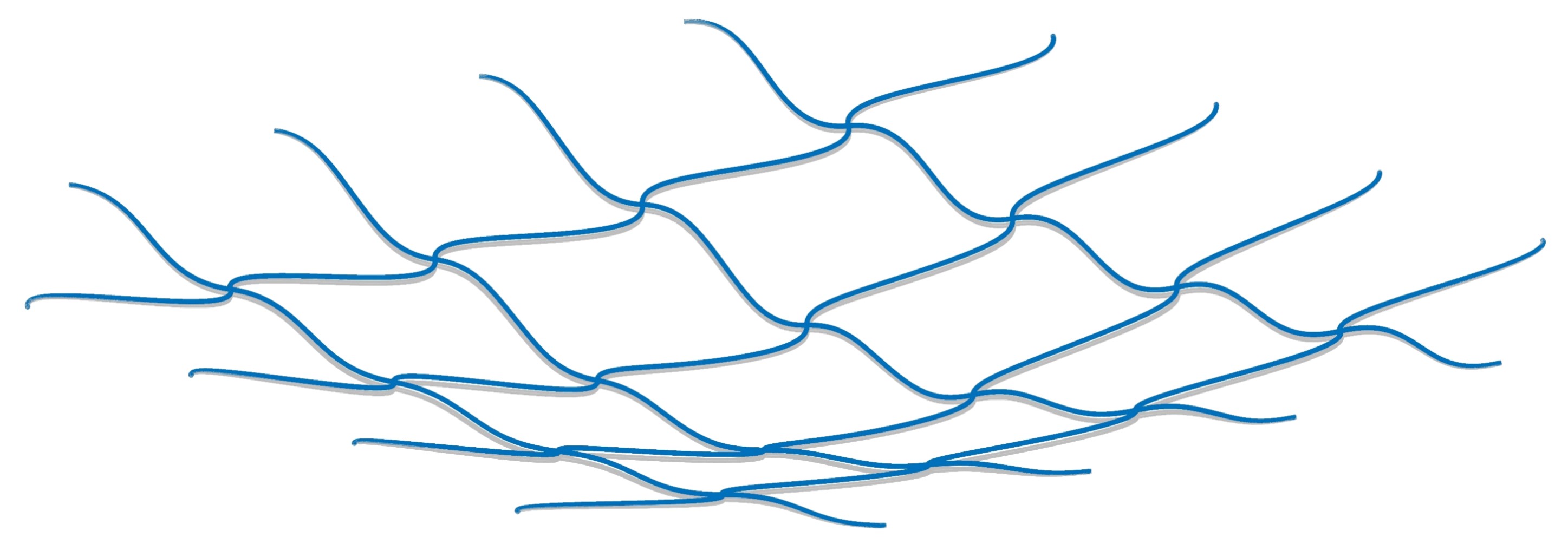}}
\caption{3D view of the lattice centroid for $t=\SI{1.5}{s}$ with elastic (solid grey) and viscoelastic (solid blue) materials.\label{fig:curved_lattice_3D}}
\end{figure}

The displacement in the $x_3$ direction of the top-right node of the central cell (red dot in Figure~\ref{fig:Net_configs}) is reported in Figure~\ref{fig:curved_lattice_u}, while the deformed shapes at $t=\SI{1.5}{s}$ for the three geometries analyzed are shown in Figure~\ref{fig:curved_lattice_3D}. 
As can be seen, the curvature of the beams play a crucial role in determining some features of the dynamic response of the structure. Controlled by the parameter $\psi$, it significantly increases the out-of-plane flexibility of the lattice. Moreover, while for the cases $\psi=0$ (Figure~\ref{fig:curved_lattice_u_a}) and $\psi=\pi/6$ (Figure ~\ref{fig:curved_lattice_u_b}) the free oscillations occurring from $\SI{2}{s}$ on present similar amplitudes, the lattice with $\psi=\pi/3$ (Figure~\ref{fig:curved_lattice_u_c}) undergoes a noticeable increase of the amplitudes. 
Due to the viscoelastic effects, it is also noted a progressive shifting of the system natural frequencies due to the relaxation of the viscoelastic moduli. 
This numerical test demonstrates how, with the proposed approach, the dynamic reponse of a complex nonlinear system can be controlled, and possibly programmed, by tuning simultaneously geometric and topological parameters, in addition to the material properties. 

\r{
\subsection{Auxetic meta-material}
In the last numerical application, we study a three-dimensional auxetic meta-material (Figure~\ref{fig:Auxetic_metamat}). The inner cell structure of these meta-materials results in unique global properties, such as negative thermal expansion coefficient and negative Poisson's ratio. 
Following the same strategy adopted in Section~\ref{sec:lattice}, the cell is made of elements with tunable curvature controlled by rotating of an angle $\psi$ the first and the last control points of the diagonal patches (see Figures~\ref{fig:metamat_3D_cell} and~\ref{fig:metamat_2D_cell}).
The same material properties adopted for the spiral beam and the curved lattice are here employed. The structure consists in 120 patches, featuring a circular cross-section of diameter $d=\SI{0.25}{mm}$.
The system is simply supported in the $x_3$-direction at the bottom nodes located at $x_3=\SI{0}{}$ (see Figures~\ref{fig:Metamat_undef_pi0}-\ref{fig:Metamat_undef_pi6}), and forced by distributed vertical loads acting on the beams located at $x_3=\SI{12}{mm}$. 
The load time history, similar in shape to the one shown in Figure~\ref{fig:Net_load}, is $q_3(t)=\bar{q}_3/2(1-\textnormal{sin}(4\pi t/\SI{0.25}{s}+\pi/2))$ for $t\leqslant\SI{0.25}{s}$, $q_3(t)=0$ for $t>\SI{0.25}{s}$, with peaks value $\bar{q}_3=\SI{-0.0241} {N/mm}$ (approximately 10000 times the self-weight per unit-length of the structure).

\begin{figure}
\centering
\subfigure[3D view of the meta-material cell.\label{fig:metamat_3D_cell}]
{
\begin{overpic}
[clip,width=.45\textwidth]{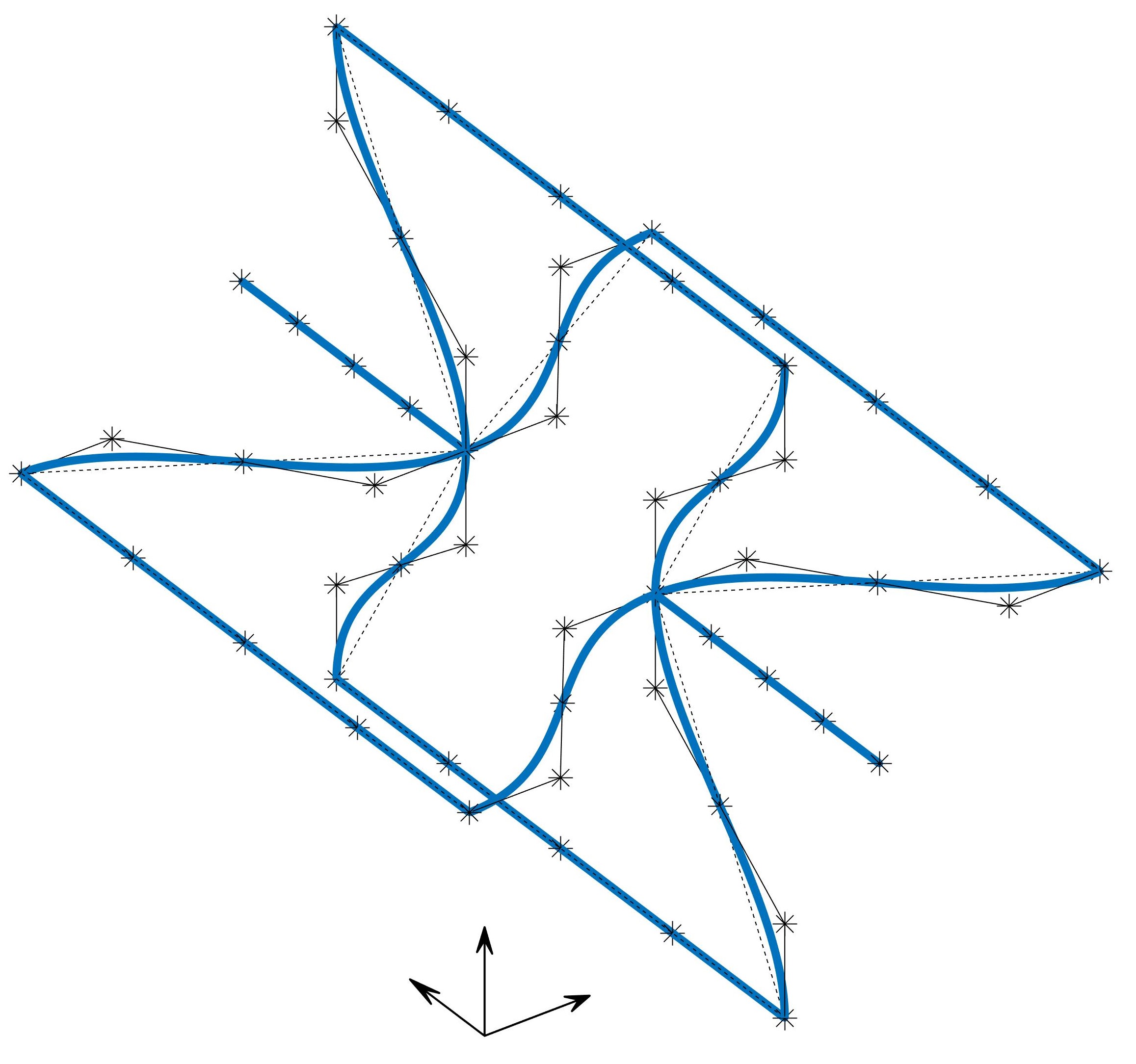}
\put(515,20){$x_1$}
\put(300,60){$x_2$}
\put(440,110){$x_3$}
\end{overpic}
}\hspace{0.5cm}
\subfigure[Planar ($x_2$-$x_3$) view of the curved cell (units are in $\SI{}{mm}$).\label{fig:metamat_2D_cell}]
{
\begin{overpic}
[clip,width=0.45
\textwidth]{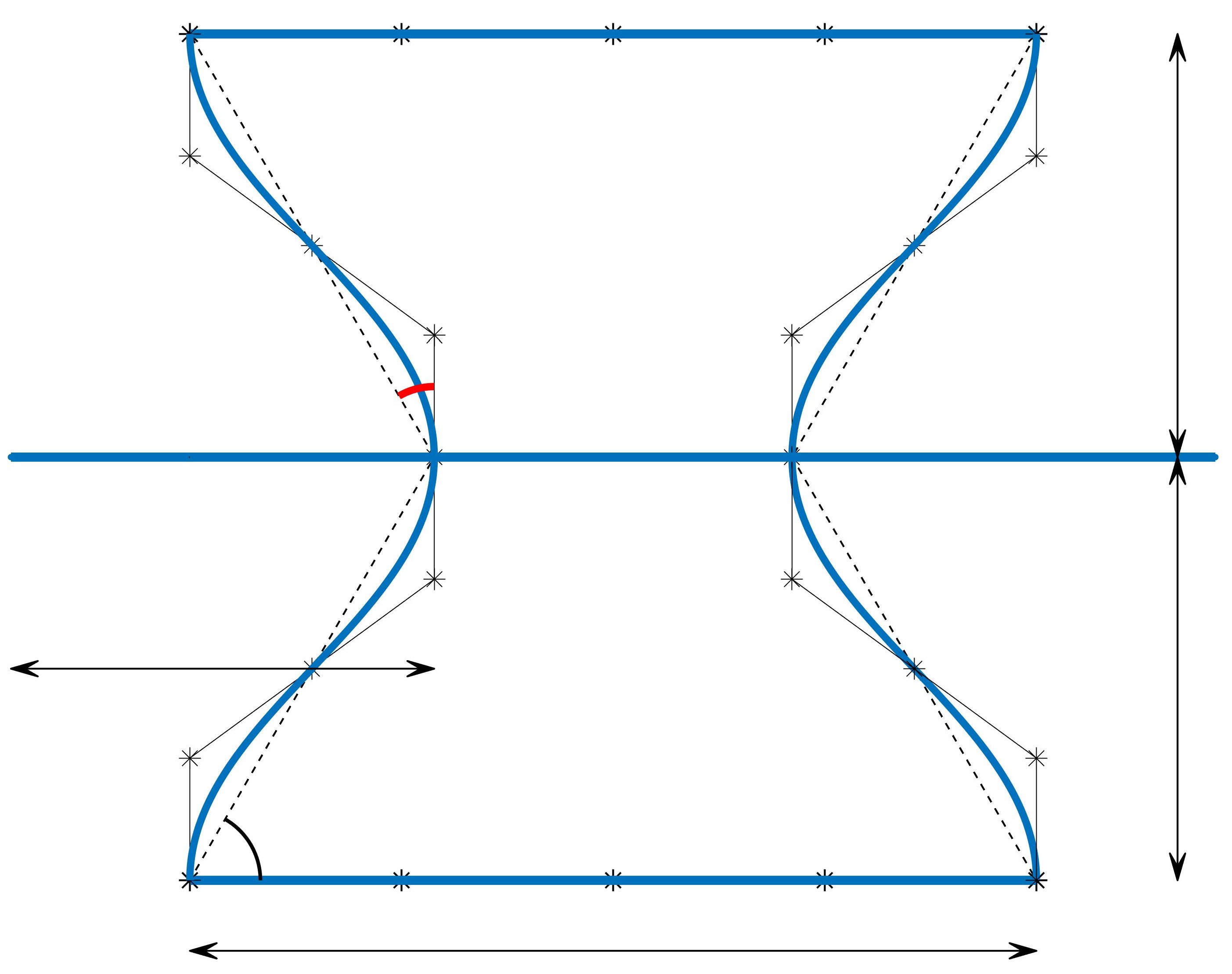}
\put(245,470){{\textcolor{red}{{$\psi$}}}}
\put(220,125){$\pi/3$}
\put(460,-7.5){\colorbox{white}{\makebox(40,50){\textcolor{black}{12}}}}
\put(150,235){\colorbox{white}{\makebox(30,30){\textcolor{black}{6}}}}
\put(925,235){\colorbox{white}{\makebox(30,30){\textcolor{black}{6}}}}
\put(925,555){\colorbox{white}{\makebox(30,30){\textcolor{black}{6}}}}
\end{overpic}
}
\centering
\subfigure[Meta-material structure with $\psi=0$.\label{fig:Metamat_undef_pi0}]
{\includegraphics[width=0.40\textwidth]{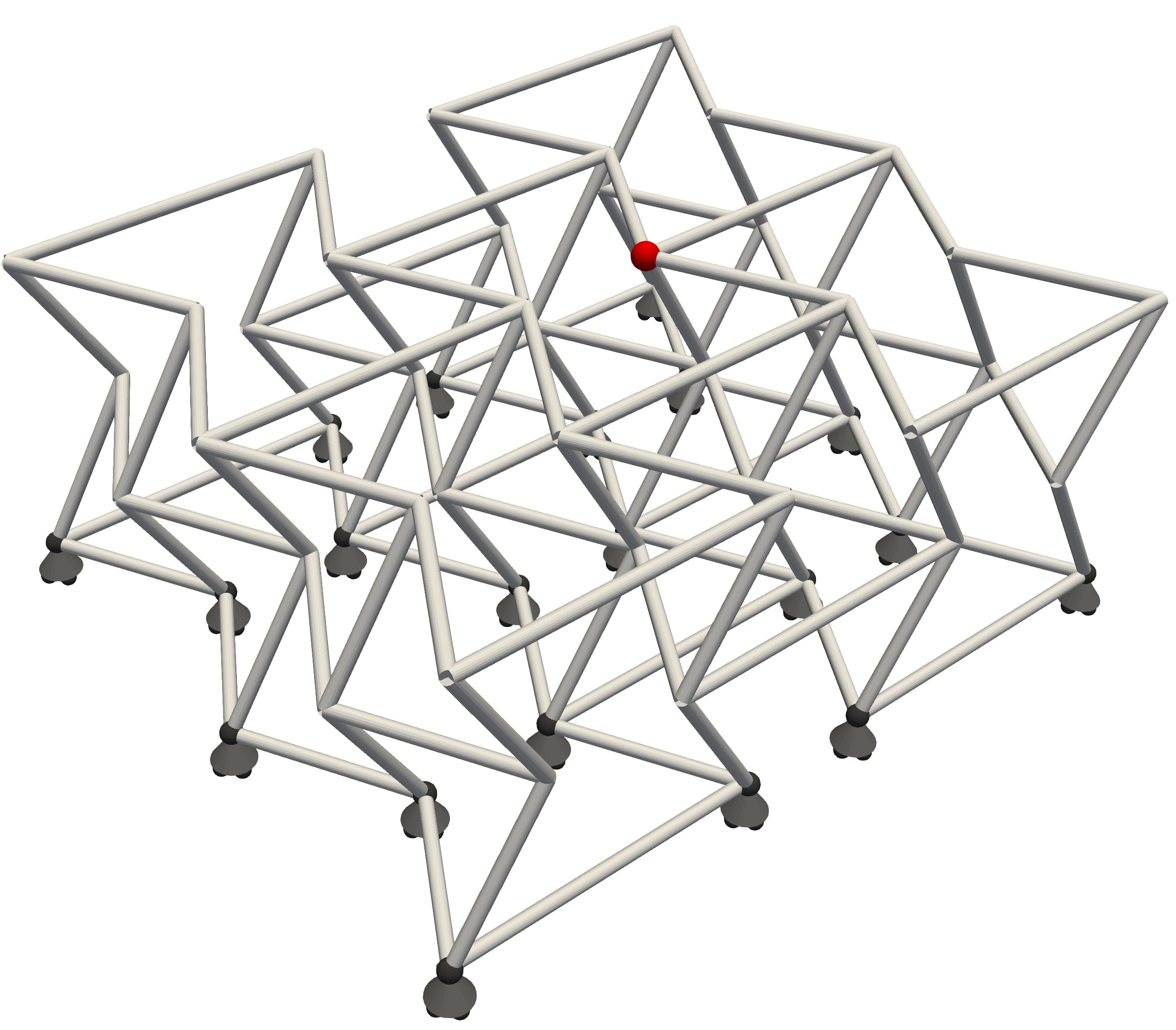}}
\subfigure[Meta-material structure with $\psi=\pi/6$.\label{fig:Metamat_undef_pi6}]
{\includegraphics[width=0.40\textwidth]{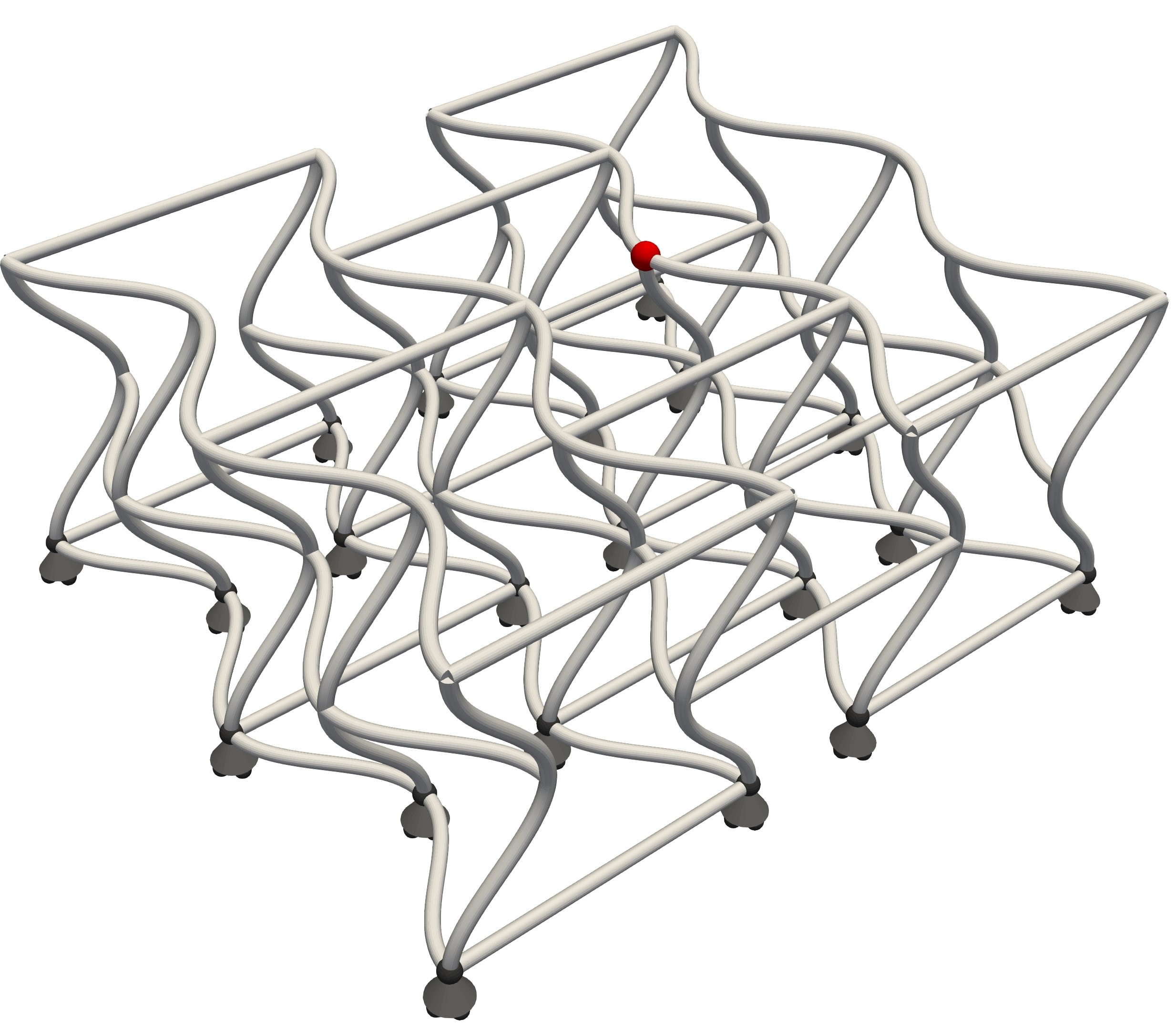}}
\centering
\subfigure[Distributed load applied to the structures.\label{fig:q3_meta}]
{\includegraphics[width=1\textwidth]{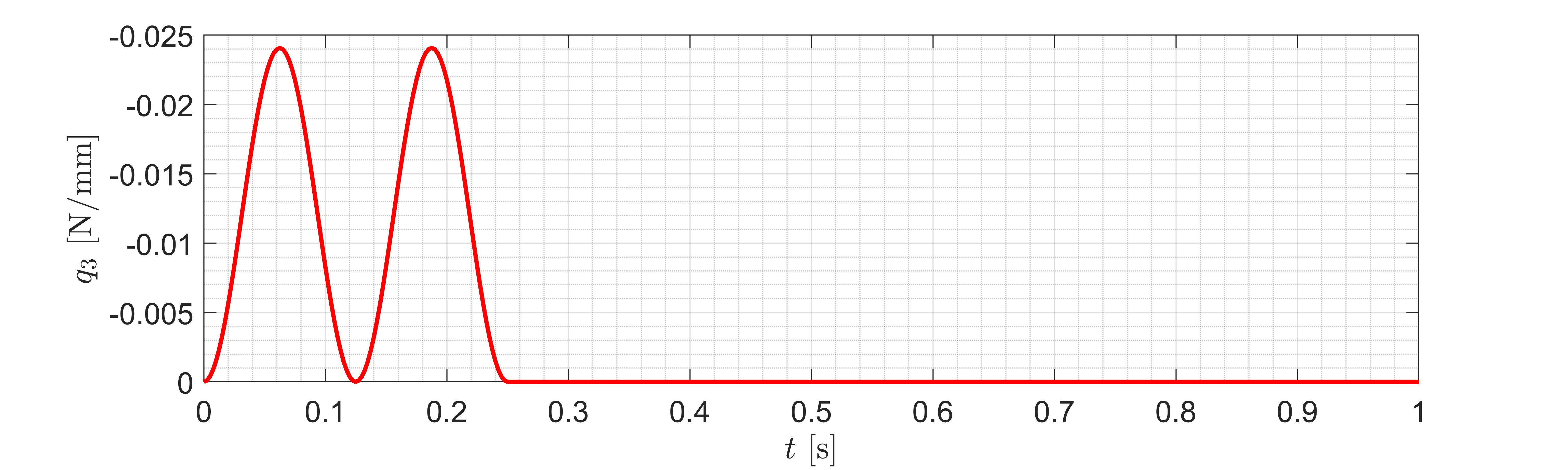}}
\caption{Auxetic meta-material.\label{fig:Auxetic_metamat}}
\end{figure}
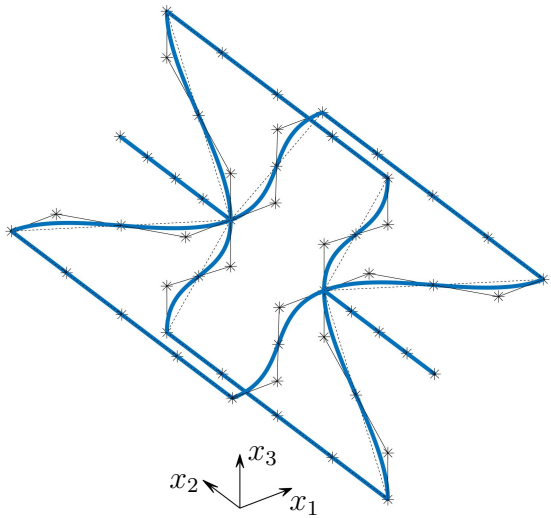
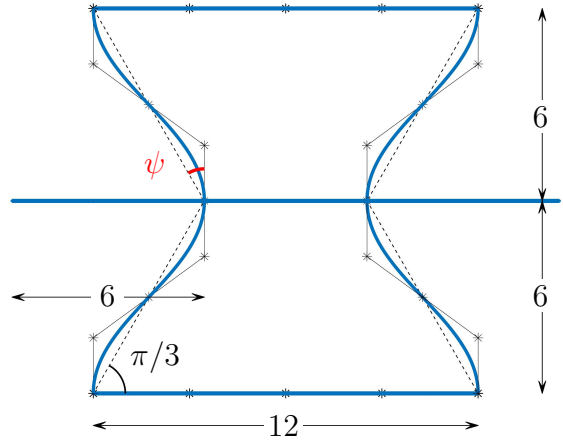
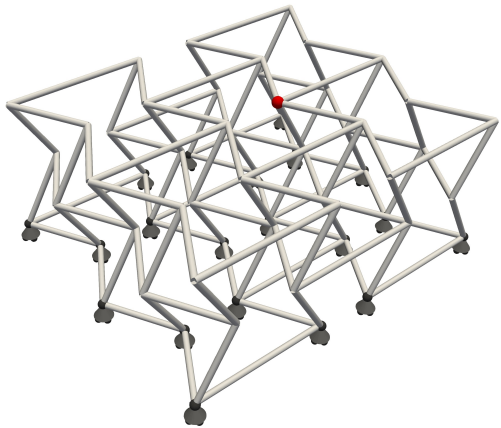
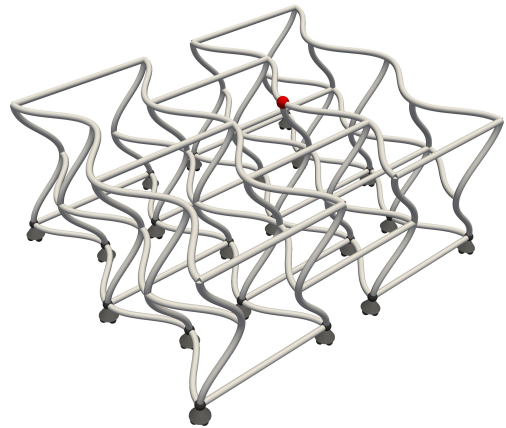
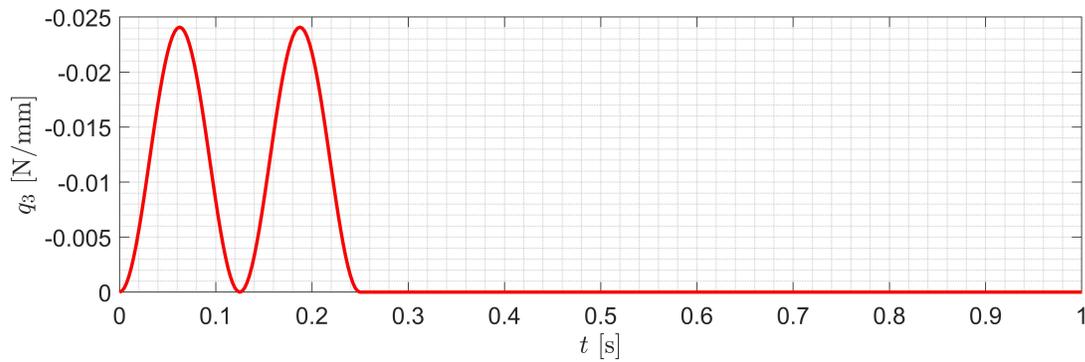

The time histories of the three displacement components of the central upper node of the meta-material (red dot in Figure~\ref{fig:Metamat_undef_pi0} and~\ref{fig:Metamat_undef_pi6}) are reported in Figure~\ref{fig:U_metamat}. Figures~\ref{fig:metamat_RETT_U1},~\ref{fig:metamat_RETT_U2} and~\ref{fig:metamat_RETT_U3} refer to the meta-material with $\psi=0$, whereas Figures~\ref{fig:metamat_CURVO_U1},~\ref{fig:metamat_CURVO_U2} and~\ref{fig:metamat_CURVO_U3} to the one with $\psi=\pi/6$. A significant effect of the cells curvature is noted, especially for the $u_2$ and $u_3$ components, with larger motions in the main peaks occurring for this case compared to the one with straight cells. 
Importantly, comparing the cases with $\psi=0$ and $\psi=\pi/6$, we observe that the curvature induces a shift of the response frequencies. This confirms our assumption about the possibility to design new meta-materials with predefined dynamic properties exploiting a larger design space that includes curvature of the cells, in addition to topology and material properties. 
The influence of the viscoelastic material on the dynamic response of the meta-materials is also clearly observed in terms of damping capabilities compared to the purely elastic response. 

In Figure~\ref{fig:metamat_deformed_plot}, we also show the deformed configurations of the two meta-materials (viscoelastic case only) at $t=\SI{0.1875}{s}$. The characteristic auxetic behaviour can be noticed especially from the planar views in Figures~\ref{fig:metamat_RETT_plane} and~\ref{fig:metamat_CURVO_plane}.

\begin{figure}
\centering
\subfigure[Displacement component along $x_1$-axis for the meta-material with $\psi=0$.\label{fig:metamat_RETT_U1}]
{\includegraphics[width=0.47\textwidth]{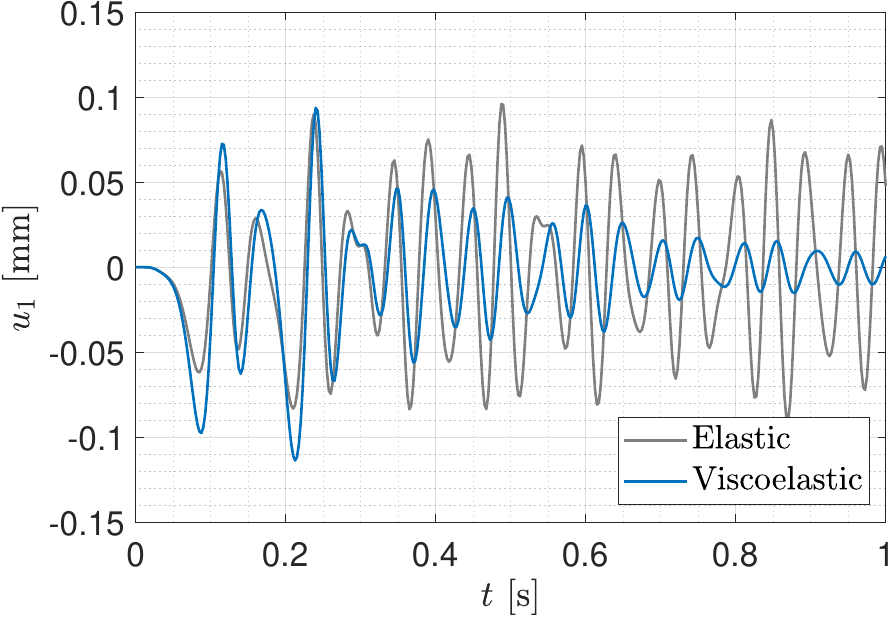}}
\hspace{0.3cm}
\subfigure[Displacement component along $x_1$-axis for the meta-material with $\psi=\pi/6$.\label{fig:metamat_CURVO_U1}]
{\includegraphics[width=0.47\textwidth]{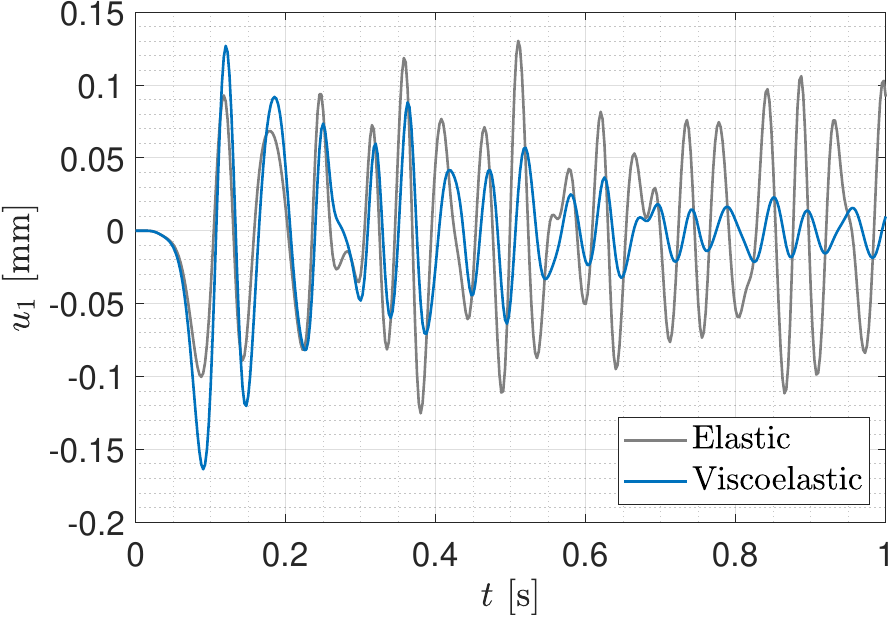}}
\centering
\subfigure[Displacement component along $x_2$-axis for the meta-material with $\psi=0$.\label{fig:metamat_RETT_U2}]
{\includegraphics[width=0.47\textwidth]{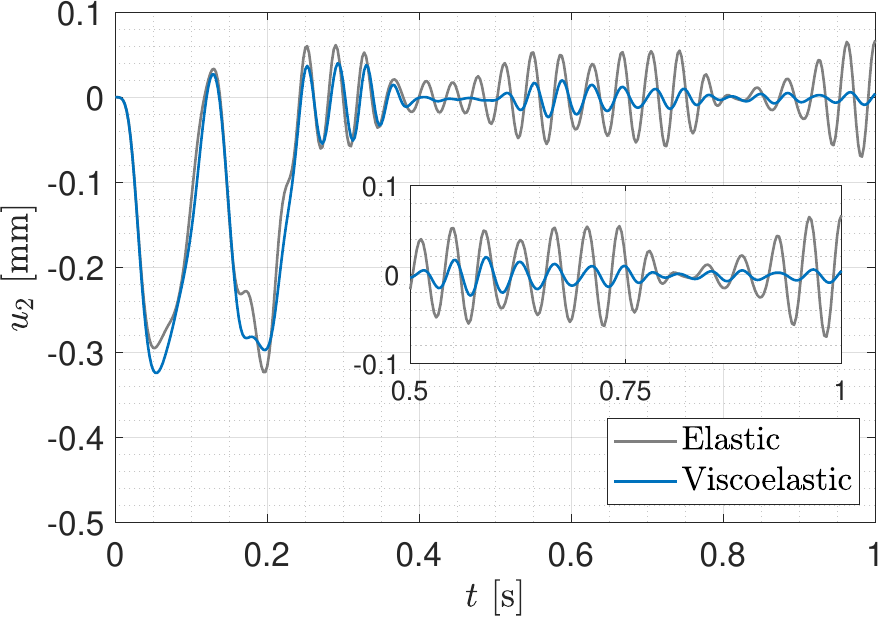}}
\hspace{0.3cm}
\subfigure[Displacement component along $x_2$-axis for the meta-material with $\psi=\pi/6$.\label{fig:metamat_CURVO_U2}]
{\includegraphics[width=0.47\textwidth]{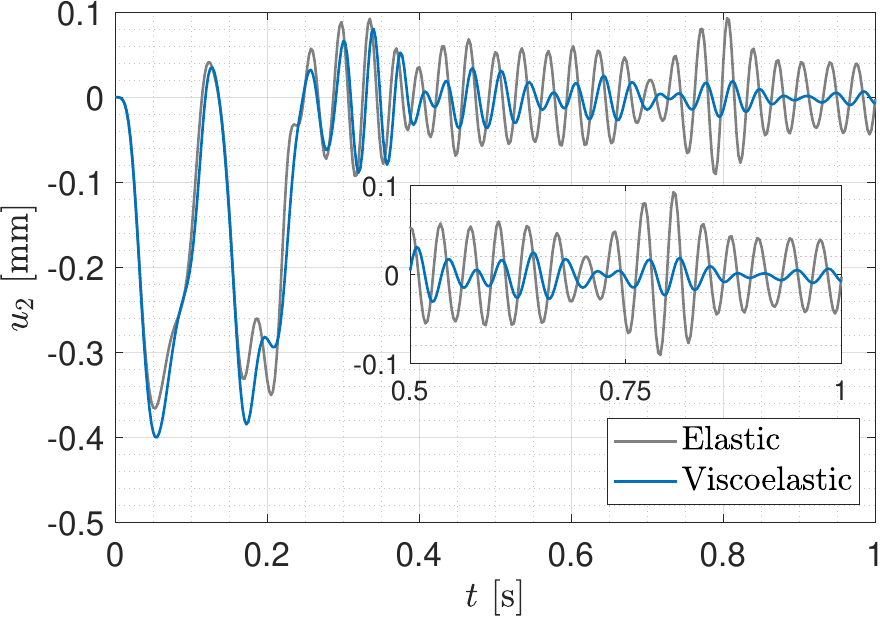}}
\centering
\subfigure[Displacement component along $x_3$-axis for the meta-material with $\psi=0$.\label{fig:metamat_RETT_U3}]
{\includegraphics[width=0.47\textwidth]{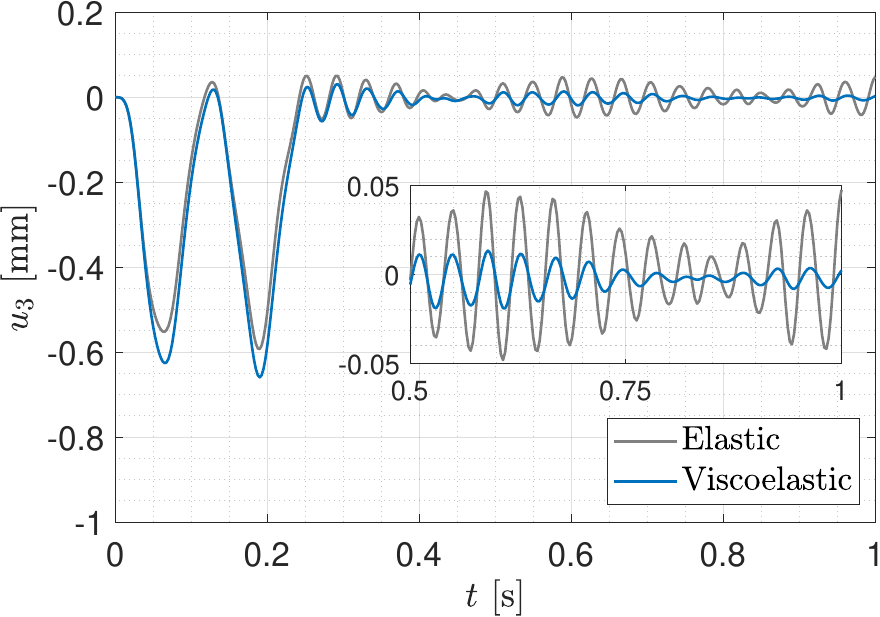}}
\hspace{0.3cm}
\subfigure[Displacement component along $x_3$-axis for the meta-material with $\psi=\pi/6$.\label{fig:metamat_CURVO_U3}]
{\includegraphics[width=0.47\textwidth]{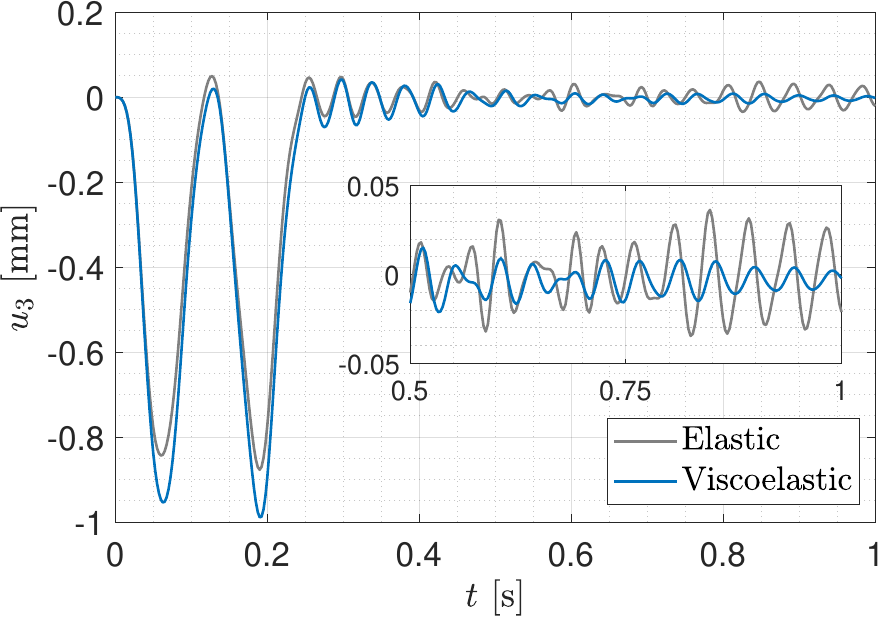}}
\caption{Comparison of the displacements of the auxetic meta-materials made of elastic (solid grey) and viscoelastic (solid blue) materials.\label{fig:U_metamat}}
\end{figure}

\begin{figure}
\centering
\subfigure[3D view for the case $\psi=0$.\label{fig:metamat_RETT_3D}]
{\includegraphics[width=0.47\textwidth]{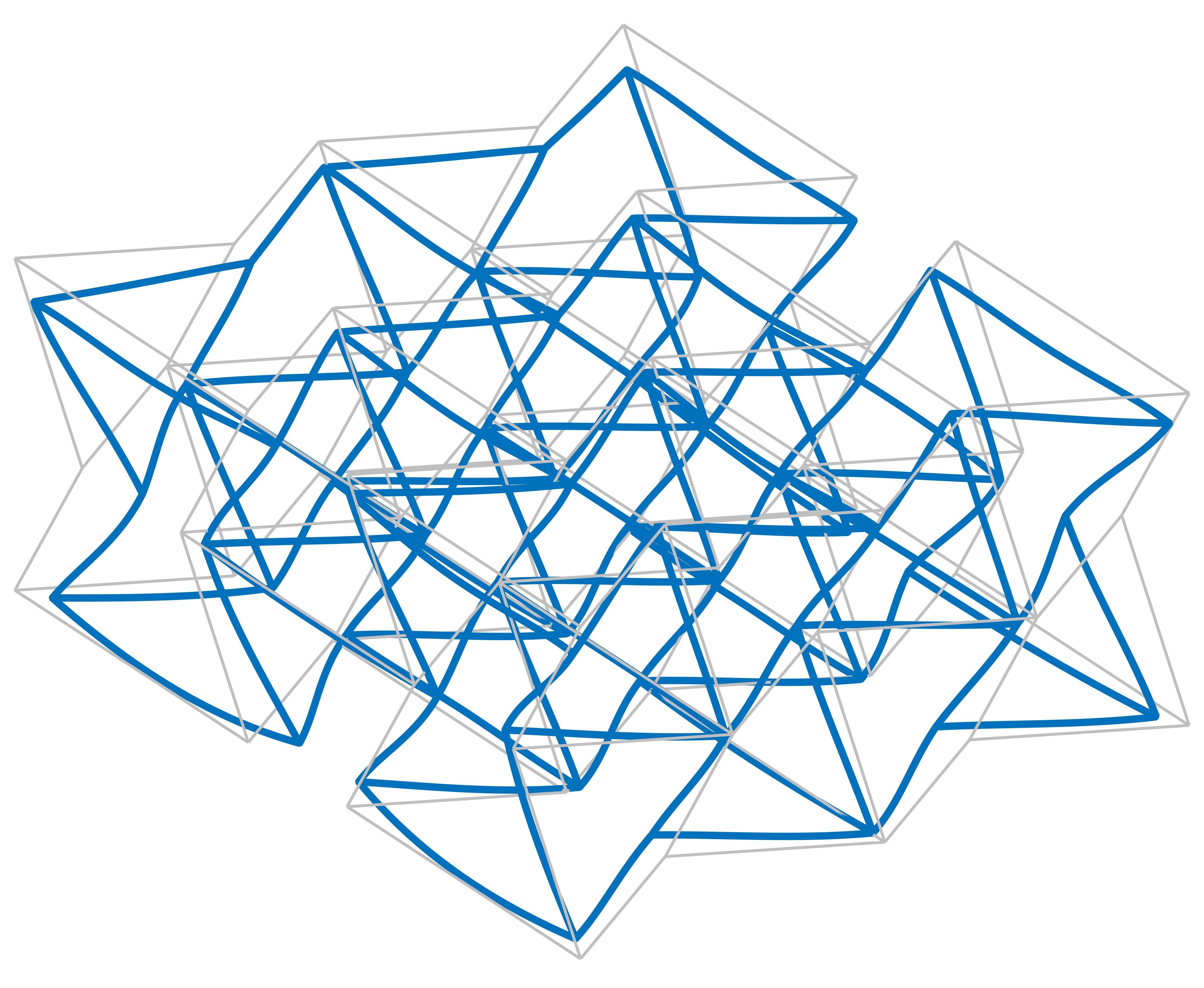}}
\hspace{0.3cm}
\subfigure[Planar $(x_1-x_2)$ view for the case $\psi=0$.\label{fig:metamat_RETT_plane}]
{\includegraphics[width=0.43\textwidth]{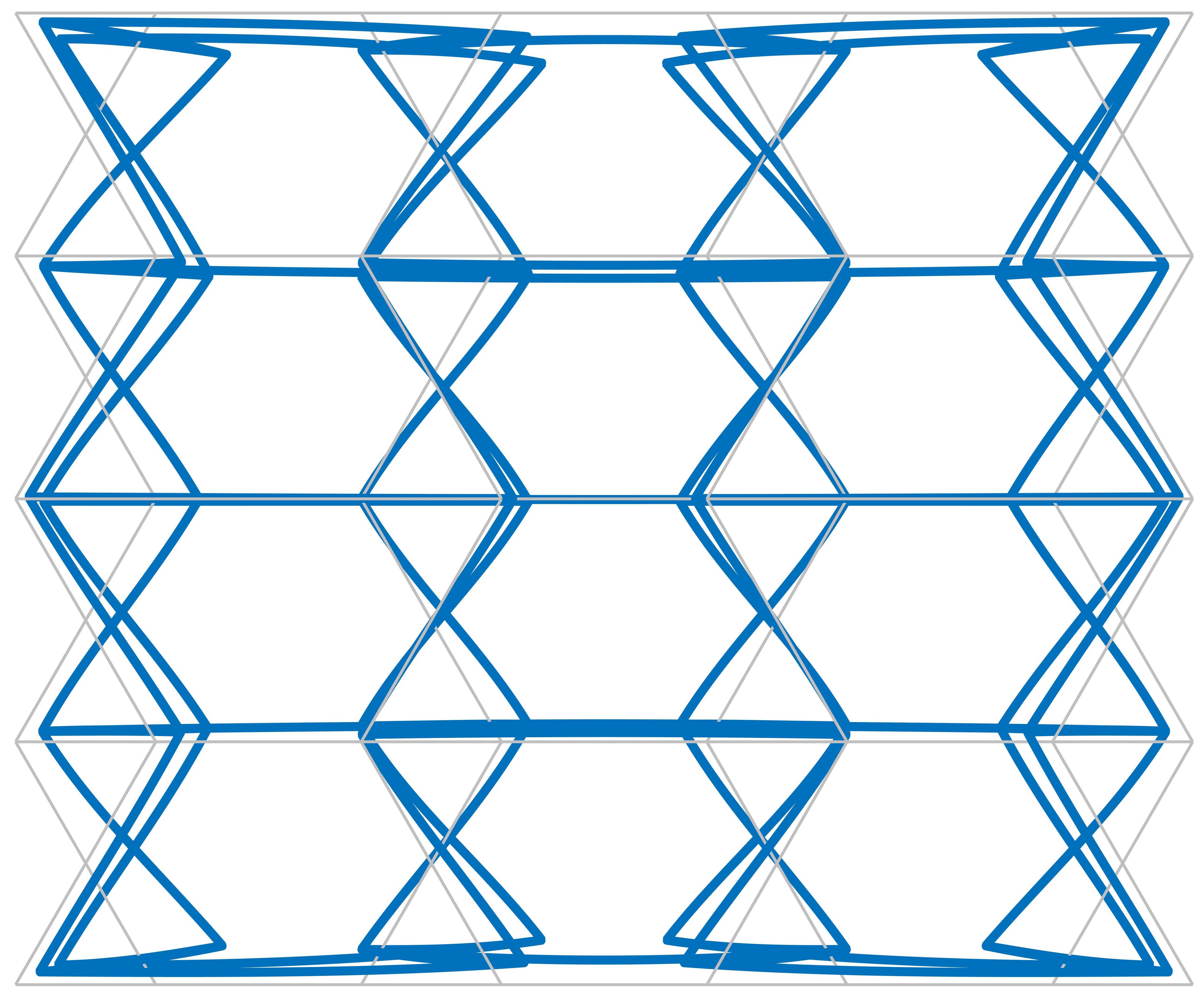}}
\centering
\subfigure[3D view for the case $\psi=\pi/6$.\label{fig:metamat_CURVO_3D}]
{\includegraphics[width=0.47\textwidth]{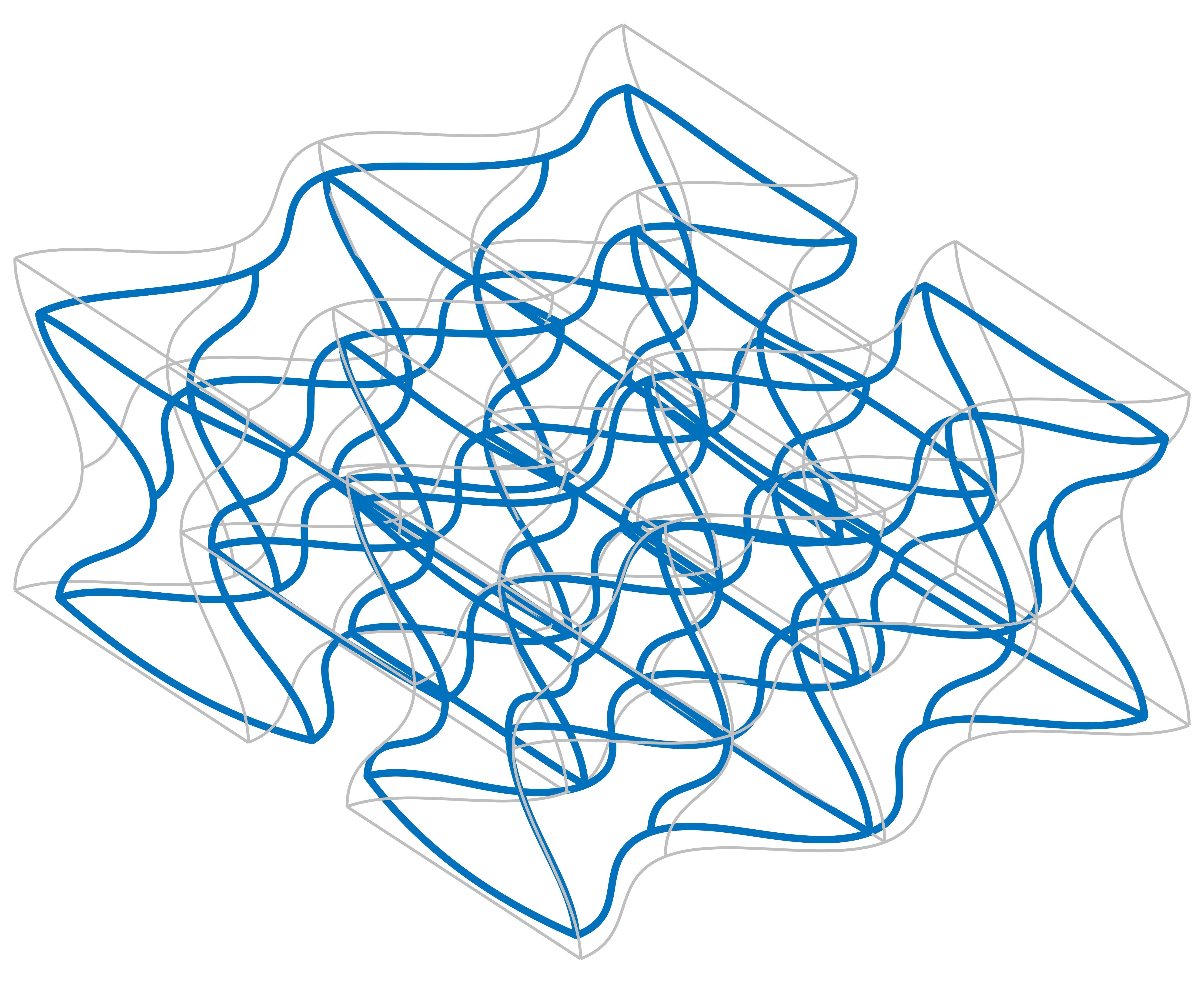}}
\hspace{0.3cm}
\subfigure[Planar $(x_1-x_2)$ view for the case $\psi=\pi/6$.\label{fig:metamat_CURVO_plane}]
{\includegraphics[width=0.43\textwidth]{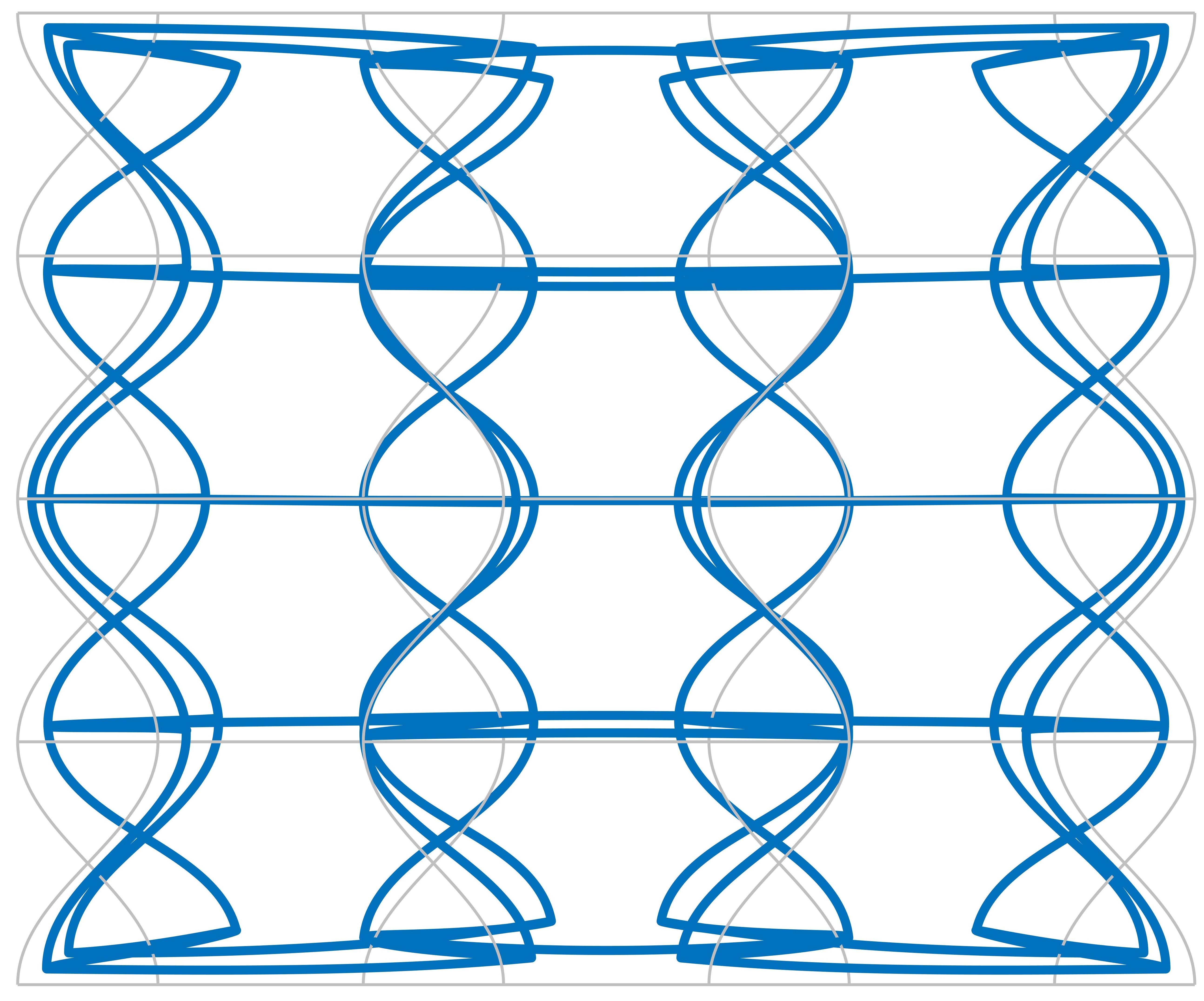}}
\caption{Deformed at $t=\SI{0.1875}{s}$ (blue) and undeformed (light grey) configurations for the viscoelastic meta-material with $\psi=0$ (top) and $\psi=\pi/6$ (bottom).
\label{fig:metamat_deformed_plot}}
\end{figure}
}

\section{Conclusions\label{sec:conclusions}}
In response to the increasing interest in the design of mechanical meta-materials and any other programmable engineering device whose dynamic response is mainly determined by the tunable topology and geometry of the building inner cells, in addition to the constituent material properties, we propose an efficient and flexible computational formulation for the dynamics of complex-shaped spatial shear-deformable beams and beam structures made of viscoelastic material. 
The method consists in an extension of the quasi-static model to the dynamic case, where an $\SO3$-consistent implicit time-integration scheme is used both for accelerations and rate-dependent viscoelastic strain measures. 
Viscoelasticity is modeled through the generalized Maxwell model with a tunable number of rheological elements. High efficiency and high-order space accuracy are achieved by employing an isogeometric collocation (IGA-C) method to discretize the differential equations. The method requires only one evaluation point per unknown and completely bypasses numerical integration. Computational cost is further reduced thanks to a displacement-based formulation that allows to minimize the number of unknowns. 
Several numerical applications are provided to showcase all the expected capabilities of the proposed method. In particular, we demonstrate the capabilities to manage beams with an arbitrarily curved initial geometry and that the high-order space accuracy is preserved in dynamics. 
Finally, \r{a parametric study on two meta-materials is presented to test the proposed formulation on complex multi-patch systems. Theses are a planar lattice and a three-dimensional auxetic meta-material. For both cases, straight and curved cells are considered.} With these applications, we prove that, combining the capabilities to represent and control any three-dimensional curved geometry with rate-dependent materials, the parametric design space can be considerably expanded unlocking unprecedented potentialities in the inverse design of programmable materials and devices.

\section*{Acknowledgments}
EM was partially supported by the European Union - Next Generation EU, in the context of The National Recovery and Resilience Plan, Investment 1.5 Ecosystems of Innovation, Project Tuscany Health Ecosystem (THE). (CUP: B83C22003920001).

EM was also partially supported by the National Centre for HPC, Big Data and Quantum Computing funded by the European Union within the Next Generation EU recovery plan. (CUP B83C22002830001). 

EM and GF were partially supported by the UniFI project IGA4Stent - ``Patient-tailored stents: an innovative computational isogeoemtric analysis approach for 4D printed shape changing devices''. (CUP B55F21007810001).

These supports are gratefully acknowledged.

\clearpage
\r{
\section*{Appendix A}
In this section, we provide further details about the generalized Maxwell model for one-dimensional solids. 
Starting from Eq.~\eqref{stress_viscoel}, the dissipative internal stresses are given by
\bEq
\f N_{\alp}=\B H^v_{N\alp}\dot{\f\Gam}_{N\alp} \sepr{and} \f M_{\alp}=\B H^v_{M\alp}\dot{\f K }_{M\alp}\,,\label{if_visco}\nonumber
\eEq
where $\B H^v_{N\alp}$ and $\B H^v_{M\alp}$ are diagonal viscosity matrices associated with the $\alp$th Maxwell element, while $\dot{\f\Gam}_{N\alp}$ and $\dot{\f K }_{M\alp}$ are the viscous strain rate vectors. Recalling the evolution laws (Eq.~\eqref{evol_f_m}), and assuming that the elastic response is linear, the dissipative internal stresses can be expressed as 
\bEq
\f N_{\alp}=\fr{\B H^v_{N\alp}}{\tau_\alp}(\f\Gam_N-\f\Gam_{N\alp})=\CNalp(\f\Gam_N-\f\Gam_{N\alp})\,,\nonumber
\eEq
\bEq
\f M_{\alp}=\fr{\B H^v_{M\alp}}{\tau_\alp}(\f{K}_M-\f{K}_{M\alp})=\CMalp(\f{K}_M-\f{K}_{M\alp})\,,\nonumber
\eEq
where $\CNalp$ and $\CMalp$ are the elasticity tensors associated with the $\alp$th Maxwell element introduced in Section 2.2. 
To properly compute the viscous stresses at the time $t^{n+1}=nh$, $\f N_{\alp}^{n+1}$ and $\f M_{\alp}^{n+1}$, Eq.~\eqref{evol_f_m} must be integrated in time. As discussed in Section 3.1, we perform this task using the trapezoidal rule as follows
\bAl
\f\Gam_{N\alp}^{n+1} &=\f\Gam_{N\alp}^{n}+\fr{h}{2\tau_\alp} [\f\Gam_N^{n+1}-\f\Gam_{N\alp}^{n+1}+\f\Gam_N^{n}-\f\Gam_{N\alp}^{n}]\,, \nonumber\\
\left(\fr{2\tau_\alp+h}{2\tau_\alp}\right)\f\Gam_{N\alp}^{n+1} & =\fr{h}{2\tau_\alp}\f\Gam_N^{n+1}+\left(\fr{2\tau_\alp-h}{2\tau_\alp}\right)\f\Gam_{N\alp}^{n}+\fr{h}{2\tau_\alp}\f\Gam_N^{n}\,,\nonumber\\
\f\Gam_{N\alp}^{n+1} &= \left(\fr{h}{2\tau_\alp+h}\right)\f\Gam_N^{n+1}+\left(\fr{2\tau_\alp-h}{2\tau_\alp+h}\right)\f\Gam_{N\alp}^{n}+\left(\fr{h}{2\tau_\alp+h}\right)\f\Gam_N^{n}\,,\nonumber
\end{align}
which, after collecting terms, delivers Eq.~\eqref{g_trapz}.

In a very similar way, we obtain the time-discretized viscous strain measure $\f K_{M\alp}^{n+1}$. Starting from
\beq
\f K_{M\alp}^{n+1}=\f K_{M\alp}^{n}+\fr{h}{2\tau_\alp} [\f K_M^{n+1}-\f K_{M\alp}^{n+1}+\f K_M^{n}-\f K_{M\alp}^{n}]\,,
\eeq
and following the same steps as for $\f\Gam_{N\alp}^{n+1}$, it leads to Eq.~\eqref{k_trapz}. 

In this way, the current viscous strains, $\f\Gam_{N\alp}^{n+1}$ and $\f K_{M\alp}^{n+1}$, are expressed in terms of current total strains, $\f\Gam_N^{n+1}$ and $\f K_{M}^{n+1}$, and in terms of quantities known from the previous time step, $\betgp$ and $\betkp$. That means that, at time $t^{n+1}$, only $\f\Gam_N^{n+1}$ and $\f K_{M}^{n+1}$ must be linearized, allowing to use the same strategy employed for the linear elastic formulation. For the sake of completeness, these linearizations are given in the following.
\begin{align*}
L[\f\Gam_{N\alp}^{n+1}]&=\haf\Gam_{N\alp}^{n+1}+\fr{d}{d\veps}(\f\Gam_{N\alp\veps}^{n+1})_{\veps=0}=\haf\Gam_{N\alp}^{n+1}+\left(\fr{h}{2\tau_\alp+h}\right)\fr{d}{d\veps}(\f\Gam_{N\veps}^{n+1})_{\veps=0}=\\ \nonumber
&=\haf\Gam_{N\alp}^{n+1}+\left(\fr{h}{2\tau_\alp+h}\right)\left[\hat{\fR R}\Tra{^{n+1}}\delta\f\eta,_s^{n+1}+ (\wti{\hfR R\Tra{^{n+1}}\haf c,_s^{n+1}})\delta\f\vTht^{n+1} \right]\,,\nonumber
\end{align*}

\begin{align*}
L[\f K_{M\alp}^{n+1}] &=\haf K_{M\alp}^{n+1}+\fr{d}{d\veps}(\f K_{M\alp\veps}^{n+1})_{\veps=0}=\haf K_{M\alp}^{n+1}+\left(\fr{h}{2\tau_\alp+h}\right)\fr{d}{d\veps}(\haf K_{M\veps}^{n+1})_{\veps=0}=\\ 
& =\haf K_{M\alp}^{n+1}+\left(\fr{h}{2\tau_\alp+h}\right)\left[\hti{\f K}{^{n+1}}\delta\f\vTht^{n+1}+\delta\f\vTht,_s^{n+1} \right]\,,\nonumber
\end{align*}
where the operator $\fr{d}{d\veps}(\cdot)_{\veps=0}$ performs the directional derivative with respect to the configuration increment, $\veps\in\Rn$ \cite{Marino2016a}. 
Further linearization can be obtained applying the chain rule in combination with other directional derivatives available in \cite{Marino2016a} and \cite{Marino2019b}. 
}

\clearpage
\bibliographystyle{elsarticle-num}
\bibliography{mylibrary_new_merge_final}

\begin{thebibliography}{10}
\expandafter\ifx\csname url\endcsname\relax
  \def\url#1{\texttt{#1}}\fi
\expandafter\ifx\csname urlprefix\endcsname\relax\def\urlprefix{URL }\fi
\expandafter\ifx\csname href\endcsname\relax
  \def\href#1#2{#2} \def\path#1{#1}\fi

\bibitem{Bertoldi_etal2017}
K.~Bertoldi, V.~Vitelli, J.~Christensen, M.~van Hecke, {Flexible mechanical
  metamaterials}, Nature Reviews Materials 2~(11) (2017) 17066.

\bibitem{Xue_etal2023}
T.~Xue, S.~Adriaenssens, S.~Mao, {Learning the nonlinear dynamics of mechanical
  metamaterials with graph networks}, International Journal of Mechanical
  Sciences 238 (2023) 107835.

\bibitem{Deng_etal2021JAP}
B.~Deng, J.~R. Raney, K.~Bertoldi, V.~Tournat, {Nonlinear waves in flexible
  mechanical metamaterials}, Journal of Applied Physics 130~(4) (2021) 40901.

\bibitem{Deng_etal2021JMPS}
B.~Deng, J.~Li, V.~Tournat, P.~K. Purohit, K.~Bertoldi, {Dynamics of mechanical
  metamaterials: A framework to connect phonons, nonlinear periodic waves and
  solitons}, Journal of the Mechanics and Physics of Solids 147 (2021) 104233.

\bibitem{Zheng_etal2023}
X.~Zheng, X.~Zhang, T.-T. Chen, I.~Watanabe, X.~Zheng, I.~Watanabe, X.~Zhang,
  T.-T. Chen, {Deep Learning in Mechanical Metamaterials: From Prediction and
  Generation to Inverse Design}, Advanced Materials 35~(45) (2023) 2302530.

\bibitem{Karathanasopoulos_etal2017}
N.~Karathanasopoulos, H.~Reda, J.~francois Ganghoffer, {Designing
  two-dimensional metamaterials of controlled static and dynamic properties},
  Computational Materials Science 138 (2017) 323--332.

\bibitem{Zhu_etal2021}
R.~Zhu, B.~Mao, Q.~Zhao, Z.~Wang, X.~Han, Y.~Yang, H.~Li, {Dynamic
  characteristics of Mn-Cu high damping alloy subjected to impact load},
  Advances in Mechanical Engineering 13~(4) (2021) 16878140211013616.

\bibitem{Portela_etal2021}
C.~M. Portela, B.~W. Edwards, D.~Veysset, Y.~Sun, K.~A. Nelson, D.~M. Kochmann,
  J.~R. Greer, {Supersonic impact resilience of nanoarchitected carbon}, Nature
  Materials 20~(11) (2021) 1491--1497.

\bibitem{Simo1985}
J.~C. Simo, {A finite strain beam formulation. The three-dimensional dynamic
  problem. Part I}, Computer Methods in Applied Mechanics and Engineering
  49~(1) (1985) 55--70.

\bibitem{SimoVu-Quoc1986}
J.~C. Simo, L.~Vu-Quoc, {A three-dimensional finite-strain rod model. Part II:
  Computational aspects}, Computer Methods in Applied Mechanics and Engineering
  58~(1) (1986) 79--116.

\bibitem{Simo&Vu-Quoc1988}
J.~C. Simo, L.~Vu-Quoc, {On the dynamics in space of rods undergoing large
  motions — A geometrically exact approach}, Computer Methods in Applied
  Mechanics and Engineering 66~(2) (1988) 125--161.

\bibitem{Cardona1988}
A.~Cardona, M.~Geradin, {A beam finite element non-linear theory with finite
  rotations}, International Journal for Numerical Methods in Engineering
  26~(September 1987) (1988) 2403--2438.

\bibitem{IbraMikdad1998}
A.~Ibrahimbegovi{\'{c}}, M.~A.~L. Mikdad, {Finite rotations in dynamics of
  beams and implicit time-stepping schemes}, International Journal for
  Numerical Methods in Engineering 41~(November 1996) (1998) 781--814.

\bibitem{Jelenic1998}
G.~Jelenic, M.~A. Crisfield, {Interpolation of Rotational Variables in
  Nonlinear Dynamics of 3D Beams}, International Journal for Numerical Methods
  in Engineering 1222~(February 1997) (1998) 1193--1222.

\bibitem{Jelenic1999}
G.~Jeleni{\'{c}}, M.~Crisfield, {Geometrically exact 3D beam theory:
  implementation of a strain-invariant finite element for statics and
  dynamics}, Computer Methods in Applied Mechanics and Engineering 171~(1-2)
  (1999) 141--171.

\bibitem{Makinen2001}
J.~M{\"{a}}kinen, {Critical study of Newmark-scheme on manifold of finite
  rotations}, Computer Methods in Applied Mechanics and Engineering 191 (2001)
  817--828.

\bibitem{Romero&Armero2002}
I.~Romero, F.~Armero, {An objective finite element approximation of the
  kinematics of geometrically exact rods and its use in the formulation of an
  energy-momentum conserving scheme in dynamics}, International Journal for
  Numerical Methods in Engineering 54~(12) (2002) 1683--1716.

\bibitem{Makinen2007}
J.~M{\"{a}}kinen, {Total Lagrangian Reissner's geometrically exact beam element
  without singularities}, International Journal for Numerical Methods in
  Engineering 70~(October 2006) (2007) 1009--1048.

\bibitem{Makinen2008}
J.~M{\"{a}}kinen, {Rotation manifold SO(3) and its tangential vectors},
  Computational Mechanics 42~(6) (2008) 907--919.

\bibitem{PimentaCampelloWriggers2008}
P.~M. Pimenta, E.~M.~B. Campello, P.~Wriggers, {An exact conserving algorithm
  for nonlinear dynamics with rotational DOFs and general hyperelasticity. Part
  1: Rods}, Computational Mechanics 42~(5) (2008) 715--732.

\bibitem{Lang_etal2011}
H.~Lang, J.~Linn, M.~Arnold, {Multi-body dynamics simulation of geometrically
  exact Cosserat rods}, Multibody System Dynamics 25~(3) (2011) 285--312.

\bibitem{Bruls_etal2012}
O.~Br{\"{u}}ls, A.~Cardona, M.~Arnold, {Lie group generalized-$\alpha$ time
  integration of constrained flexible multibody systems}, Mechanism and Machine
  Theory 48 (2012) 121--137.

\bibitem{Zupan_etal2012}
E.~Zupan, M.~Saje, D.~Zupan, {Quaternion-based dynamics of geometrically
  nonlinear spatial beams using the Runge–Kutta method}, Finite Elements in
  Analysis and Design 54 (2012) 48--60.

\bibitem{Zupan_etal2013}
E.~Zupan, M.~Saje, D.~Zupan, {Dynamics of spatial beams in quaternion
  description based on the Newmark integration scheme}, Computational Mechanics
  51~(1) (2013) 47--64.

\bibitem{Sonneville_etal2014}
V.~Sonneville, A.~Cardona, O.~Br{\"{u}}ls, {Geometrically exact beam finite
  element formulated on the special Euclidean group SE(3)}, Computer Methods in
  Applied Mechanics and Engineering 268~(3) (2014) 451--474.

\bibitem{Thanh-Nam_etal2014}
T.-N. Le, J.-M. Battini, M.~Hjiaj, {A consistent 3D corotational beam element
  for nonlinear dynamic analysis of flexible structures}, Computer Methods in
  Applied Mechanics and Engineering 269 (2014) 538--565.

\bibitem{Almonacid2015}
P.~M. Almonacid, {Explicit symplectic momentum-conserving time-stepping scheme
  for the dynamics of geometrically exact rods}, Finite Elements in Analysis
  and Design 96 (2015) 11--22.

\bibitem{Weeger_etal2017}
O.~Weeger, B.~Narayanan, M.~L. Dunn, {Isogeometric collocation for nonlinear
  dynamic analysis of Cosserat rods with frictional contact}, Nonlinear
  Dynamics (2017) 1--15.

\bibitem{Zupan&Zupan2018}
E.~Zupan, D.~Zupan, {On conservation of energy and kinematic compatibility in
  dynamics of nonlinear velocity-based three-dimensional beams}, Nonlinear
  Dynamics 95~(2) (2019) 1379--1394.

\bibitem{Marino2019a}
E.~Marino, J.~Kiendl, L.~{De Lorenzis}, {Explicit isogeometric collocation for
  the dynamics of three-dimensional beams undergoing finite motions}, Computer
  Methods in Applied Mechanics and Engineering 343 (2019) 530--549.

\bibitem{Marino2019b}
E.~Marino, J.~Kiendl, L.~{De Lorenzis}, {Isogeometric collocation for implicit
  dynamics of three-dimensional beams undergoing finite motions}, Computer
  Methods in Applied Mechanics and Engineering 356 (2019) 548--570.

\bibitem{Chen_etal2022}
J.~Chen, Z.~Huang, Q.~Tian, {A multisymplectic Lie algebra variational
  integrator for flexible multibody dynamics on the special Euclidean group SE
  (3)}, Mechanism and Machine Theory 174 (2022) 104918.

\bibitem{Leyendecker_etal2006}
S.~Leyendecker, P.~Betsch, P.~Steinmann, {Objective energy–momentum
  conserving integration for the constrained dynamics of geometrically exact
  beams}, Computer Methods in Applied Mechanics and Engineering 195~(19-22)
  (2006) 2313--2333.

\bibitem{Galvanetto1996}
U.~Galvanetto, M.~A. Crisfield, {An energy-conserving co-rotational procedure
  for the dynamics of planar beam structures}, International Journal for
  Numerical Methods in Engineering (1996).

\bibitem{Boyer2011}
F.~Boyer, G.~{De Nayer}, A.~Leroyer, M.~Visonneau, {Geometrically Exact
  Kirchhoff Beam Theory: Application to Cable Dynamics}, Journal of
  Computational and Nonlinear Dynamics 6~(4) (2011) 041004.

\bibitem{Arena_2016}
A.~Arena, A.~Pacitti, W.~Lacarbonara, {Nonlinear response of elastic cables
  with flexural-torsional stiffness}, International Journal of Solids and
  Structures 87 (2016) 267--277.

\bibitem{Stroehle&Betsch2022}
T.~Str{\"{o}}hle, P.~Betsch, {A simultaneous space-time discretization approach
  to the inverse dynamics of geometrically exact strings}, International
  Journal for Numerical Methods in Engineering 123~(11) (2022) 2573--2609.

\bibitem{Schubert_etal2023}
M.~Schubert, R.~T. {Sato Mart{\'{i}}n de Almagro}, K.~Nachbagauer,
  S.~Ober-Bl{\"{o}}baum, S.~Leyendecker, {Discrete adjoint method for
  variational integration of constrained ODEs and its application to optimal
  control of geometrically exact beam dynamics}, Multibody System Dynamics
  60~(3) (2024) 447--474.

\bibitem{Firouzi_etal2024}
N.~Firouzi, S.~Lenci, M.~Amabili, T.~Rabczuk, {Nonlinear free vibrations of
  Timoshenko–Ehrenfest beams using finite element analysis and direct
  scheme}, Nonlinear Dynamics (2024).

\bibitem{Lang&Arnold2012}
H.~Lang, M.~Arnold, {Numerical aspects in the dynamic simulation of
  geometrically exact rods}, Applied Numerical Mathematics 62~(10) (2012)
  1411--1427.

\bibitem{Linn_etal2013}
J.~Linn, H.~Lang, A.~Tuganov, {Geometrically exact Cosserat rods with
  Kelvin–Voigt type viscous damping}, Mechanical Sciences 4~(1) (2013)
  79--96.

\bibitem{Giusteri_etal2021}
G.~G. Giusteri, E.~Miglio, N.~Parolini, M.~Penati, R.~Zambetti, {Simulation of
  viscoelastic Cosserat rods based on the geometrically exact dynamics of
  special Euclidean strands}, International Journal for Numerical Methods in
  Engineering 123~(2) (2022) 396--410.

\bibitem{Zhang_etal2009}
Y.~Zhang, Q.~Tian, L.~Chen, J.~Yang, {Simulation of a viscoelastic flexible
  multibody system using absolute nodal coordinate and fractional derivative
  methods}, Multibody System Dynamics 21~(3) (2009) 281--303.

\bibitem{Mohamed&Shabana_2011}
A.~N.~A. Mohamed, A.~A. Shabana, {A nonlinear visco-elastic constitutive model
  for large rotation finite element formulations}, Multibody System Dynamics
  26~(1) (2011) 57--79.

\bibitem{Bauchau&Nemani2021}
O.~A. Bauchau, N.~Nemani, {Modeling viscoelastic behavior in flexible multibody
  systems}, Multibody System Dynamics 51~(2) (2021) 159--194.

\bibitem{Audoly_etal2013}
B.~Audoly, N.~Clauvelin, P.~T. Brun, M.~Bergou, E.~Grinspun, M.~Wardetzky, {A
  discrete geometric approach for simulating the dynamics of thin viscous
  threads}, Journal of Computational Physics 253 (2013) 18--49.

\bibitem{Lestringant_etal2020}
C.~Lestringant, B.~Audoly, D.~M. Kochmann, {A discrete, geometrically exact
  method for simulating nonlinear, elastic and inelastic beams}, Computer
  Methods in Applied Mechanics and Engineering 361 (2020) 112741.

\bibitem{Glaesener_etal2021}
R.~N. Glaesener, J.-H. Bastek, F.~Gonon, V.~Kannan, B.~Telgen,
  B.~Sp{\"{o}}ttling, S.~Steiner, D.~M. Kochmann, {Viscoelastic truss
  metamaterials as time-dependent generalized continua}, Journal of the
  Mechanics and Physics of Solids 156 (2021) 104569.

\bibitem{Amabili_etal2022}
M.~Amabili, G.~Ferrari, M.~H. Ghayesh, C.~Hameury, H.~{Hena Zamal}, {Nonlinear
  vibrations and viscoelasticity of a self-healing composite cantilever beam:
  Theory and experiments}, Composite Structures 294 (2022) 115741.

\bibitem{Marino_etal2020}
E.~Marino, S.~F. Hosseini, A.~Hashemian, A.~Reali, {Effects of parameterization
  and knot placement techniques on primal and mixed isogeometric collocation
  formulations of spatial shear-deformable beams with varying curvature and
  torsion}, Computers and Mathematics with Applications 80~(11) (2020)
  2563--2585.

\bibitem{Ignesti_etal2023}
D.~Ignesti, G.~Ferri, F.~Auricchio, A.~Reali, E.~Marino, {An improved
  isogeometric collocation formulation for spatial multi-patch shear-deformable
  beams with arbitrary initial curvature}, Computer Methods in Applied
  Mechanics and Engineering 403 (2023) 115722.

\bibitem{Ferri_etal2023}
G.~Ferri, D.~Ignesti, E.~Marino, {An efficient displacement-based isogeometric
  formulation for geometrically exact viscoelastic beams}, Computer Methods in
  Applied Mechanics and Engineering 417 (2023) 116413.

\bibitem{Auricchio2010}
F.~Auricchio, L.~{Beir{\~{a}}o Da Veiga}, T.~J.~R. Hughes, A.~Reali,
  G.~Sangalli, {Isogeometric Collocation Methods}, Mathematical Models and
  Methods in Applied Sciences 20~(11) (2010) 2075--2107.

\bibitem{Auricchio2012}
F.~Auricchio, L.~{Beir{\~{a}}o da Veiga}, T.~J.~R. Hughes, A.~Reali,
  G.~Sangalli, {Isogeometric collocation for elastostatics and explicit
  dynamics}, Computer Methods in Applied Mechanics and Engineering 249-252
  (2012) 2--14.

\bibitem{Fahrendorf_etal2022}
F.~Fahrendorf, S.~Shivanand, B.~V. Rosic, M.~S. Sarfaraz, T.~Wu, L.~{De
  Lorenzis}, H.~G. Matthies, {Collocation Methods and Beyond in Non-linear
  Mechanics}, Springer International Publishing, Cham, 2022, pp. 449--504.

\bibitem{Hughes2005a}
T.~Hughes, J.~Cottrell, Y.~Bazilevs, {Isogeometric analysis: CAD, finite
  elements, NURBS, exact geometry and mesh refinement}, Computer Methods in
  Applied Mechanics and Engineering 194~(39-41) (2005) 4135--4195.

\bibitem{Cottrell2009}
J.~A. Cottrell, T.~J.~R. Hughes, Y.~Bazilevs, {Isogeometric Analysis: Toward
  Integration of CAD and FEA}, Wiley, 2009.

\bibitem{Schillinger2013}
D.~Schillinger, J.~Evans, A.~Reali, M.~Scott, T.~J.~R. Hughes, {Isogeometric
  collocation: Cost comparison with Galerkin methods and extension to adaptive
  hierarchical NURBS discretizations}, Computer Methods in Applied Mechanics
  and Engineering 267 (2013) 170--232.

\bibitem{Gomez2014}
H.~Gomez, A.~Reali, G.~Sangalli, {Accurate, efficient, and (iso)geometrically
  flexible collocation methods for phase-field models}, Journal for
  Computational Physics 262 (2014) 153--171.

\bibitem{DeLorenzis2015}
L.~{De Lorenzis}, J.~Evans, T.~Hughes, A.~Reali, {Isogeometric collocation:
  Neumann boundary conditions and contact}, Computer Methods in Applied
  Mechanics and Engineering 284 (2015) 21--54.

\bibitem{Kruse2015}
R.~Kruse, N.~Nguyen-Thanh, L.~{De Lorenzis}, T.~Hughes, {Isogeometric
  collocation for large deformation elasticity and frictional contact
  problems}, Computer Methods in Applied Mechanics and Engineering 296 (2015)
  73--112.

\bibitem{Gomez&DeLorenzis2016}
H.~Gomez, L.~{De Lorenzis}, {The variational collocation method}, Computer
  Methods in Applied Mechanics and Engineering 309 (2016) 152--181.

\bibitem{Auricchio2013}
F.~Auricchio, L.~{Beir{\~{a}}o da Veiga}, J.~Kiendl, C.~Lovadina, A.~Reali,
  {Locking-free isogeometric collocation methods for spatial Timoshenko rods},
  Computer Methods in Applied Mechanics and Engineering 263 (2013) 113--126.

\bibitem{Kiendl2015a}
J.~Kiendl, F.~Auricchio, T.~Hughes, A.~Reali, {Single-variable formulations and
  isogeometric discretizations for shear deformable beams}, Computer Methods in
  Applied Mechanics and Engineering 284 (2015) 988--1004.

\bibitem{KiendlAuricchioReali2017}
J.~Kiendl, F.~Auricchio, A.~Reali, {A displacement-free formulation for the
  Timoshenko beam problem and a corresponding isogeometric collocation
  approach}, Meccanica (2017) 1--11.

\bibitem{Reali2015}
A.~Reali, H.~Gomez, {An isogeometric collocation approach for Bernoulli-Euler
  beams and Kirchhoff plates}, Computer Methods in Applied Mechanics and
  Engineering 284 (2015) 623--636.

\bibitem{Kiendl2015}
J.~Kiendl, F.~Auricchio, L.~{Beir{\~{a}}o da Veiga}, C.~Lovadina, A.~Reali,
  {Isogeometric collocation methods for the Reissner-Mindlin plate problem},
  Computer Methods in Applied Mechanics and Engineering 284 (2015) 489--507.

\bibitem{KiendlMarinoDeLorenzis2017}
J.~Kiendl, E.~Marino, L.~{De Lorenzis}, {Isogeometric collocation for the
  Reissner-Mindlin shell problem}, Computer Methods in Applied Mechanics and
  Engineering 325 (2017) 645--665.

\bibitem{Maurin_elal2018}
F.~Maurin, F.~Greco, L.~Coox, D.~Vandepitte, W.~Desmet, {Isogeometric
  collocation for Kirchhoff-Love plates and shells}, Computer Methods in
  Applied Mechanics and Engineering 329 (2018) 396--420.

\bibitem{Maurin_etal2018b}
F.~Maurin, F.~Greco, S.~Dedoncker, W.~Desmet, {Isogeometric analysis for
  nonlinear planar Kirchhoff rods: Weighted residual formulation and
  collocation of the strong form}, Computer Methods in Applied Mechanics and
  Engineering (2018).

\bibitem{Evans_etal2018}
J.~A. Evans, R.~R. Hiemstra, T.~J.~R. Hughes, A.~Reali, {Explicit higher-order
  accurate isogeometric collocation methods for structural dynamics}, Computer
  Methods in Applied Mechanics and Engineering 338 (2018) 208--240.

\bibitem{Fahrendorf_etal2020}
F.~Fahrendorf, S.~Morganti, A.~Reali, T.~J. Hughes, L.~D. Lorenzis, {Mixed
  stress-displacement isogeometric collocation for nearly incompressible
  elasticity and elastoplasticity}, Computer Methods in Applied Mechanics and
  Engineering 369 (2020) 113112.

\bibitem{Torre_etal2023}
M.~Torre, S.~Morganti, A.~Nitti, M.~D. de~Tullio, F.~S. Pasqualini, A.~Reali,
  {Isogeometric mixed collocation of nearly-incompressible electromechanics in
  finite deformations for cardiac muscle simulations}, Computer Methods in
  Applied Mechanics and Engineering 411 (2023) 116055.

\bibitem{Marino2016a}
E.~Marino, {Isogeometric collocation for three-dimensional geometrically exact
  shear-deformable beams}, Computer Methods in Applied Mechanics and
  Engineering 307 (2016) 383--410.

\bibitem{Weeger2017}
O.~Weeger, S.-K. Yeung, M.~L. Dunn, {Isogeometric collocation methods for
  Cosserat rods and rod structures}, Computer Methods in Applied Mechanics and
  Engineering 316 (2017) 100--122.

\bibitem{Marino2017b}
E.~Marino, {Locking-free isogeometric collocation formulation for
  three-dimensional geometrically exact shear-deformable beams with arbitrary
  initial curvature}, Computer Methods in Applied Mechanics and Engineering 324
  (2017) 546--572.

\bibitem{Weeger_etal2022}
O.~Weeger, D.~Schillinger, R.~M{\"{u}}ller, {Mixed isogeometric collocation for
  geometrically exact 3D beams with elasto-visco-plastic material behavior and
  softening effects}, Computer Methods in Applied Mechanics and Engineering 399
  (2022) 115456.

\bibitem{Bishop1975}
R.~L. Bishop, {There is More than One Way to Frame a Curve}, The American
  Mathematical Monthly 82~(3) (1975) 246.

\bibitem{Christensen2013}
R.~M. Christensen, {Theory of Viscoelasticity: Second Edition}, Dover Civil and
  Mechanical Engineering, Dover Publications, 2013.

\bibitem{Simo&Wong1991}
J.~C. Simo, K.~K. Wong, {Unconditionally stable algorithms for rigid body
  dynamics that exactly preserve energy and momentum}, International Journal
  for Numerical Methods in Engineering 31~(1) (1991) 19--52.

\bibitem{Gravouil&Combescure2001}
A.~Gravouil, A.~Combescure, {Multi-time-step explicit-implicit method for
  non-linear structural dynamics}, International Journal for Numerical Methods
  in Engineering 50~(1) (2001) 199--225.

\bibitem{Wan_etal2022}
M.~Wan, K.~Yu, H.~Sun, {4D printed programmable auxetic metamaterials with
  shape memory effects}, Composite Structures 279 (2022) 114791.

\end{thebibliography}
\end{document}